\documentclass[aip,graphicx,eqsecnum]{revtex4-1}

\usepackage{amssymb}
\usepackage{amsthm}
\usepackage{amsmath}
\usepackage[mathscr]{eucal}
\usepackage{graphicx}
\usepackage{epsfig}
\usepackage{graphics}
\newcommand{\Sumprime}{\mathop{{\sum}^{\prime}}}

\draft 

\begin{document}


\title{(Pre-)Hilbert spaces in twistor quantization} 



\author{Shinichi Deguchi$\,$}
\email[]{deguchi@phys.cst.nihon-u.ac.jp}
\affiliation{Institute of Quantum Science, College of Science and Technology, 
Nihon University, Tokyo 101-8308, Japan}

\author{Jun-ichi Note$\,$}
\email[]{note@phys.cst.nihon-u.ac.jp}
\affiliation{Graduate School of Quantum Science and Technology, 
Nihon University, Tokyo 101-8308, Japan}

\date{\today}

\begin{abstract}
In twistor theory, the canonical quantization procedure, called twistor quantization, is 
performed with the twistor operators represented  
as $\hat{Z}^{A}=Z^{A}\;\!(\in \Bbb{C})$ 
and $\hat{\bar{Z}}{}_{A} =-\partial/\partial Z^{A}$.  
However, it has not been clarified what kind of function spaces this representation 
is valid in. 
In the present paper, we intend to find appropriate (pre-)Hilbert spaces  
in which the above representation is realized as an adjoint pair of operators. 
To this end, we define an inner product for the helicity eigenfunctions 
by an integral over the product space of the circular space $S^{1}$ 
and the upper half of projective twistor space. 
Using this inner product, we define a Hilbert space in some particular case and 
indefinite-metric pre-Hilbert spaces in other particular cases, 
showing that the above-mentioned representation is valid in these spaces. 
It is also shown that only the Penrose transform in the first particular case 
yields positive-frequency massless fields without singularities, 
while the Penrose transforms in the other particular cases yield 
positive-frequency massless fields with singularities. 
\end{abstract}

\pacs{03.65.-w, 02.40.Tt, 03.65.Pm}

\maketitle 


\section{Introduction}

Twistor theory was first proposed by Penrose in 1967 as a novel approach 
to finding a unified framework for general relativity and quantum physics, 
aiming at establishing a theory of quantum gravity.\cite{Pen1}   
In twistor theory,\cite{PM, PR, HT, Tak, Hug, Pen2, Pen3}  
a complex space called twistor space is considered to be 
a primary object for expressing physics, 
while 4-dimensional space-time is treated as a secondary object. 
One of the common motivations in early studies on twistor theory is thus to describe 
4-dimensional space-time, gravity, and even the elementary particles in an equal footing 
on the basis of the complex geometry of twistor space. 
Such an ambitious attempt in twistor theory has been summarized by Penrose himself as 
the {\em twistor programme}.\cite{Pen2, Pen3}

Although twistor theory has provided various interesting ideas, it cannot be said that 
this theory has succeeded at physics in accordance with the twistor programme. 
From the viewpoint of physics, 
recent impressive progress related to twistor theory is only  
the discovery of a twistor string theory by Witten,\cite{Wit} 
which leads to the twistor approach to explaining scattering amplitudes in 
Yang-Mills theory.\cite{CSW1, CSW2}  
(An earlier twistor approach to Yang-Mills scattering amplitudes was considered by Nair.\cite{Nai}) 
On the other hand, 
twistor theory has yielded skillful geometrical tools for solving 
nonlinear partial differential equations such as the anti-self-dual Yang-Mills equation,\cite{ADHM, AHS} 
the anti-self-dual equations for gravity \cite{Pen4} and the Bogomolny equation.\cite{Hit} 
Also, there have been many other mathematical developments in twistor theory; 
see, e.g., Refs. 17-20, and 24.

It seems that one of the reasons why twistor theory has not developed well in physics 
and therefore the twistor programme has not been accomplished is 
that the quantum-theoretical framework of twistor theory 
has not yet been established sufficiently. 
In fact, in comparison with the standard quantum theory, 
mathematical settings such as Hilbert spaces have not been investigated satisfactorily 
in the canonical quantization procedure in twistor theory, called twistor quantization.   
To be precise, in twistor quantization, 
the twistor operators $\hat{Z}^{A}$ ($A=0, 1, 2, 3$) 
and their adjoint twistor operators $\hat{\bar{Z}}_{A}$ 
are naively represented  
as $\hat{Z}^{A}=Z^{A}$ and $\hat{\bar{Z}}{}_{A} =-\partial/\partial Z^{A}$, 
with the twistor variables $Z^{A}$\:\! $(\in \Bbb{C})$. 
However, it has not been clarified what kind of function spaces this representation is valid in. 
One of the main purposes of this paper is to find appropriate function spaces 
(or more precisely, appropriate (pre-)Hilbert spaces)  
in which the above representation holds true as an adjoint pair of operators.

Until now, there have been a few attempts to define Hilbert spaces in twistor quantization. 
In fact, Penrose gave an inner product of two holomorphic functions of $Z^{A}$ 
that have the same degree of homogeneity.\cite{Pen5, PM}
With this inner product, Penrose defined a Hilbert space and showed that the representation 
$\hat{Z}^{A}=Z^{A}$, $\hat{\bar{Z}}{}_{A} =-\partial/\partial Z^{A}$ 
is valid on this space.  
However, in his argument, the details on the inner product, such as the finiteness of the inner product,  
are unclear. Hence, there is room to doubt the presence of the Hilbert space. 
Penrose's inner product was modified by himself 
so that it can directly be derived from the scalar product between two massless fields 
in 4-dimensional space-time.\cite{Pen6} 
Even after the modification, 
the representation $\hat{Z}^{A}=Z^{A}$, $\hat{\bar{Z}}{}_{A} =-\partial/\partial Z^{A}$  
holds, but the details on the inner product still remain unclear.

A mathematically elegant method for defining an inner product in twistor theory  
has been studied by Eastwood and co-workers.\cite{EG, BE}   
This approach uses cohomologies skillfully in such a manner that 
consistency with the Penrose transform is manifestly ensured. 
Using the twistor elementary states and their density, Eastwood and Pilato showed 
positive definiteness of the $\mathrm{U}(p,q)$-invariant inner product 
in the cohomological formulation.\cite{EP}  
Another cohomological approach was given by M\"{u}ller to 
obtain a $\mathrm{SU}(2,2)$-invariant inner product.\cite{Mul}  
In these cohomological approaches, however, 
representations of the twistor operators $\hat{Z}^{A}$ and $\hat{\bar{Z}}{}_{A}$ are not still considered.

In the present paper, we propose an {\em alternative} inner product of two holomorphic functions of $Z^{A}$.  
Here the two functions may have different degrees of homogeneity.  
Essentially, our approach follows the construction of ordinary quantum mechanics, 
without referring to cohomologies.  
To define the inner product, we first construct linear combinations of $\hat{Z}^{A}$, 
denoted later by $\hat{a}{}^{A}$,  
and linear combinations of $\hat{\bar{Z}}{}_{A}$, denoted later by $\hat{\bar{a}}{}^{\dot{A}}$,  
in such a manner that $\hat{a}{}^{A}$ and $\hat{\bar{a}}{}^{\dot{A}}$ 
satisfy a Weyl-Heisenberg algebra of indefinite-metric type. 
The commutation relations of this algebra are unitarily equivalent to 
what $\hat{Z}^{A}$ and $\hat{\bar{Z}}{}_{A}$ satisfy.  
Next, we provide a coherent state \cite{Kur, NO} defined as a simultaneous eigenstate of  
the operators $\hat{\bar{a}}{}^{\dot{A}}$ 
and consider the helicity eigenvalue equation written in terms of 
$\hat{a}{}^{A}$ and $\hat{\bar{a}}{}^{\dot{A}}$. 
This equation can easily be solved in the coherent-state basis to 
obtain the helicity eigenvalues and their corresponding eigenfunctions.  
In the present paper, 
we assume that the holomorphic parts of the helicity eigenfunctions are transformed  
into positive-frequency massless fields in complexified Minkowski space via  
the Penrose transform.\cite{PM, PR, HT, Tak}   
(The holomorphic parts are precisely the so-called twistor functions.) 
This assumption is realized if the holomorphic parts are functions on the upper half of 
twistor space. Taking into account this fact, 
we define an inner product of two arbitrary helicity eigenfunctions 
by an integral over the product space of the circular space $S^{1}$ 
and the upper half of projective twistor space. 
(This inner product can also be regarded as the one defined for the corresponding 
holomorphic parts.)  
Carrying out the integration in the inner product, we obtain an expression  
that includes the orthogonality condition for the helicity eigenfunctions 
and a multiplicative factor consisting of gamma functions. 
The multiplicative factor is evaluated by making use of the method of analytic continuation 
for the gamma function. 
We particularly examine the inner product for the helicity eigenfunctions 
each of whose holomorphic parts has singularities on two hyperplanes in twistor space. 
Such holomorphic parts are especially important in twistor theory 
from a practical viewpoint related to twistor diagrams.\cite{PM, Pen6, HH, Hod1, Hod2}  
It is then shown that the helicity eigenfunctions in a particular case can be normalized  
to unity, while the helicity eigenfunctions in other particular cases can be normalized  
to either $1$ or $-1$.

In our approach, 
a Hilbert space for twistor quantization is defined as a set of  
the linear combinations of the helicity eigenfunctions in the first particular case mentioned above. 
In each of the other particular cases, it is possible to 
define an indefinite-metric pre-Hilbert space (or an indefinite inner product space) as a set of 
the finite linear combinations of the relevant helicity eigenfunctions.  
We show that the twistor operators represented as 
$\hat{Z}^{A}=Z^{A}$ and 
$\hat{\bar{Z}}{}_{A} =-\partial/\partial Z^{A} +\bar{Z}_{A}/2$ 
are realized, in each of the (pre-)Hilbert spaces, as an adjoint pair of operators. 
Then, it is seen that $\hat{\bar{Z}}{}_{A} =-\partial/\partial Z^{A}$ 
is recognized as the adjoint operator of $\hat{Z}^{A}=Z^{A}$ 
by choosing the holomorphic parts of the helicity eigenfunctions 
to be basis functions, instead of the helicity eigenfunctions themselves. 
In this way, we can define (pre-)Hilbert spaces appropriate for twistor quantization.

We also perform the Penrose transforms 
\cite{PM, PR, HT, Tak, Hug} of twistor functions 
in each of the particular cases to find 
the corresponding positive-frequency massless fields in complexified Minkowski space. 
We point out that only the massless fields derived in the first particular case 
have no singularities, while those derived in the other particular cases have singularities.

The present paper is organized as follows.  
In Sec. II, we briefly review twistor quantization by following popular literature on 
twistor theory. 
Section III provides a coherent state for twistor operators and gives the representation of 
twistor operators with respect to the coherent-state basis. 
In Sec. IV, we consider the helicity eigenvalue equation and solve it in the coherent-state basis. 
It is verified there that the helicity eigenfunctions are simultaneous eigenfunctions of  
the Cartan generators of $\mathrm{SU}(2,2)$. 
In Sec. V, we propose an inner product defined for the helicity eigenfunctions 
and examine it in particular cases after using the method of analytic continuation 
for the gamma function. 
In Sec. VI, we define (pre-)Hilbert spaces in twistor quantization and show that the adjointness 
relations between twistor operators are valid in these spaces.  
In Sec. VII, we perform the Penrose transforms of the simplest twistor functions in each of the 
particular cases and investigate singularities of the massless fields derived by these transforms. 
Finally, Sec. VIII is devoted to a summary and discussion. 
Appendix A provides the Schwinger representation of the $\mathrm{SU}(2,2)$ Lie algebra. 
Appendix B demonstrates the Penrose transform of a general twistor function in the first 
particular case.

\section{Brief review of twistor quantization}

In this section, we briefly review the twistor quantization procedure  
explained in Refs. 2, 3, and 6-8.

Let $Z^{A}$ ($A=0, 1, 2, 3$) be a twistor and $\bar{Z}_{A}$ its dual twistor. 
In terms of 2-component spinors, $Z^{A}$ and $\bar{Z}_{A}$ are expressed as   
\begin{align}
Z{}^{A} = (\omega^{\alpha}, \pi_{\dot{\alpha}}) \, , \quad\;
\bar{Z}{}_{A} = (\bar{\pi}_{\alpha}, \bar{\omega}^{\dot{\alpha}}) \, , 
\label{2.1}
\end{align}
where $\omega{}^{\alpha}$ ($\alpha=0, 1$) and $\pi{}_{\dot{\alpha}}$ ($\dot{\alpha}=\dot{0}, \dot{1}$) 
are 2-component Weyl spinors, and 
$\bar{\omega}{}^{\dot{\alpha}}$ and $\bar{\pi}{}_{\alpha}$ are their complex conjugate spinors. 
The spinors $\omega{}^{\alpha}$ and $\pi{}_{\dot{\alpha}}$ are related by 
$\omega^{\alpha}=i z^{\alpha \dot{\alpha}} \pi_{\dot{\alpha}}$, where 
the $z^{\alpha \dot{\alpha}}$ constitute coordinates of  
a point in complexified compactified Minkowski space $\mathbb{C}\mathbf{M}^{\sharp}$. 
(Here, $\mathbf{M}$ denotes 4-dimensional Minkowski space.) 
The space coordinatized by $(Z{}^{A})$ is called twistor space and is denoted by $\mathbf{T}$. 
Twistor space is a normed complex vector space with 
the pseudo-Hermitian norm squared $\bar{Z}{}_{A} Z{}^{A}$ of signature $(2,2)$. 
With this norm squared, the helicity of a massless particle propagating in  
4-dimensional Minkowski space $\mathbf{M}$ is simply represented as 
\begin{align}
s=\frac{1}{2} \bar{Z}{}_{A} Z{}^{A} . 
\label{2.1.1}
\end{align}
The conformal group of $\mathbf{M}$ is represented linearly in $\mathbf{T}$ 
as the linear group $\mathrm{SU}(2,2)$.\cite{Yao, Pen7}  
Then, the conformal invariance of the helicity 
is evident from Eq. (\ref{2.1.1}), because $\bar{Z}{}_{A} Z{}^{A}$ is invariant 
under the $\mathrm{SU}(2,2)$ transformations. 
It can be said that twistors 
are $\mathrm{SU}(2,2)$ spinors for the conformal group of $\mathbf{M}$.

In quantizing the classical system of twistors, 
$Z^{A}$ and $\bar{Z}_{A}$ are replaced by the corresponding twistor operators 
$\hat{Z}{}^{A}$ and $\hat{\bar{Z}}{}_{A}$ satisfying the commutation relations 
\begin{subequations}
\label{2.2}
\begin{align}
\big[\hat{Z}{}^{A}, \hat{\bar{Z}}{}_{B} \big] &= \delta^{A}_{B}\,, 
\label{2.2a}
\\
\big[\hat{Z}{}^{A}, \hat{Z}{}^{B \:\!} \big] &=\big[\hat{\bar{Z}}{}_{A},\hat{\bar{Z}}{}_{B} \big]=0\,. 
\label{2.2b}
\end{align}
\end{subequations}
So-called twistor quantization is carried out on the basis of the commutation relations 
(\ref{2.2a}) and (\ref{2.2b}). 
[The expression (\ref{2.1.1}), as well as the commutation relations (\ref{2.2a}) and (\ref{2.2b}), 
can systematically be derived from the gauged Shirafuji action.\cite{Bar, BP, DEN}]  
By analogy with standard quantum mechanics, we can naively take the representation 
in which $\hat{Z}{}^{A}$ reduces to $Z^{A}$: 
\begin{align}
\hat{Z}{}^{A} \doteq Z{}^{A}, \quad \;
\hat{\bar{Z}}{}_{A} \doteq -\dfrac{\partial}{\partial Z{}^{A}}. 
\label{2.3}
\end{align}
(Here, the symbol $\doteq$ stands for ``is represented by''.) 
This representation has actually been introduced in popular literature on twistor theory. 
A wave function appropriate for the representation (\ref{2.3}) is to be holomorphic 
in $Z^{A}$. Such a wave function, $f(Z)$, is referred to as 
the twistor (wave) function. 
In the twistor quantization procedure, the helicity $s$ is also treated as an operator.  
After having considered the Weyl ordering, the helicity operator reads    
\begin{align} 
\hat{s} = \frac{1}{4}\!
\left( \hat{\bar{Z}}{}_{A} \hat{Z}{}^{A} + \hat{Z}{}^{A} \hat{\bar{Z}}{}_{A}\right) .
\label{2.4}
\end{align}
The eigenvalue equation $\hat{s}f=sf$ can be written in the representation (\ref{2.3}) as 
\begin{align}
-\dfrac{1}{2} \bigg( Z{}^{A} \dfrac{\partial}{\partial Z{}^{A}} +2 \bigg) f(Z) =sf(Z) \,, 
\label{2.4.1}
\end{align}
where $s$ is understood as a helicity eigenvalue. 
Obviously, Eq. (\ref{2.4.1}) is satisfied by a homogeneous twistor function of degree $-2s-2$. 
This degree must be an integer so that $f$ can be a single-valued function on $\mathbf{T}$. 
In this way, $s$ is restricted to integer and half-integer values.

\section{A coherent state representation of twistor operators}

From Eq. (\ref{2.1}), we see that the complex conjugate of $Z^{A}$, 
i.e., $\bar{Z}^{\dot{A}}:=\overline{Z^{A}}$  ($\dot{A}=\dot{0}, \dot{1}, \dot{2}, \dot{3}$),  
is related to $\bar{Z}_{A}$ by $\bar{Z}^{\dot{A}}=\bar{Z}_{B} J^{B\dot{A}}$, 
with the inverse metric $J^{A\dot{B}}$ on $\mathbf{T}$ defined by
\begin{align}
\big( J^{A\dot{B}} \big):=
\begin{pmatrix}
\: 0 &\; I_{2} \: \\
\: I_{2} &\; 0 \
\end{pmatrix} . 
\label{3.0}
\end{align}
Here, $I_{2}$ denotes the $2\times2$ unit matrix. 
In the twistor quantization procedure, $\bar{Z}^{\dot{A}}$ is replaced by 
the adjoint operator $\hat{\bar{Z}}{}^{\dot{A}}=\hat{\bar{Z}}_{B} J^{B\dot{A}}$ of $\hat{Z}{}^{A}$  
satisfying $[\hat{Z}{}^{A}, \hat{\bar{Z}}{}^{\dot{B}} \:\!] = J^{A\dot{B}}$. 
(At present, the adjointness relation between $\hat{Z}{}^{A}$ and $\hat{\bar{Z}}{}^{\dot{A}}$ 
is just a formality.) 
This commutation relation is, of course, essential for twistor quantization. 
However, it is inconvenient for our analysis, because $(J^{A\dot{B}})$ is not a diagonal matrix. 
Desirable commutation relations are provided for the operators 
\begin{gather}
\begin{aligned}
\hat{a}{}^{0} &:=\frac{1}{\sqrt{2}} \!\left(\hat{Z}{}^{0} + \hat{Z}{}^{2} \right) , \\
\hat{a}{}^{2} &:=\frac{1}{\sqrt{2}} \!\left(-\hat{Z}{}^{0} + \hat{Z}{}^{2} \right) , 
\end{aligned}
\quad \;
\begin{aligned}
\hat{a}{}^{1} &:=\frac{1}{\sqrt{2}} \!\left(\hat{Z}{}^{1} + \hat{Z}{}^{3} \right) , \\
\hat{a}{}^{3} &:=\frac{1}{\sqrt{2}} \!\left(-\hat{Z}{}^{1} + \hat{Z}{}^{3} \right) , 
\end{aligned}
\label{3.1}
\end{gather}
and their adjoint operators 
\begin{gather}
\begin{aligned}
\hat{\bar{a}}{}^{\dot{0}} &:=\frac{1}{\sqrt{2}} \!\left(\hat{\bar{Z}}{}_{0} + \hat{\bar{Z}}{}_{2} \right) , \\
\hat{\bar{a}}{}^{\dot{2}} &:=\frac{1}{\sqrt{2}} \!\left(\hat{\bar{Z}}{}_{0} - \hat{\bar{Z}}{}_{2} \right) , 
\end{aligned}
\quad \;
\begin{aligned}
\hat{\bar{a}}{}^{\dot{1}} &:=\frac{1}{\sqrt{2}} \!\left(\hat{\bar{Z}}{}_{1} + \hat{\bar{Z}}{}_{3} \right) , \\
\hat{\bar{a}}{}^{\dot{3}} &:=\frac{1}{\sqrt{2}} \!\left(\hat{\bar{Z}}{}_{1} - \hat{\bar{Z}}{}_{3} \right) . 
\end{aligned}
\label{3.2}
\end{gather}
In fact, using the commutation relations (\ref{2.2a}) and (\ref{2.2b}), 
we can show that 
\begin{subequations}
\label{3.3}
\begin{align}
\big[ \hat{a}{}^{A},\hat{\bar{a}}{}^{\dot{B}\:\!} \big] &= I^{A \dot{B}}\,, 
\label{3.3a}
\\
\big[\hat{a}{}^{A}, \hat{a}{}^{B \:\!} \big] &
=\big[\hat{\bar{a}}{}^{\dot{A}}, \hat{\bar{a}}{}^{\dot{B} \:\!} \big]=0\,, 
\label{3.3b}
\end{align}
\end{subequations}
where $I^{A\dot{B}}$ is the diagonal inverse metric of the form  
$(I^{A\dot{B}})= \text{diag}(1,1,-1,-1)$. 
From this, we see that $\mathbf{T}$ possesses 
the pseudo-Hermitian metric of signature $(2,2)$, defined by  
$(I_{\dot{A} B})= \text{diag}(1,1,-1,-1)$. 
The commutation relations (\ref{3.3a}) and (\ref{3.3b}) constitute a 
Weyl-Heisenberg algebra of indefinite-metric type.

Now, we construct a coherent state that is defined to be a simultaneous eigenstate of 
the operators $\hat{\bar{a}}{}^{\dot{A}}$. 
For this purpose, we first introduce the unitary operator \cite{Kur} 
\begin{align}
\hat{U} (\alpha, \bar{\alpha}) 
:= \exp\!
\left( 
\hat{\bar{a}}{}^{\dot{A}} I_{\dot{A} B} \alpha{}^{B}
- \bar{\alpha}{}^{\dot{B}}  I_{\dot{B} A} \hat{a}{}^{A}
\right) , 
\label{3.4} 
\end{align}
where $\alpha^{A}$ are complex numbers and $\bar{\alpha}{}^{\dot{A}}$ are 
their complex conjugates. 
The operator $\hat{U} = \hat{U} (\alpha, \bar{\alpha})$ 
generates translations of $\hat{a}{}^{A}$ and $\hat{\bar{a}}{}^{\dot{A}}$
in the following manner:
\begin{align}
\hat{U}{}^{\dagger}  \hat{a}{}^{A} \hat{U} 
= \hat{a}{}^{A} + \alpha{}^{A}
,\quad\;
\hat{U}{}^{\dagger}  \hat{\bar{a}}{}^{\dot{A}} \hat{U} 
= \hat{\bar{a}}{}^{\dot{A}} + \bar{\alpha}{}^{\dot{A}}.
\label{3.5}
\end{align}
We introduce the vacuum state $|\bar{0} \rangle$ specified by
\begin{align}
\hat{\bar{a}}{}^{\dot{A}} |\bar{0} \rangle = 0 \,,
\quad \langle \bar{0} | \bar{0} \rangle = 1\,.
\label{3.6}
\end{align}
Then, it is readily verified that the vector 
$|\bar{\alpha} \rangle :=\hat{U}| \bar{0} \rangle$  
fulfills the eigenvalue equation 
$\hat{\bar{a}}{}^{\dot{A}} |\bar{\alpha} \rangle 
=\bar{\alpha}{}^{\dot{A}} |\bar{\alpha} \rangle$. 
This demonstrates that $|\bar{\alpha} \rangle$ is actually a coherent state 
for the operators $\hat{\bar{a}}{}^{\dot{A}}$. 
The normalization condition $\langle \bar{\alpha} |\bar{\alpha} \rangle=1$ 
is guaranteed by the unitarity of $\hat{U}$. 
By using the Campbell-Baker-Hausdorff formula, 
$|\bar{\alpha} \rangle$ can be expressed as 
\begin{align}
| \bar{\alpha} \rangle 
=\exp \!\left( \dfrac{1}{2} \| \alpha \|^{2} \right) 
\exp \!\left( -\bar{\alpha}{}^{\dot{A}} I{}_{\dot{A} B} \hat{a}{}^{B}  \right) \!|\bar{0} \rangle \,, 
\label{3.7}  
\end{align}
where $\| \alpha \|^{2}:=\bar{\alpha}{}^{\dot{A}} I{}_{\dot{A} B} \alpha{}^{B}
=|\alpha^{0}|^{2}+|\alpha^{1}|^{2}-|\alpha^{2}|^{2}-|\alpha^{3}|^{2}$. 
(When two operators $X$ and $Y$ commute with $[X, Y]$, 
the Campbell-Baker-Hausdorff formula reads  
$ e^{X+Y}=e^{X} e^{Y} e^{-\frac{1}{2} [X, Y]}$.)  
Using $\langle \bar{0} |\hat{a}{}^{A}=0$, 
we can also show that the dual vector 
$\langle \bar{\alpha} |= \langle \bar{0} | \hat{U}{}^{\dagger}$ satisfies \cite{NO}  
\begin{subequations}
\label{3.8}
\begin{align}
\langle \bar{\alpha} | \hat{a}{}^{A} &= \alpha^{A} \langle \bar{\alpha} | \,,
\label{3.8a}
\\
\langle \bar{\alpha} | \hat{\bar{a}}{}^{\dot{A}}
&= \left( - \dfrac{\partial}{\partial \alpha{}^{B}} I{}^{B \dot{A}}
+ \dfrac{1}{2} \bar{\alpha}{}^{\dot{A}} \right) \!
\langle \bar{\alpha} | \,.
\label{3.8b}
\end{align}
\end{subequations}

The complex numbers $\alpha^{A}$ and the twistor variables  
$Z^{A}=(\omega^{\alpha}, \pi_{\dot{\alpha}})$ are related 
by the relations that are obtained by replacing $\hat{a}^{A}$ and $\hat{Z}^{A}$ in 
Eq, (\ref{3.1}) with $\alpha^{A}$ and $Z^{A}$, respectively: 
\begin{gather}
\begin{aligned}
\alpha^{0} &=\frac{1}{\sqrt{2}} \!\left(Z^{0} +Z^{2} \right) ,  \\
\alpha^{2} &=\frac{1}{\sqrt{2}} \!\left(-Z^{0} + Z^{2} \right) ,
\end{aligned}
\quad \; 
\begin{aligned}
\alpha^{1} &=\frac{1}{\sqrt{2}} \!\left(Z^{1} + Z^{3} \right) ,  \\
\alpha^{3} &=\frac{1}{\sqrt{2}} \!\left(-Z^{1} + Z^{3} \right) ,  
\end{aligned}
\label{3.9}
\end{gather}
or equivalently, 
\begin{gather}
\begin{aligned}
\alpha^{0} &=\frac{1}{\sqrt{2}} \!\left(\omega^{0} +\pi_{\dot{0}} \right) ,  \\
\alpha^{2} &=\frac{1}{\sqrt{2}} \!\left(-\omega^{0} +\pi_{\dot{0}} \right) ,
\end{aligned}
\quad \; 
\begin{aligned}
\alpha^{1} &=\frac{1}{\sqrt{2}} \!\left(\omega^{1} +\pi_{\dot{1}} \right) ,  \\
\alpha^{3} &=\frac{1}{\sqrt{2}} \!\left(-\omega^{1} +\pi_{\dot{1}}  \right) .   
\end{aligned}
\label{3.10}
\end{gather}
With these relations, it is easy to see that $\|\alpha \|^{2}=\bar{Z}_{A} Z^{A}
=\bar{\pi}_{\alpha} \omega^{\alpha} +\bar{\omega}^{\dot{\alpha}} \pi_{\dot{\alpha}}$. 
Since $\alpha^{A}$ are related to $Z^{A}$ by a unitary transformation specified by Eq. (\ref{3.9}), 
we may call $\alpha^{A}$ a twistor (defined with respect to another basis of $\mathbf{T}$). 
Correspondingly, we may call $\hat{a}{}^{A}$ and $\hat{\bar{a}}{}^{\dot{A}}$ twistor operators.

In terms of the twistor variables $Z^{A}$ and $\bar{Z}_{A}$ 
and their corresponding operators, Eq. (\ref{3.8}) can be written as   
\begin{align}
\langle \bar{Z} | \hat{Z}{}^{A} = Z^{A} \langle \bar{Z} |\,, 
\quad \;
\langle \bar{Z} | \hat{\bar{Z}}{}_{A}
= \left( -\dfrac{\partial}{\partial Z{}^{A}} + \frac{1}{2} \bar{Z}_{A} \right) \!
\langle \bar{Z} | \,, 
\label{3.11}
\end{align}
where $\langle \bar{Z}| 
:=\langle \bar{0}|\exp \!\big(-\hat{\bar{Z}}{}_{A} Z^{A} \big)
\exp \!\big( \frac{1}{2}\bar{Z}_{A} Z^{A} \big) 
\left(=\langle \bar{\alpha}| \:\!\right)$. 
Apart from the additive factor $\frac{1}{2} \bar{Z}_{A}$, 
Eq. (\ref{3.11}) leads to the representation given in Eq. (\ref{2.3}). 
If $\langle \bar{Z}|$ is defined by 
$\langle \bar{Z}|=\langle \bar{0}|\exp \!\big(-\hat{\bar{Z}}{}_{A} Z^{A} \big)$ 
without the multiplicative factor $\exp \!\big( \frac{1}{2}\bar{Z}_{A} Z^{A} \big)$,  
we have 
$\langle \bar{Z} | \hat{\bar{Z}}{}_{A}= -\partial \langle \bar{Z}|/ \partial Z^{A}$,  
and hence immediately find Eq. (\ref{2.3}). 
However, in general, this $\langle \bar{Z}|$ is not a unit vector, because it satisfies  
$\langle \bar{Z}|\bar{Z} \rangle =\exp \!\big( -\bar{Z}_{A} Z^{A} \big)$.

\section{Simultaneous eigenfunctions for the helicity operator 
and the Cartan generators of $\mathbf{SU}\boldsymbol{(2,2)}$}

The procedure in Sec. III is a mere formality at present,  
because function spaces in which Eq. (\ref{3.8}) is realized are still unclear. 
Therefore, we now try to find functions suitable for defining desirable function spaces 
that can be shown to be (pre-)Hilbert spaces.

In terms of $\hat{a}{}^{A}$ and $\hat{\bar{a}}{}^{\dot{A}}$,  
the helicity operator (\ref{2.4}) can be written as 
\begin{align}
\hat{s} &= \dfrac{1}{4} \!\left( 
\hat{\bar{a}}{}^{\dot{A}} I{}_{\dot{A} B} \hat{a}{}^{B}
+ \hat{a}{}^{B} \hat{\bar{a}}{}^{\dot{A}} I{}_{\dot{A} B}
\right) 
\nonumber 
\\
&=\frac{1}{2} \!\left(
\hat{\bar{a}}{}^{\dot{0}} \hat{a}{}^{0} + \hat{\bar{a}}{}^{\dot{1}} \hat{a}^{1}
- \hat{\bar{a}}{}^{\dot{2}} \hat{a}{}^{2} - \hat{\bar{a}}{}^{\dot{3}} \hat{a}^{3} \right) 
+1\,.
\label{4.1}
\end{align}
With this form of $\hat{s}$, we consider the helicity eigenvalue equation 
\begin{align}
\hat{s} | \varPhi \rangle = s | \varPhi \rangle \,, 
\label{4.2}
\end{align}
where $s$ is a helicity eigenvalue and $|\varPhi \rangle$ is its corresponding helicity eigenvector. 
Multiplying both sides of Eq. $(\ref{4.2})$ by $\langle \bar{\alpha} |$ on the left and using 
Eq. (\ref{3.8}), we have 
\begin{align}  
\left(
\alpha{}^{A} \dfrac{\partial}{\partial \alpha{}^{A} } - \dfrac{1}{2}
 \| \alpha \|^{2}
\right)
\varPhi (\alpha) = ( -2s-2 ) \varPhi (\alpha) \,, 
\label{4.3}
\end{align}
where $\varPhi (\alpha)$ is the helicity eigenfunction defined by 
$\varPhi (\alpha) := \langle \bar{\alpha} | \varPhi \rangle$. 
This equation can easily be solved to yield the particular solution 
\begin{align} 
\varPhi_{k,l ,m,n} (\alpha) = f_{k,l ,m,n} (\alpha) 
\exp \!\left( \frac{1}{2} \| \alpha \|^{2} 
\right) ,
\label{4.4}
\end{align}
with the holomorphic function 
\begin{align}
f_{k,l,m,n} (\alpha)
:= C_{k,l,m,n} (\alpha{}^{0}){}^{k} (\alpha{}^{1}){}^{l} 
(\alpha{}^{2}){}^{m} (\alpha{}^{3}){}^{n} \,.
\label{4.5}
\end{align}
Here, $C{}_{k,l ,m,n}$ is an undetermined coefficient, and 
$k$, $l$, $m$ and $n$ are constants satisfying 
\begin{align} 
s=-\frac{1}{2}(k + l + m + n)-1 \,.
\label{4.6}
\end{align}
Clearly, $f_{k,l,m,n}$ is a homogeneous twistor function of degree $-2s-2$. 
The single-valuedness of $\varPhi_{k,l ,m,n}$, or equivalently that of $f_{k,l,m,n}$, 
is valid if and only if $k$, $l$, $m$ and $n$ are integers. 
Then, from Eq. (\ref{4.6}), 
the helicity eigenvalue $s$ is determined to be either integer or half-integer values. 
The helicity of a massless particle is thus quantized as a result of twistor quantization.

Now, we note that the helicity operator $\hat{s}$ commutes with all the generators 
of $\mathrm{SU}(2,2)$ represented as Eq. (\ref{A.6}); see Appendix A. 
The Lie group $\mathrm{SU}(2,2)$ has rank 3, and 
in the Schwinger representation (\ref{A.6}), its Cartan generators are given by  
\begin{equation}
\begin{split}
\hat{\varLambda}_{3} &= \dfrac{1}{2}
\!\left( \hat{\bar{a}}{}^{\dot{0}} \hat{a}{}^{0} - \hat{\bar{a}}{}^{\dot{1}} \hat{a}{}^{1}
\right) ,
\quad 
\hat{\varLambda}_{6} = -\dfrac{1}{2}
\!\left( \hat{\bar{a}}{}^{\dot{2}} \hat{a}{}^{2} - \hat{\bar{a}}{}^{\dot{3}} \hat{a}{}^{3}
\right) ,
\\
\hat{\varLambda}_{15} &= \dfrac{1}{2 \sqrt{2}} 
\!\left(
\hat{\bar{a}}{}^{\dot{0}} \hat{a}{}^{0} + \hat{\bar{a}}{}^{\dot{1}} \hat{a}^{1}
+ \hat{\bar{a}}{}^{\dot{2}} \hat{a}{}^{2} + \hat{\bar{a}}{}^{\dot{3}} \hat{a}^{3}
\right) .
\end{split}
\label{4.7}
\end{equation}
Because $\hat{s}$, $\hat{\varLambda}_{3}$, $\hat{\varLambda}_{6}$, and 
$\hat{\varLambda}_{15}$ commute with each other, they have 
a simultaneous eigenfunction $\varPhi (\alpha)$ satisfying 
\begin{subequations}
\label{4.8}
\begin{align}   
\langle \bar{\alpha} | \hat{s} | \varPhi \rangle 
&= s\:\! \varPhi (\alpha) \,,
\label{4.8a}
\\
\langle \bar{\alpha} | \hat{\varLambda}{}_{3} | \varPhi \rangle
&= K \varPhi (\alpha) \,,
\label{4.8b}
\\
\langle \bar{\alpha} | \hat{\varLambda}{}_{6} | \varPhi \rangle
&= L \varPhi (\alpha) \,,
\label{4.8c}
\\
\langle \bar{\alpha} | \hat{\varLambda}{}_{15} | \varPhi \rangle
&= \dfrac{1}{\sqrt{2}}M \varPhi (\alpha) \,, 
\label{4.8d}
\end{align}
\end{subequations}
where $s$, $K$, $L$ and $M/\sqrt{2}$ are eigenvalues of 
$\hat{s}$, $\hat{\varLambda}{}_{3}$, $\hat{\varLambda}{}_{6}$  
and $\hat{\varLambda}{}_{15}$, respectively. 
(Equation (\ref{4.8a}) is identical with Eq. (\ref{4.2}) multiplied by $\langle \bar{\alpha} |$.)  
Using Eq. (\ref{3.8}), we can show that $\varPhi_{k,l,m,n}(\alpha)$ is  
a solution of the simultaneous equations (\ref{4.8a})-(\ref{4.8d}) 
provided that Eq. (\ref{4.6}) and 
\begin{align}
K = \dfrac{1}{2} (- k + l )\,, \quad
L = \dfrac{1}{2} (- m + n )\,, \quad
M = \dfrac{1}{2} (- k - l + m + n) 
\label{4.9}
\end{align}
are fulfilled. 
In this way, $\varPhi_{k,l,m,n}$ is confirmed to be a simultaneous eigenfunction 
for $\hat{s}$, $\hat{\varLambda}_{3}$, $\hat{\varLambda}_{6}$ and 
$\hat{\varLambda}_{15}$. 
From Eq. (\ref{4.9}), it follows that $K$, $L$, $M$, as well as $s$, take  
integer and half-integer values. 
The set of Eqs. (\ref{4.6}) and (\ref{4.9}) can inversely be solved as 
\begin{gather}
\begin{aligned}
k &= -\dfrac{1}{2} (s+2K+M+1) \,, \\
m &= -\dfrac{1}{2} (s+2L-M+1) \,, 
\end{aligned}
\quad 
\begin{aligned}
l &= \dfrac{1}{2} (-s+2K-M-1) \,, \\
n &= \dfrac{1}{2} (-s+2L+M-1) \,. 
\end{aligned}
\label{4.10}
\end{gather}
This fact demonstrates that the combination of eigenvalues $(s, K, L, M/\sqrt{2} \;\!)$ is 
in bijective correspondence with the combination of integers $(k, l, m, n)$. 
For this reason, we can use $(s, K, L, M)$ to uniquely specify the simultaneous eigenfunction 
$\varPhi_{k,l,m,n}$, which fact enables us to denote $\varPhi_{k,l,m,n}$ as $\varPhi_{s, K, L, M}$,   
and correspondingly $f_{k,l,m,n}$ as $f_{s, K, L, M}$, namely, 
\begin{align}
\varPhi_{s, K, L, M}:=\varPhi_{k,l,m,n} \,,  
\; \quad 
f_{s, K, L, M}:=f_{k,l,m,n} \,.
\label{4.10.5}
\end{align}
The helicity eigenvalue $s$ labels an irreducible representation of $\mathrm{SU}(2,2)$. 
This can be understood from the fact that the eigenvalue of the quadratic Casimir operator 
of $\mathrm{SU}(2,2)$ is determined to be $3({{s}}^2 -1)/2$, as seen from Eq. (\ref{A.10}) in Appendix A. 
The eigenfunction $\varPhi_{s, K, L, M}$ is classified into the irreducible representation 
of $\mathrm{SU}(2,2)$ labeled by $s$ 
and is completely specified by the remaining eigenvalues $K$, $L$, and $M$.\cite{Yao}

Now, we consider application of the operators $\hat{a}{}^{A}$ to 
$\varPhi_{k,l,m,n}$. 
This can be evaluated by using Eq. (\ref{3.8a}) as follows: 
\begin{subequations}
\label{4.11}
\begin{align}
\langle \bar{\alpha} | \hat{a}{}^{0} | \varPhi_{k,l,m,n} \rangle
&= \alpha{}^{0} \varPhi{}_{k,l,m,n}(\alpha) 
= \dfrac{C_{k,l,m,n}}{C_{k+1,l,m,n}} \:\!\varPhi_{k+1,l,m,n}(\alpha) \,, 
\label{4.11a}
\\
\langle \bar{\alpha} | \hat{a}{}^{1} | \varPhi_{k,l,m,n} \rangle
&= \alpha{}^{1} \varPhi{}_{k,l,m,n}(\alpha) 
= \dfrac{C_{k,l,m,n}}{C_{k,l+1,m,n}} \:\!\varPhi_{k,l+1,m,n}(\alpha) \,,  
\label{4.11b}
\\
\langle \bar{\alpha} | \hat{a}{}^{2} | \varPhi_{k,l,m,n} \rangle
&= \alpha{}^{2} \varPhi{}_{k,l,m,n}(\alpha) 
= \dfrac{C_{k,l,m,n}}{C_{k,l,m+1,n}} \:\!\varPhi_{k,l,m+1,n}(\alpha) \,,  
\label{4.11c}
\\
\langle \bar{\alpha} | \hat{a}{}^{3} | \varPhi_{k,l,m,n} \rangle
&= \alpha{}^{3} \varPhi{}_{k,l,m,n}(\alpha) 
= \dfrac{C_{k,l,m,n}}{C_{k,l,m,n+1}} \:\!\varPhi_{k,l,m,n+1}(\alpha) \,.  
\label{4.11d}
\end{align}
\end{subequations}
Also, using Eq. (\ref{3.8b}), we can evaluate application of the operators 
$\hat{\bar{a}}{}^{\dot{A}}$ to $\varPhi_{k,l,m,n}\;\!$:  
\begin{subequations}
\label{4.12}
\begin{align}
\langle \bar{\alpha} | \hat{\bar{a}}{}^{\dot{0}} | \varPhi_{k,l,m,n} \rangle 
&=-e^{  \| \alpha \|^{2} /2} \:\!\frac{\partial f_{k,l,m,n}}{\partial \alpha^{0}} 
= -\dfrac{C_{k,l,m,n}}{C_{k-1,l,m,n}} \:\! k \:\!\varPhi_{k-1,l,m,n}(\alpha) \,,
\label{4.12a}
\\
\langle \bar{\alpha} | \hat{\bar{a}}{}^{\dot{1}} | \varPhi_{k,l,m,n} \rangle 
&=-e^{  \| \alpha \|^{2} /2} \:\!\frac{\partial f_{k,l,m,n}}{\partial \alpha^{1}} 
= -\dfrac{C_{k,l,m,n}}{C_{k,l-1,m,n}} \:\! l\:\!\varPhi_{k,l-1,m,n}(\alpha) \,,
\label{4.12b}
\\
\langle \bar{\alpha} | \hat{\bar{a}}{}^{\dot{2}} | \varPhi_{k,l,m,n} \rangle 
&=e^{  \| \alpha \|^{2} /2} \:\!\frac{\partial f_{k,l,m,n}}{\partial \alpha^{2}} 
= \dfrac{C_{k,l,m,n}}{C_{k,l,m-1,n}} \:\! m\:\!\varPhi_{k,l,m-1,n}(\alpha) \,,
\label{4.12c}
\\
\langle \bar{\alpha} | \hat{\bar{a}}{}^{\dot{3}} | \varPhi_{k,l,m,n} \rangle 
&=e^{  \| \alpha \|^{2} /2} \:\!\frac{\partial f_{k,l,m,n}}{\partial \alpha^{3}} 
= \dfrac{C_{k,l,m,n}}{C_{k,l,m,n-1}} \:\! n\:\!\varPhi_{k,l,m,n-1}(\alpha) \,.
\label{4.12d}
\end{align}
\end{subequations}
It is seen from Eqs. (\ref{4.11}) and (\ref{4.12}) that the $\hat{a}{}^{A}$ behave 
as creation operators, while the $\hat{\bar{a}}{}^{\dot{A}}$ behave as 
annihilation operators.

\section{An appropriate inner product for the eigenfunctions $\boldsymbol{\varPhi_{s,K,L,M}}$} 

If we follow the argument of coherent states,\cite{Kur, NO} 
it is quite natural to naively define the inner product of two arbitrary eigenfunctions, 
$\varPhi_{s,K,L,M}$ and $\varPhi_{s^{\prime},K^{\prime},L^{\prime},M^{\prime}}$,  
as 
\begin{align}
&\langle \varPhi_{s,K,L,M} | \varPhi_{s^{\prime},K^{\prime},L^{\prime},M^{\prime}} \rangle
_{\mathrm{naive}} 
\nonumber
\\
&: = \int_{\mathbf{T}} 
\overline{\varPhi_{s,K,L,M} (\alpha)} 
\, \varPhi_{s^{\prime}, K^{\prime}, L^{\prime}, M^{\prime}} (\alpha) 
\:\!d^{\;\!8} \!\!\:\mu (\alpha, \bar{\alpha}) 
\label{5.1}
\\
&\; = \int_{\mathbf{T}} 
\overline{f_{s,K,L,M} (\alpha)} 
\;\! f_{s^{\prime}, K^{\prime}, L^{\prime}, M^{\prime}} (\alpha) \exp\:\!\! \|\alpha \|^{2} 
\:\!d^{\;\! 8}  \!\!\: \mu (\alpha, \bar{\alpha}) \,, 
\nonumber 
\end{align}
where 
\begin{align}
d^{\;\!8} \!\!\: \mu (\alpha, \bar{\alpha}):=
d \alpha^{0} \wedge d \alpha^{1} \wedge d \alpha^{2} \wedge d \alpha^{3}
\wedge d \bar{\alpha}^{\dot{0}} \wedge d \bar{\alpha}^{\dot{1}} 
\wedge d \bar{\alpha}^{\dot{2}} \wedge d \bar{\alpha}^{\dot{3}} \,.
\label{5.2}
\end{align}
Obviously, $d^{\;\!8} \!\!\: \mu$ is invariant under 
$\mathrm{SU}(2,2)$ transformations. 
Although Eq. (\ref{5.1}) might seem to be well-defined, actually,  
$\langle \varPhi_{s,K,L,M} | \varPhi_{s,K,L,M} \rangle_{\mathrm{naive}}$  
diverges because the integrand contains  
the multiplicative factor $\exp (|\alpha^{0}|{}^{2}+|\alpha^{1}|{}^{2})$, 
which strictly increases fast as  
$|\alpha^{0}|\rightarrow \infty$ or as $|\alpha^{1}|\rightarrow \infty$. 
Hence, the inner product (\ref{5.1}) is not well-defined and 
we cannot use it in our approach.

Before providing an appropriate inner product, we recall that in twistor theory, 
{\em projective twistors} are considered to be more essential than twistors themselves. 
From this point of view, it is sufficient to define an inner product of 
$\varPhi_{s,K,L,M}$ and $\varPhi_{s^{\prime},K^{\prime},L^{\prime},M^{\prime}}$ 
in such a manner that projective twistors are taken to be integration variables. 
For a nonzero twistor $\alpha^{A}$, the projective twistor $[ \alpha^{A} ]$ is defined as 
the proportionality class $[ \alpha^{A} ] :=
\big\{ \upsilon \alpha^{A} \big|\, \upsilon \in \Bbb{C}\setminus\{0\} \big\}$.  
The projective twistor space $\mathbf{PT}\:\!(\;\!\cong \mathbb{C}\mathbf{P}^{3})$ 
is a 3-dimensional complex space coordinatized by 
$[ ( \alpha^{A} ) ] :=\big\{ ( \upsilon \alpha^{A} ) \big|\, 
\upsilon\in \Bbb{C}\setminus\{0\} \big\}$, that is,   
$\mathbf{PT}:= \big\{\:\! [ ( \alpha^{A} ) ]\, \big|\, ( \alpha^{A} ) \in \mathbf{T} 
\setminus \{\boldsymbol{0} \} \big\}$. 
Clearly, the projective twistor $[ \alpha^{A} ]$ is invariant under the complexified scale transformation 
$\alpha^{A} \rightarrow \upsilon \alpha^{A}$, and hence the functions of $[ \alpha^{A} ]$ 
remain invariant under this transformation. 
Conversely, the functions of $\alpha^{A}$  
that are invariant under the complexified scale transformation 
can be treated as functions of $[ \alpha^{A} ]$, that is, functions on $\mathbf{PT}$. 
Similarly, the functions of $\alpha^{A}$  
that are invariant under the (pure) scale transformation $\alpha^{A} \rightarrow |\upsilon| \alpha^{A}$ 
can be treated as functions on $S^{1} \times\mathbf{PT}$.  
Here, $S^{1}$ denotes the circular space parametrized by a phase common to the 
twistor variables.

Now, we assume that the twistor functions 
$f_{s,K,L,M}$ and $f_{s^{\prime},K^{\prime},L^{\prime},M^{\prime}}$ 
are transformed into positive-frequency massless fields in 
complexified Minkowski space $\mathbb{C}\mathbf{M}$ 
via the Penrose transform.\cite{PM, PR, HT, Tak}   
In this case, 
$f_{s,K,L,M}$ and $f_{s^{\prime},K^{\prime},L^{\prime},M^{\prime}}$ are realized as functions on 
the upper half of twistor space, namely 
$\mathbf{T}^{+}:= \big\{ ( \alpha^{A} ) \in \mathbf{T} \, \big|\, \| \alpha \|^{2} >0 \big\}$. 
Considering this, we propose the following inner product: 
\begin{subequations}
\label{5.3}
\begin{align}
&\langle \varPhi_{s,K,L,M} | \varPhi_{s^{\prime},K^{\prime},L^{\prime},M^{\prime}} \rangle 
\nonumber \\
& :=
\lim_{\epsilon \rightarrow +0}
\dfrac{-\epsilon}{\varGamma( s + s^{\prime} + 2 \epsilon - 1)}
\int_{S^{1} \times \mathbf{PT}^{+}} 
\overline{\varPhi_{s+\epsilon, K,L,M} (\alpha)} 
\,\varPhi_{s^{\prime}+\epsilon, K^{\prime}, L^{\prime}, M^{\prime}} (\alpha) 
\nonumber \\
& 
\qquad \times \big( \|\alpha \|^{2} \big){}^{s + s^{\prime} + 2\epsilon + 2}
\exp \! \big(-\|\alpha \|^{2} \big) \:\! 
d^{\;\! 7} \!\!\: \mu (\alpha, \bar{\alpha}) 
\label{5.3a}
\\
&\; =
\lim_{\epsilon \rightarrow +0}
\dfrac{-\epsilon}{\varGamma( s + s^{\prime} + 2 \epsilon - 1)}
\int_{S^{1} \times \mathbf{PT}^{+}}
\overline{f_{s+\epsilon, K,L,M} (\alpha)} 
\;\! f_{s^{\prime}+\epsilon, K^{\prime}, L^{\prime}, M^{\prime}} (\alpha) 
\nonumber \\
& 
\qquad \times \big( \|\alpha \|^{2} \big){}^{s + s^{\prime} + 2\epsilon + 2} \:\!
d^{\;\! 7} \!\!\: \mu (\alpha, \bar{\alpha}) \,, 
\label{5.3b}
\end{align}
\end{subequations}
with the 7-form 
\begin{align}
& d^{\;\! 7} \!\!\: \mu(\alpha, \bar{\alpha}) 
\nonumber 
\\
& :=\frac{1}{12 \big( \|\alpha \|^{2} \big){}^{4} }
\Big( 
d \alpha{}^{0} \wedge d \alpha{}^{1} \wedge d \alpha{}^{2} \wedge d \alpha{}^{3}
\wedge \varepsilon_{\dot{A} \dot{B} \dot{C} \dot{D}} 
\bar{\alpha}{}^{\dot{A}} d \bar{\alpha}{}^{\dot{B}} \wedge d \bar{\alpha}{}^{\dot{C}}
\wedge d \bar{\alpha}{}^{\dot{D}} 
\nonumber \\
& 
\quad\:\: + \varepsilon_{{A} {B} {C} {D}} 
{\alpha}{}^{{A}} d {\alpha}{}^{{B}} \wedge d {\alpha}{}^{{C}}
\wedge d {\alpha}{}^{{D}} \wedge
d \bar{\alpha}{}^{\dot{0}} \wedge d \bar{\alpha}{}^{\dot{1}} 
\wedge d \bar{\alpha}{}^{\dot{2}} \wedge d \bar{\alpha}{}^{\dot{3}} \Big) \,, 
\label{5.4}
\end{align}
where 
$\mathbf{PT}^{+}:=\big\{\;\! [( \alpha^{A} )] \in \mathbf{PT}  \, \big|\, \| \alpha \|^{2} >0 \big\}$,   
and $\varepsilon_{0123}=\varepsilon_{\dot{0} \dot{1} \dot{2} \dot{3}} =1$. 
The integrand 
$\overline{f_{s,K,L,M}(\alpha)} \;\! f_{s^{\prime}, K^{\prime}, L^{\prime}, M^{\prime}}(\alpha) 
\big( \|\alpha \|^{2} \big){}^{s+s^{\prime}+2}$  
can be regarded as a function on 
the product space $S^{1} \times \mathbf{PT}^{+}$, 
because it remains invariant under the (pure) scale transformation 
$\alpha^{A} \rightarrow |\upsilon| \alpha^{A}$, and also 
$f_{s,K,L,M}$ and $f_{s^{\prime}, K^{\prime}, L^{\prime}, M^{\prime}}$ are functions on $\mathbf{T}^{+}$. 
(When $s=s^{\prime}$, this integrand remains invariant under the complexified scale transformation 
$\alpha^{A} \rightarrow \upsilon \alpha^{A}$ and therefore is treated as a function on $\mathbf{PT}^{+}$.)  
Adding the infinitesimal positive number $\epsilon$ to $s$ and $s^{\prime}$  
is necessary for making Eq. (\ref{5.3}) well-defined. 
The 7-form $d^{\;\! 7} \!\!\:\mu$ is invariant under $\mathrm{SU}(2,2)$ transformations 
and under the complexified scale transformation; 
thus, $d^{\;\! 7} \!\!\; \mu$ is recognized as an integration measure on $S^{1} \times \mathbf{PT}^{+}$. 
From the facts stated above, it is clear that the inner product 
$\langle \varPhi_{s,K,L,M} | \varPhi_{s^{\prime},K^{\prime},L^{\prime},M^{\prime}} \rangle$ 
is invariant under the (pure) scale transformation. 
[The projective twistor subspace $\mathbf{PT}^{+}$ is isomorphic to the coset space 
$\mathrm{SU}(2,2)/  \mathrm{S} \big(\mathrm{U}(2,1) \times \mathrm{U}(1) \big)$.\cite{Wel}  
Similarly, $S^{1} \times\mathbf{PT}^{+}$ is isomorphic to $\mathrm{SU}(2,2)/ \mathrm{SU}(2,1)$.]

To see more precisely that $d^{\;\! 7} \!\!\; \mu$ is an integration measure on $S^{1} \times \mathbf{PT}^{+}$, 
here we introduce the inhomogeneous coordinates $(\zeta^{1}, \zeta^{2}, \zeta^{3})$ of $\mathbf{PT}^{+}$ 
defined by 
\begin{align}
\zeta^{1} := \dfrac{\alpha^{1}}{\alpha^{0}} \,, \quad\: 
\zeta^{2} := \dfrac{\alpha^{2}}{\alpha^{0}} \,, \quad\:
\zeta^{3} := \dfrac{\alpha^{3}}{\alpha^{0}} \,. 
\label{5.5}
\end{align}
[Strictly speaking, 
$(\zeta^{1}, \zeta^{2}, \zeta^{3})$ is a local coordinate system on the open set 
$\mathbf{PT}^{+}_{0} := \big\{ \:\! [(\alpha^{A})] \in \mathbf{PT}^{+} \, \big|\,\alpha^{0} \neq 0 \big\}$.]   
With these coordinates, $d^{\;\! 7} \!\!\: \mu$ can be expressed as 
\begin{align}
d^{\;\! 7} \!\!\: \mu =d\theta \wedge d^{\;\! 6} \!\!\: \mu \,, 
\label{5.6}
\end{align}
where $\theta$ is the phase common to the twistor variables 
\begin{align}
\theta:=-\frac{1}{4} i \log \! \left( \prod_{A=0}^{3} \frac{\alpha^{A}}{|\alpha^{A}|} \right) 
=-i \log \! \left( \frac{\alpha^{0}}{|\alpha^{0}|} \right) 
-\frac{1}{4} i \log \! \left( \prod_{A=1}^{3} \frac{\zeta^{A}}{|\zeta^{A}|} \right) . 
\label{5.7}
\end{align}
The range of $\theta$ is determined to be $[\:\! 0, 2\pi)$.  
The 6-form $d^{\;\! 6} \!\!\: \mu$ is given by 
\begin{align}
d^{\;\! 6} \!\!\: \mu :=\dfrac{i}{K^{4}}
d\zeta{}^{1} \wedge d\zeta{}^{2} \wedge d\zeta{}^{3} \wedge 
d\bar{\zeta}{}^{\:\!\dot{1}} \wedge d\bar{\zeta}{}^{\:\!\dot{2}} \wedge d\bar{\zeta}{}^{\:\!\dot{3}} \,, 
\label{5.8}
\end{align}
with $K:=1 + |\zeta{}^{1}|^{2} - |\zeta{}^{2}|^{2} - |\zeta{}^{3}|^{2}$. 
The 6-form $d^{\;\! 6} \!\!\: \mu$ is precisely the volume element of 
$\mathbf{PT}^{+}$. [This volume element can be 
derived from the K\"{a}hler form $\varOmega := i \partial \bar{\partial} \log K$ 
as $d^{\;\! 6} \!\!\; \mu=(3!)^{-1} \varOmega \wedge \varOmega \wedge \varOmega$. 
\cite{Nak, IK} 
The K\"{a}hler form $\varOmega$ itself is independent of the choice of local coordinate system,     
so that $d^{\;\! 6} \!\!\: \mu$ is also coordinate independent.]    
In this way, Eq. (\ref{5.6}) demonstrates that $d^{\;\! 7} \!\!\: \mu$ is indeed 
an integration measure on $S^{1} \times \mathbf{PT}^{+}$.

For evaluating the inner product (\ref{5.3}), it is convenient to use 
the hyperbolic polar coordinates  
$(\|\alpha\|, \eta, \chi, \psi, \theta, \vartheta, \phi, \varphi)$ defined by 
$\|\alpha\|:=\sqrt{ \|\alpha\|{}^{2}}$ and 
\begin{subequations}
\label{5.9}
\begin{align}
\alpha^{0} &=\|\alpha\| \:\! e^{i(\theta+\vartheta+\phi)} \cosh \eta \cos \chi \,, 
\label{5.9a}
\\
\alpha^{1} &=\|\alpha\| \:\! e^{i(\theta+\vartheta-\phi)} \cosh \eta \sin \chi \,, 
\label{5.9b}
\\
\alpha^{2} &=\|\alpha\| \:\! e^{i(\theta-\vartheta+\varphi)} \sinh \eta \cos \psi \,, 
\label{5.9c}
\\
\alpha^{3} &=\|\alpha\| \:\! e^{i(\theta-\vartheta-\varphi)} \sinh \eta \sin \psi \,. 
\label{5.9d}
\end{align}
\end{subequations}
Here, by virtue of $\| \alpha \|^{2} >0$, 
it follows that $\|\alpha\|$ takes values in the coordinate range $0< \|\alpha\| <\infty$. 
The other coordinate ranges are determined to be  
\begin{align}
0 \leq \eta < \infty \,, \quad\! 
0 \leq \chi, \:\! \psi \leq \frac{\pi}{2} \,, \quad \!
0 \leq \theta < 2\pi \,, \quad\! 
-\pi < \vartheta,\:\! \phi, \:\! \varphi <\pi \,. 
\label{5.10}
\end{align}
Equations (\ref{5.9}) and (\ref{5.10}) 
can be found through the polar decomposition 
$\alpha^{A}=|\alpha^{A}| e^{i\theta_{A}}$ ($\:\! 0\leq |\alpha^{A}| < \infty, \;
0 \leq \theta_{A} < 2\pi$).  
In the hyperbolic polar coordinate system, $d^{\:\! 7} \!\!\: \mu$ is expressed as 
\begin{align}
d^{\:\! 7} \!\!\: \mu =d\theta \wedge 2(\sinh 2\eta)^{3} \sin 2\chi \sin 2\psi 
\;\! d\eta \wedge d\chi \wedge d\psi \wedge d\vartheta \wedge d\phi \wedge d\varphi \,. 
\label{5.11}
\end{align}
Substituting Eqs. (\ref{5.9}) and (\ref{5.11}) into Eq. (\ref{5.3b}) and carrying out 
the integration over $S^{1} \times \mathbf{PT}^{+}$, we obtain 
\begin{align}
&\langle \varPhi_{s,K,L,M} | \varPhi_{s^{\prime},K^{\prime},L^{\prime},M^{\prime}} \rangle 
\notag
\\
& =\frac{1}{2} (4 \pi)^{4} |C_{s,K,L,M}|{}^{2} 
\delta_{ss^{\prime}} \delta_{KK^{\prime}} \delta_{LL^{\prime}} \delta_{MM^{\prime}} 
\notag
\\
& \, \quad \times \lim_{\epsilon \rightarrow +0} 
\dfrac{- \epsilon }
{\varGamma ( s+M+\epsilon ) \:\!
 \varGamma ( -s-M+1-\epsilon )}
\notag
\\
&
\, \quad \times
\varGamma \bigg( \dfrac{-s+2K-M+1-\epsilon}{2} \bigg) 
\varGamma \bigg( \dfrac{-s-2K-M+1-\epsilon}{2} \bigg)
\notag
\\
&
\, \quad \times 
\varGamma \bigg( \dfrac{-s+2L+M+1-\epsilon}{2} \bigg) 
\varGamma \bigg( \dfrac{-s-2L+M+1-\epsilon}{2} \bigg) \:\!. 
\label{5.12}
\end{align}
Here, the integration formulas 
\begin{subequations}
\label{5.13}
\begin{align}
\int_{0}^{\infty} 
(\sinh \eta)^{2x-1} (\cosh \eta)^{2y-1} d\eta 
&=\frac{\varGamma(x) \varGamma(-x-y+1)}{2 \varGamma(-y+1)} \,, 
\label{5.13a}
\\
& \:\:\quad  \Re{(x)}>0 \,, \; \Re{(x+y)}<1 \,, 
\notag
\\
\int_{0}^{\pi/2} 
(\sin \chi)^{2x-1} (\cos \chi)^{2y-1} d\chi 
&=\frac{\varGamma(x) \varGamma(y)}{2 \varGamma(x+y)}  \,, 
\label{5.13b}
\\
& \:\:\quad  \Re{(x)}>0 \,, \; \Re{(y)}>0 \;\! 
\notag
\end{align}
\end{subequations}
have been used. Also, an analytic continuation of the gamma function 
$\varGamma( 2s + 2 \epsilon - 1)$ that occurs in calculating the inner product  
has been considered. 
The orthogonality denoted by $\delta_{ss^{\prime}}$ has been found 
by the integration over $S^{1}$;  
thus, $\langle \varPhi_{s,K,L,M} | \varPhi_{s,K^{\prime},L^{\prime},M^{\prime}} \rangle$,  
the inner product restricted within the subspace specified by $s$,    
turns out to be given as an integral over $\mathbf{PT}^{+}$. 
In terms of $(k, l, m, n)$, Eq. (\ref{5.12}) can be written as 
\begin{align}
&\langle \varPhi_{k,l,m,n} | \varPhi_{k^{\prime},l^{\prime},m^{\prime},n^{\prime}} \rangle 
\notag
\\
& 
=  (4 \pi)^{4} |C_{k, l, m, n}|{}^{2}
\delta_{kk^{\prime}} \delta_{l l^{\prime}} \delta_{mm^{\prime}} \delta_{nn^{\prime}} 
\notag
\\ 
& \, \quad \times \lim_{\varepsilon \rightarrow +0} 
\dfrac{-\varepsilon 
\varGamma ( k + 1 - \varepsilon ) \varGamma ( l + 1 - \varepsilon )
\varGamma ( m + 1 - \varepsilon ) \varGamma ( n + 1 - \varepsilon )}
{\varGamma ( - k - l - 1 + 2\varepsilon ) \varGamma ( k + l + 2 - 2\varepsilon )} \,, 
\label{5.14}
\end{align}
where $\varepsilon:=\epsilon/2$. 
In deriving Eqs (\ref{5.12}) and (\ref{5.14}), 
the arguments of the gamma functions have been assumed to be positive. 
Accordingly, it follows that 
Eq. (\ref{5.14}) is valid only for the small region 
\begin{align}
\{ (k, l, m, n)\:\! |\, k, l, m, n > -1+\varepsilon, \:
-2+2\varepsilon <k+l<-1+2\varepsilon \} \,. 
\label{5.15}
\end{align}
In this region, $k$ and $l$ can never be integers, and 
$m$ and $n$ can be only natural numbers including 0. 
In order that $k$, $l$, $m$ and $n$ can be integers,  
now we perform an analytic continuation of Eq. (\ref{5.14}) by 
using the formula \cite{Ryd}
\begin{align}
\varGamma(-n\pm\varepsilon)
=\frac{(-1)^{n}}{n!} \bigg[ \pm\frac{1}{\varepsilon} +\psi_{1}(n+1) +\mathcal{O}(\varepsilon) \bigg] \:\!,  
\quad n \in \Bbb{N}_0 \;\!, 
\label{5.16}
\end{align}
with $\psi_{1}(n+1):=\sum_{p=1}^{n} p^{-1} -\gamma$. 
Here, $\gamma$ is the Euler-Mascheroni constant. 
[We can also perform  
an analytic continuation of Eq. (\ref{5.14}) by using the reflection formula 
$\varGamma(x) \varGamma(1-x)=\pi/\sin \pi x$ 
($x \in \Bbb{C} \setminus \Bbb{Z}$).     
This analytic continuation leads to the same results as those obtained 
by using Eq. (\ref{5.16}).]  
Applying Eq. (\ref{5.16}) to either of the two gamma functions in 
the denominator of Eq. (\ref{5.14}), 
we can simplify Eq. (\ref{5.14}) as 
\begin{align}
&\langle \varPhi_{k,l,m,n} | \varPhi_{k^{\prime},l^{\prime},m^{\prime},n^{\prime}} \rangle 
\notag
\\
& 
=  2(4 \pi)^{4} (-1)^{k+l} \:\!|C_{k, l, m, n}|{}^{2}
\delta_{kk^{\prime}} \delta_{l l^{\prime}} \delta_{mm^{\prime}} \delta_{nn^{\prime}} 
\notag
\\ 
& \, \quad \times \lim_{\varepsilon \rightarrow +0} 
\varepsilon^{2}  
\varGamma ( k + 1 - \varepsilon ) \varGamma ( l + 1 - \varepsilon )
\varGamma ( m + 1 - \varepsilon ) \varGamma ( n + 1 - \varepsilon ) \,.  
\label{5.17}
\end{align}
Then, by applying Eq. (\ref{5.16}) to Eq. (\ref{5.17}), it becomes possible to evaluate 
$\langle \varPhi_{k,l,m,n} | \varPhi_{k^{\prime},l^{\prime},m^{\prime},n^{\prime}} \rangle$ 
when some or all of $k$, $l$, $m$ and $n$ are negative integers.  
Obviously, the orthogonality condition for the eigenfunctions $\varPhi_{k,l,m,n}$ 
is fulfilled in Eq. (\ref{5.17}).

Now, we focus our attention on the cases in which two of 
$k$, $l$, $m$ and $n$ in the twistor function $f_{k,l,m,n}$ take negative integer values, 
while the other two take non-negative integer values. 
These cases are especially important in the practical sense that   
the Penrose transform of such a twistor function yields a massless field in $\mathbb{C}\mathbf{M}$  
that is referred to in the literature on twistor theory as an elementary state.  
\cite{PM, Pen6, HH, Hod1, Hod2}  
(Sometimes such a twistor function itself is referred to as an elementary state.) 
Also, twistor functions of this form fit well into the framework of twistor diagrams. 
\cite{PM, Pen6, HH, Hod1, Hod2} 
Before evaluating  
$\langle \varPhi_{k,l,m,n} | \varPhi_{k^{\prime},l^{\prime},m^{\prime},n^{\prime}} \rangle$, 
we note that the combination of gamma functions in Eq. (\ref{5.14}) 
is unchanged under the interchange of $k$ and $l$ and under that of $m$ and $n$. 
This interchange symmetry originates in the fact that $\alpha^{0}$ and $\alpha^{1}$ 
have been assigned the same metric signature ^^ ^^ $+$", while $\alpha^{2}$ and $\alpha^{3}$ 
have been assigned the same metric signature ^^ ^^ $-$",      
as seen from the definition of $\| \alpha \|^{2}$ given under Eq. (\ref{3.7}).   
By virtue of the interchange symmetry, 
it is sufficient for the moment if the following three of the possible six cases are investigated: 
(a) $k, l \in  \Bbb{Z}^{-}$, $m, n \in \Bbb{N}_{0}\:\!$, 
(b) $k, l \in  \Bbb{N}_{0}\:\!$, $m, n \in \Bbb{Z}^{-}$, and 
(c1) $k, n \in \Bbb{Z}^{-}$, $l, m \in  \Bbb{N}_{0}\:\!$. 
The remaining three cases, that is, 
(c2) $l, m \in  \Bbb{Z}^{-}$, $k, n \in \Bbb{N}_{0}\:\!$, 
(c3) $k, m \in  \Bbb{Z}^{-}$, $l, n \in \Bbb{N}_{0}\:\!$, and 
(c4) $l, n \in \Bbb{Z}^{-}$, $k, m \in  \Bbb{N}_{0}\:\!$,  
can be immediately found from the case (c1) by the interchange of $k$ and $l$ 
and/or that of $m$ and $n$. It should be stressed here that 
in all of the six cases, the helicity eigenvalue $s$ can take arbitrary integer and half-integer values. 
In the following, we examine Eq. (\ref{5.17}) in each of the cases (a), (b) and (c1) individually.

\subsection{Case (a)}
In this case, $k$, $l$, $m$, and $n$ take the values 
\begin{align}
k, l  = -1, -2, -3, \cdots, \quad m, n = 0, 1, 2, \cdots.
\label{5.18}
\end{align}
Applying Eq. (\ref{5.16}) to 
$\varGamma(k+1-\varepsilon)$ and $\varGamma(l+1-\varepsilon)$ 
contained in Eq. (\ref{5.17}), we have 
\begin{align}
&
\langle \varPhi_{k,l,m,n} | \varPhi_{k^{\prime},l^{\prime},m^{\prime},n^{\prime}} \rangle 
\notag
\\
& 
= 2 (4 \pi)^{4} |C_{k, l, m, n}|{}^{2}
\delta_{kk^{\prime}} \delta_{l l^{\prime}} \delta_{mm^{\prime}} \delta_{nn^{\prime}} 
\frac{m! \:\! n!}{(-k-1)! \:\! (-l-1)!} \,.
\label{5.19}
\end{align}
By choosing $C_{k, l, m, n}$ to be 
\begin{align}
C_{k, l, m, n}=\frac{1}{(4\pi){}^2} \sqrt{ \frac{(-k-1)! \:\! (-l-1)!}{2m! \:\! n!} } \,,  
\label{5.20}
\end{align}
the eigenfunction $\varPhi_{k,l,m,n}$ is normalized to unity and 
Eq. (\ref{5.19}) reduces to the orthonormality condition 
\begin{align}
&
\langle \varPhi_{k,l,m,n} | \varPhi_{k^{\prime},l^{\prime},m^{\prime},n^{\prime}} \rangle 
=\delta_{kk^{\prime}} \delta_{l l^{\prime}} \delta_{mm^{\prime}} \delta_{nn^{\prime}} \,.
\label{5.21}
\end{align}
\subsection{Case (b)}
In this case, $k$, $l$, $m$, and $n$ take the values 
\begin{align}
k, l = 0, 1, 2, \cdots, \quad m, n  = -1, -2, -3, \cdots.
\label{5.22}
\end{align}
Applying Eq. (\ref{5.16}) to 
$\varGamma(m+1-\varepsilon)$ and $\varGamma(n+1-\varepsilon)$  
contained in Eq. (\ref{5.17}) and using Eq. (\ref{4.6}), we have 
\begin{align}
&
\langle \varPhi_{k,l,m,n} | \varPhi_{k^{\prime},l^{\prime},m^{\prime},n^{\prime}} \rangle 
\notag
\\
& 
=  2(4 \pi)^{4} |C_{k, l, m, n}|{}^{2}
\delta_{kk^{\prime}} \delta_{l l^{\prime}} \delta_{mm^{\prime}} \delta_{nn^{\prime}} 
\frac{(-1)^{2s} k! \:\! l!}{(-m-1)! \:\! (-n-1)!} \,. 
\label{5.23}
\end{align}
By choosing $C_{k, l, m, n}$ to be  
\begin{align}
C_{k, l, m, n}=\frac{1}{(4\pi){}^2} \sqrt{ \frac{(-m-1)! \:\! (-n-1)!}{2k! \:\! l!} } \,, 
\label{5.24}
\end{align}
Eq. (\ref{5.23}) reduces to the indefinite orthonormality condition 
\begin{align}
&
\langle \varPhi_{k,l,m,n} | \varPhi_{k^{\prime},l^{\prime},m^{\prime},n^{\prime}} \rangle 
=(-1)^{2s} \delta_{kk^{\prime}} \delta_{l l^{\prime}} \delta_{mm^{\prime}} \delta_{nn^{\prime}} \,.
\label{5.25}
\end{align}
Thus, $\varPhi_{k,l,m,n}$ is normalized to $1$ or $-1$ according to whether 
the helicity eigenvalue $s$ is integer or half-integer. 
\subsection{Case (c1)}
In this case, $k$, $l$, $m$, and $n$ take the values 
\begin{align}
k, n = -1, -2, -3, \cdots, \quad l, m = 0, 1, 2, \cdots.
\label{5.26}
\end{align}
Applying Eq. (\ref{5.16}) to 
$\varGamma(k+1-\varepsilon)$ and $\varGamma(n+1-\varepsilon)$  
contained in Eq. (\ref{5.17}), we have 
\begin{align}
&
\langle \varPhi_{k,l,m,n} | \varPhi_{k^{\prime},l^{\prime},m^{\prime},n^{\prime}} \rangle 
\notag
\\
& 
=  2(4 \pi)^{4} |C_{k, l, m, n}|{}^{2}
\delta_{kk^{\prime}} \delta_{l l^{\prime}} \delta_{mm^{\prime}} \delta_{nn^{\prime}} 
\frac{(-1)^{l-n}\:\! l! \:\! m!}{(-k-1)! \:\! (-n-1)!} \,. 
\label{5.27}
\end{align}
By choosing $C_{k, l, m, n}$ to be  
\begin{align}
C_{k, l, m, n}=\frac{1}{(4\pi){}^2} \sqrt{ \frac{(-k-1)! \:\! (-n-1)!}{2\:\! l! \:\! m!} } \,, 
\label{5.28}
\end{align}
Eq. (\ref{5.27}) reduces to the indefinite orthonormality condition 
\begin{align}
&
\langle \varPhi_{k,l,m,n} | \varPhi_{k^{\prime},l^{\prime},m^{\prime},n^{\prime}} \rangle 
=(-1)^{l-n} \delta_{kk^{\prime}} \delta_{l l^{\prime}} \delta_{mm^{\prime}} \delta_{nn^{\prime}} \,.
\label{5.29}
\end{align}
In this case, $\varPhi_{k,l,m,n}$ is normalized to $1$ or $-1$ according to 
the values of $l$ and $n$, even if the value of $s$ is fixed. 
It is now clear that the eigenfunctions $\varPhi_{k,l,m,n}$ in the cases (c2), (c3), and (c4) 
are also normalized to $1$ or $-1$.

\section{(Pre-)Hilbert spaces in twistor quantization}

In this section, we provide (pre-)Hilbert spaces valid for each of the cases (a), (b), and 
(c$\:\!i$) $(i=1, 2, 3, 4)$.  
These spaces are function spaces consisting of linear combinations of $\varPhi_{k,l,m,n}$ 
defined on $\mathbf{T}^{+}$. 
We also verify 
that $\hat{\bar{a}}{}^{\dot{A}}$ is represented in the (pre-)Hilbert spaces 
as the adjoint operator of $\hat{a}{}^{A}$.

\subsection{Case (a)}

In the case (a), we consider the linear combination 
\begin{align}
\varPhi^{(a)}(\alpha):= 
\sum_{k, l \in  \Bbb{Z}^{-}, \;\! m, n \in \Bbb{N}_{0}} 
c_{k,l,m,n} \varPhi_{k,l,m,n} (\alpha) \,, 
\quad c_{k,l,m,n} \in \Bbb{C} \,.
\label{6.1}
\end{align}
Then, as a set of functions of this form, we define the normed linear space 
\begin{align}
\mathsf{H}^{(a)}:=\Bigg\{ \varPhi^{(a)}(\alpha) \,\Bigg| \,
\langle \varPhi^{(a)} | \varPhi^{(a)} \rangle 
=\sum_{k, l \in  \Bbb{Z}^{-}, \;\!  m, n \in \Bbb{N}_{0}} |c_{k,l,m,n}|^{2} < \infty \Bigg\} 
\label{6.2}
\end{align}
so as to be consistent with the orthonormality condition (\ref{5.21}). 
Using the inequality $|z_{1}+z_{2}|{}^{2} \leq 2(|z_{1}|{}^{2} +|z_{2}|{}^{2})$ 
$(z_{1}, z_{2} \in \Bbb{C})$, 
we can show for all $\varPhi_{1}^{(a)}, \varPhi_{2}^{(a)} \in \mathsf{H}^{(a)}$ 
and for all $c_{1}, c_{2} \in \Bbb{C}$ that 
$c_{1} \varPhi_{1}^{(a)} +c_{2} \varPhi_{2}^{(a)} \in \mathsf{H}^{(a)}$. 
Also, using $|z_{1} z_{2}| \leq \frac{1}{2} (|z_{1}|{}^{2} +|z_{2}|{}^{2})$, 
it is readily seen that the inner product 
$\langle \varPsi^{(a)} | \varPhi^{(a)} \rangle  
=\sum_{k, l \in  \Bbb{Z}^{-}, \;\! m, n \in \Bbb{N}_{0}} \overline{b_{k,l,m,n}}\:\! c_{k,l,m,n}$  
is well-defined.  
Here, the $b_{k,l,m,n}\in \Bbb{C}$ are coefficients of $\varPsi^{(a)} \in \mathsf{H}^{(a)}$. 
Evidently  
$\overline{\langle \varPsi^{(a)} | \varPhi^{(a)} \rangle}=\langle \varPhi^{(a)} | \varPsi^{(a)} \rangle$ 
and $\langle \varPsi^{(a)} | c_{1} \varPhi_{1}^{(a)} +c_{2} \varPhi_{2}^{(a)} \rangle
=c_{1} \langle \varPsi^{(a)}|\varPhi_{1}^{(a)} \rangle +c_{2} \langle \varPsi^{(a)}|\varPsi_{2}^{(a)} \rangle$ 
are satisfied. 
Furthermore, we can prove the completeness of $\mathsf{H}^{(a)}$ 
with respect to the norm $\sqrt{\langle \varPhi^{(a)} | \varPhi^{(a)} \rangle}$, 
which vanishes if and only if $\varPhi^{(a)}=0$. 
In this way, $\mathsf{H}^{(a)}$ is established as a Hilbert space.

Application of the operators $\hat{a}{}^{A}$ to $\varPhi^{(a)}$ 
can be evaluated by using Eqs. (\ref{4.11}) and (\ref{5.20}) as
\begin{subequations}
\label{6.3}
\begin{align}
\langle \bar{\alpha} | \hat{a}{}^{0} | \varPhi^{(a)} \rangle
&=\sum_{k, l \in  \Bbb{Z}^{-}, \;\! m, n \in \Bbb{N}_{0}} \sqrt{-k} \;\! 
c_{k-1,l,m,n} \varPhi_{k,l,m,n} (\alpha) \, , 
\label{6.3a}
\\
\langle \bar{\alpha} | \hat{a}{}^{1} | \varPhi^{(a)} \rangle
&=\sum_{k, l \in  \Bbb{Z}^{-}, \;\! m, n \in \Bbb{N}_{0}} \sqrt{-l} \;\! 
c_{k,l-1,m,n} \varPhi_{k,l,m,n} (\alpha) \, , 
\label{6.3b}
\\
\langle \bar{\alpha} | \hat{a}{}^{2} | \varPhi^{(a)} \rangle
&=\sum_{k, l \in  \Bbb{Z}^{-}, \;\! m, n \in \Bbb{N}_{0}} \sqrt{m} \;\! 
c_{k,l,m-1,n} \varPhi_{k,l,m,n} (\alpha) \, , 
\label{6.3c}
\\
\langle \bar{\alpha} | \hat{a}{}^{3} | \varPhi^{(a)} \rangle
&=\sum_{k, l \in  \Bbb{Z}^{-}, \;\! m, n \in \Bbb{N}_{0}} \sqrt{n} \;\! 
c_{k,l,m,n-1} \varPhi_{k,l,m,n} (\alpha) \,. 
\label{6.3d}
\end{align}
\end{subequations}
Here, it should be noted that 
$\langle \bar{\alpha} | \hat{a}{}^{0} | \varPhi_{-1,l,m,n} \rangle
=\langle \bar{\alpha} | \hat{a}{}^{1} | \varPhi_{k,-1,m,n} \rangle=0$. 
Because each of the $\langle \bar{\alpha} | \hat{a}{}^{A} | \varPhi^{(a)} \rangle$ 
is expressed as a linear combination of the basis functions 
$\{\varPhi_{k,l,m,n}\}_{k, l \in  \Bbb{Z}^{-}, \;\! m, n \in \Bbb{N}_{0}}$ of $\mathsf{H}^{(a)}$, 
it follows that the domain of $\hat{a}{}^{A}$, denoted by $\mathsf{D}^{(a)}(\hat{a}{}^{A})$, 
is a linear subspace of $\mathsf{H}^{(a)}$. 
For example, $\mathsf{D}^{(a)}(\hat{a}{}^{0})$ is given by 
$\mathsf{D}^{(a)}(\hat{a}{}^{0}):=\big\{ \varPhi^{(a)}(\alpha) \in \mathsf{H}^{(a)} \big| 
\sum_{k, l \in  \Bbb{Z}^{-}, \;\! m, n \in \Bbb{N}_{0}} (-k)|c_{k,l,m,n}|^{2} < \infty \big\}$.   
Using Eqs. (\ref{4.11a}) and (\ref{5.20}), we see that 
$|\langle \bar{\alpha} | \hat{a}{}^{0} | \varPhi_{k,l,m,n} \rangle|{}^{2}
=-k-1\rightarrow \infty$ as $k\rightarrow -\infty$. 
This implies that $\hat{a}{}^{0}$ is an unbounded operator,\cite{Kat, RS, AE} and hence it 
cannot be defined on the whole of $\mathsf{H}^{(a)}$. 
In a similar manner, we can show that the remaining operators $\hat{a}{}^{A}$ ($A=1,2,3$)  
are also unbounded; hence, they cannot also be defined on the whole of $\mathsf{H}^{(a)}$. 
We therefore need to treat $\hat{a}{}^{A}$ as a well-defined operator on 
the subspace $\mathsf{D}^{(a)}(\hat{a}{}^{A}) \subset \mathsf{H}^{(a)}$, 
not on the whole Hilbert space $\mathsf{H}^{(a)}$. 
It is not difficult to prove that  
$\mathsf{D}^{(a)}(\hat{a}{}^{A})$ is dense (in $\mathsf{H}^{(a)}$);\cite{Kat, RS, AE}  
that is, an arbitrary element of $\mathsf{H}^{(a)}$ can be approximated by 
an element of $\mathsf{D}^{(a)}(\hat{a}{}^{A})$ to any level of accuracy.

Application of the operators $\hat{\bar{a}}{}^{\dot{A}}$ to $\varPhi^{(a)}$ 
can be evaluated by using Eqs. (\ref{4.12}) and (\ref{5.20}) as follows: 
\begin{subequations}
\label{6.4}
\begin{align}
\langle \bar{\alpha} | \hat{\bar{a}}{}^{\dot{0}} | \varPhi^{(a)} \rangle 
&=\sum_{k, l \in  \Bbb{Z}^{-}, \;\! m, n \in \Bbb{N}_{0}} \sqrt{-k-1} \;\! 
c_{k+1,l,m,n} \varPhi_{k,l,m,n} (\alpha) \, , 
\label{6.4a}
\\
\langle \bar{\alpha} | \hat{\bar{a}}{}^{\dot{1}} | \varPhi^{(a)} \rangle 
&=\sum_{k, l \in  \Bbb{Z}^{-}, \;\! m, n \in \Bbb{N}_{0}} \sqrt{-l-1} \;\! 
c_{k,l+1,m,n} \varPhi_{k,l,m,n} (\alpha) \, , 
\label{6.4b}
\\
\langle \bar{\alpha} | \hat{\bar{a}}{}^{\dot{2}} | \varPhi^{(a)} \rangle 
&=\sum_{k, l \in  \Bbb{Z}^{-}, \;\! m, n \in \Bbb{N}_{0}} \sqrt{m+1} \;\! 
c_{k,l,m+1,n} \varPhi_{k,l,m,n} (\alpha) \, , 
\label{6.4c}
\\
\langle \bar{\alpha} | \hat{\bar{a}}{}^{\dot{3}} | \varPhi^{(a)} \rangle 
&=\sum_{k, l \in  \Bbb{Z}^{-}, \;\! m, n \in \Bbb{N}_{0}} \sqrt{n+1} \;\!  
c_{k,l,m,n+1} \varPhi_{k,l,m,n} (\alpha) \, . 
\label{6.4d}
\end{align}
\end{subequations}
Here, we should note that  
$\langle \bar{\alpha} |  \hat{\bar{a}}{}^{\dot{2}} | \varPhi_{k,l,0,n} \rangle
=\langle \bar{\alpha} |  \hat{\bar{a}}{}^{\dot{3}} | \varPhi_{k,l,m,0} \rangle=0$. 
It is clear from Eq. (\ref{6.4}) that the domain of each $\hat{\bar{a}}{}^{\dot{A}}$, 
denoted by $\mathsf{D}^{(a)}(\hat{\bar{a}}{}^{\dot{A}})$, is a linear subspace of $\mathsf{H}^{(a)}$.  
As easily seen, 
$\mathsf{D}^{(a)}(\hat{\bar{a}}{}^{\dot{A}})$ is identical to $\mathsf{D}^{(a)}(\hat{a}{}^{A})$, 
i.e., $\mathsf{D}^{(a)}(\hat{\bar{a}}{}^{\dot{A}})=\mathsf{D}^{(a)}(\hat{a}{}^{A})$.  
In common with $\hat{a}{}^{A}$, the operator $\hat{\bar{a}}{}^{\dot{A}}$ is unbounded and hence is 
treated as a well-defined operator on 
the subspace $\mathsf{D}^{(a)}(\hat{a}{}^{A})$, not on the whole of $\mathsf{H}^{(a)}$.  
Since we treat all the operators $\hat{a}{}^{A}$ and $\hat{\bar{a}}{}^{\dot{A}}$ simultaneously, 
we have to consider the domain $\mathsf{D}^{(a)}:=\bigcap_{A=0}^{3} \mathsf{D}^{(a)}(\hat{a}{}^{A})$  
common to these operators. 
Evidently $\mathsf{D}^{(a)}$ is dense (in $\mathsf{H}^{(a)}$),      
and for this reason, we may regard $\hat{a}{}^{A}$ and $\hat{\bar{a}}{}^{\dot{A}}$ as  
operators on $\mathsf{H}^{(a)}$.

Using Eqs. (\ref{6.3a}) and (\ref{5.21}), we can show that  
\begin{align}
\langle  \varPsi^{(a)} | \hat{a}{}^{0} | \varPhi^{(a)} \rangle 
&=
\sum_{\substack{k^{\prime}, l^{\prime}, k, l  \in  \Bbb{Z}^{-} \\ 
m^{\prime}, n^{\prime}, m, n \in \Bbb{N}_{0}}} 
\overline{b_{k^{\prime},l^{\prime},m^{\prime},n^{\prime}}} \sqrt{-k} \;\! c_{k-1,l,m,n} 
\langle \varPhi_{k^{\prime},l^{\prime},m^{\prime},n^{\prime}} | \varPhi_{k,l,m,n}\rangle 
\notag
\\
&=\sum_{k, l \in  \Bbb{Z}^{-}, \;\! m, n \in \Bbb{N}_{0}} 
\sqrt{-k} \: \overline{b_{k,l,m,n}} \;\! c_{k-1,l,m,n} \,. 
\label{6.5}
\end{align}
Also, using Eqs. (\ref{6.4a}) and (\ref{5.21}), we have 
\begin{align}
\langle  \varPhi^{(a)} | \hat{\bar{a}}{}^{\dot{0}} | \varPsi^{(a)} \rangle 
&=
\sum_{\substack{k^{\prime}, l^{\prime}, k, l  \in  \Bbb{Z}^{-} \\ 
m^{\prime}, n^{\prime}, m, n \in \Bbb{N}_{0}}} 
\overline{c_{k^{\prime},l^{\prime},m^{\prime},n^{\prime}}} \sqrt{-k-1} \, b_{k+1,l,m,n} 
\langle \varPhi_{k^{\prime},l^{\prime},m^{\prime},n^{\prime}} | \varPhi_{k,l,m,n}\rangle 
\notag
\\
&=\sum_{k, l \in  \Bbb{Z}^{-}, \;\! m, n \in \Bbb{N}_{0}} 
\sqrt{-k} \: \overline{c_{k-1,l,m,n}}\, b_{k,l,m,n} \,. 
\label{6.6}
\end{align}
Then, it is obvious that 
$\overline{\langle  \varPsi^{(a)} | \hat{a}{}^{0} | \varPhi^{(a)} \rangle} 
=\langle  \varPhi^{(a)} | \hat{\bar{a}}{}^{\dot{0}} | \varPsi^{(a)} \rangle$. 
In addition to this, similar relations can be found for the remaining operators 
$\hat{a}{}^{A}$ and $\hat{\bar{a}}{}^{\dot{A}}$ ($A=1,2,3$). 
Thus, for $A=0, 1, 2, 3$, we have 
\begin{align}
\overline{\langle  \varPsi^{(a)} | \hat{a}{}^{A} | \varPhi^{(a)} \rangle} 
=\langle  \varPhi^{(a)} | \hat{\bar{a}}{}^{\dot{A}} | \varPsi^{(a)} \rangle \,, 
\qquad \varPhi^{(a)}, \varPsi^{(a)} \in \mathsf{D}^{(a)} . 
\label{6.7}
\end{align}
This shows that $\hat{\bar{a}}{}^{\dot{A}}$ is represented on $\mathsf{D}^{(a)}$ as 
the adjoint operator of $\hat{a}{}^{A}$.

\subsection{Case (b)}

In the case (b), we consider the linear combination 
\begin{align}
\varPhi^{(b)}(\alpha):= 
\sum_{k, l \in \Bbb{N}_{0}, \;\! m, n \in  \Bbb{Z}^{-}} 
c_{k,l,m,n} \varPhi_{k,l,m,n} (\alpha) \,, 
\quad c_{k,l,m,n} \in \Bbb{C} \,, 
\label{6.8}
\end{align}
whose norm squared is defined from Eq. (\ref{5.25}) as 
\begin{align}
\langle \varPhi^{(b)} | \varPhi^{(b)} \rangle 
=\sum_{k, l \in \Bbb{N}_{0}, \;\! m, n \in  \Bbb{Z}^{-}} (-1){}^{2s} |c_{k,l,m,n}|^{2} \,.
\label{6.9}
\end{align}
Obviously this is an indefinite norm squared. 
Following the case (a), one may naively choose 
$\mathsf{W}^{(b)}:=\big\{ \varPhi^{(b)}(\alpha) \,\big| \:\,
|\langle \varPhi^{(b)} | \varPhi^{(b)} \rangle| <\infty \big\}$ as an appropriate function space. 
However, $\mathsf{W}^{(b)}$ is not a linear space, because it does not close under 
the addition of two arbitrary elements of $\mathsf{W}^{(b)}$.  
(In fact, we can give the elements $\varPhi_{1}^{(b)}$ and $\varPhi_{2}^{(b)}$ such that  
$\langle \varPhi_{1}^{(b)} | \varPhi_{1}^{(b)} \rangle=
\langle \varPhi_{2}^{(b)} | \varPhi_{2}^{(b)} \rangle=0$, while 
$\langle \varPhi_{1}^{(b)} +\varPhi_{2}^{(b)} | \varPhi_{1}^{(b)}+\varPhi_{2}^{(b)} \rangle 
\rightarrow \infty$.) 
Instead of $\mathsf{W}^{(b)}$, now we consider the function space 
\begin{align}
\mathsf{H}^{\prime \:\! (b)} &:=
\big\{ \varPhi^{(b)}(\alpha) \,\big|\, 
\exists\:\! k_{0}\in \Bbb{N}_{0}, \:\! \exists\:\! m_{0}\in \Bbb{Z}^{-} \:\: \mbox{s.t.} 
\notag 
\\
& \qquad\; \forall\:\! k, l \geq k_{0}, \;\! \forall\:\! m, n  \leq m_{0}  \Rightarrow c_{k,l,m,n}=0 \big\}  \,. 
\label{6.10}
\end{align}
This is precisely the set of all possible {\em finite} linear combinations of 
the basis functions $\{ \varPhi_{k,l,m,n} \}_{k, l \in  \Bbb{N}_{0}, \;\!  m, n \in \Bbb{Z}^{-}}$. 
Hence, we can simply express $\mathsf{H}^{\prime \:\! (b)}$ as 
\begin{align}
\mathsf{H}^{\prime \:\! (b)}:=\Bigg\{ \varPhi^{(b)}(\alpha) \,\Bigg| \;
\varPhi^{(b)}(\alpha)= 
\Sumprime_{k, l \in \Bbb{N}_{0}, \;\! m, n \in  \Bbb{Z}^{-}} 
c_{k,l,m,n} \varPhi_{k,l,m,n} (\alpha) \Bigg\} \,, 
\label{6.11}
\end{align}
where $\Sumprime$ denotes a {\em finite} sum. 
Clearly, $\mathsf{H}^{\prime \:\! (b)}$ is a linear space.
By equipping $\mathsf{H}^{\prime \:\! (b)}$ with the inner product 
$\langle \varPsi^{(b)} | \varPhi^{(b)} \rangle  
=\Sumprime_{k, l \in \Bbb{N}_{0}, \;\! m, n \in  \Bbb{Z}^{-}}
(-1)^{2s}\;\! \overline{b_{k,l,m,n}}\;\! c_{k,l,m,n}$ compatible with 
the norm squared (\ref{6.9}),  
$\mathsf{H}^{\prime \:\! (b)}$ is established as an indefinite-metric pre-Hilbert space 
(or an indefinite inner product space).  
Here, the $b_{k,l,m,n}\in \Bbb{C}$ are coefficients of $\varPsi^{(b)} \in \mathsf{H}^{\prime \:\! (b)}$.  
To make a scrupulous analysis, we need to consider the Hilbert space that is defined as a completion 
of $\mathsf{H}^{\prime \:\! (b)}$. 
However, we cannot define such a Hilbert space by using only the indefinite norm squared (\ref{6.9}), 
because this norm squared does not lead to  
the notions of convergent sequence, Cauchy sequence and completeness. 
For this reason, we have to make do with $\mathsf{H}^{\prime \:\! (b)}$ 
for the present to proceed with our investigation in spite of the lack of strictness.  
[In addition to $\langle \varPsi^{(b)}|\varPhi^{(b)} \rangle$, 
we can equip $\mathsf{H}^{\prime \:\! (b)}$ with the positive-definite inner product 
$(\varPsi^{(b)} | \varPhi^{(b)}): =\langle \varPsi^{(b)} |\mathcal{J}^{(b)}| \varPhi^{(b)} \rangle  
=\Sumprime_{k, l \in \Bbb{N}_{0}, \;\! m, n \in  \Bbb{Z}^{-}} \overline{b_{k,l,m,n}}\;\! c_{k,l,m,n}$,   
provided that the metric $\mathcal{J}^{(b)}$ defined by   
$\langle \varPhi_{k,l,m,n} |\mathcal{J}^{(b)}|\varPhi_{k^{\prime},l^{\prime},m^{\prime},n^{\prime}} \rangle 
=(-1)^{2s} \langle \varPhi_{k,l,m,n} | \varPhi_{k^{\prime},l^{\prime},m^{\prime},n^{\prime}} \rangle$ 
is given to $\mathsf{H}^{\prime \:\! (b)}$. 
The indefinite-metric pre-Hilbert space $\mathsf{H}^{\prime \:\! (b)}$ with $\mathcal{J}^{(b)}$ is recognized  
as a {\em Krein space}.\cite{Kre, AI, Rod}   
With the aid of $\mathcal{J}^{(b)}$, we can define a Hilbert space $\mathsf{H}^{(b)}$ as 
the completion of $\mathsf{H}^{\prime \:\! (b)}$ with respect to the norm  
$\sqrt{(\varPhi^{(b)} | \varPhi^{(b)})}$. Then it becomes possible to 
develop an argument similar to that in the case (a).]

By using Eqs. (\ref{4.11}) and (\ref{5.24}), 
application of the operators $\hat{a}{}^{A}$ to $\varPhi^{(b)}$ can be evaluated as 
\begin{subequations}
\label{6.12}
\begin{align}
\langle \bar{\alpha} | \hat{a}{}^{0} | \varPhi^{(b)} \rangle
&=\Sumprime_{k, l \in \Bbb{N}_{0}, \;\! m, n \in  \Bbb{Z}^{-}} \sqrt{k} \;\! 
c_{k-1,l,m,n} \varPhi_{k,l,m,n} (\alpha) \, , 
\label{6.12a}
\\
\langle \bar{\alpha} | \hat{a}{}^{1} | \varPhi^{(b)} \rangle
&=\Sumprime_{k, l \in \Bbb{N}_{0}, \;\! m, n \in  \Bbb{Z}^{-}} \sqrt{l} \;\! 
c_{k,l-1,m,n} \varPhi_{k,l,m,n} (\alpha) \, , 
\label{6.12b}
\\
\langle \bar{\alpha} | \hat{a}{}^{2} | \varPhi^{(b)} \rangle
&=\Sumprime_{k, l \in \Bbb{N}_{0}, \;\! m, n \in  \Bbb{Z}^{-}} 
\sqrt{-m} \;\!  c_{k,l,m-1,n} \varPhi_{k,l,m,n} (\alpha) \, , 
\label{6.12c}
\\
\langle \bar{\alpha} | \hat{a}{}^{3} | \varPhi^{(b)} \rangle
&=\Sumprime_{k, l \in \Bbb{N}_{0}, \;\! m, n \in  \Bbb{Z}^{-}} 
\sqrt{-n} \;\!  c_{k,l,m,n-1} \varPhi_{k,l,m,n} (\alpha) \,. 
\label{6.12d}
\end{align}
\end{subequations}
Here, we should note that  
$\langle \bar{\alpha} | \hat{a}{}^{2} | \varPhi_{k,l,-1,n} \rangle
=\langle \bar{\alpha} | \hat{a}{}^{3} | \varPhi_{k,l,m,-1} \rangle=0$. 
Similarly, using Eqs. (\ref{4.12}) and (\ref{5.24}), we obtain 
\begin{subequations}
\label{6.13}
\begin{align}
\langle \bar{\alpha} | \hat{\bar{a}}{}^{\dot{0}} | \varPhi^{(b)} \rangle 
&=\Sumprime_{k, l \in \Bbb{N}_{0}, \;\! m, n \in  \Bbb{Z}^{-}} 
(-1)\sqrt{k+1} \;\!  c_{k+1,l,m,n} \varPhi_{k,l,m,n} (\alpha) \, , 
\label{6.13a}
\\
\langle \bar{\alpha} | \hat{\bar{a}}{}^{\dot{1}} | \varPhi^{(b)} \rangle 
&=\Sumprime_{k, l \in \Bbb{N}_{0}, \;\! m, n \in  \Bbb{Z}^{-}} 
(-1) \sqrt{l+1} \;\!  c_{k,l+1,m,n} \varPhi_{k,l,m,n} (\alpha) \, , 
\label{6.13b}
\\
\langle \bar{\alpha} | \hat{\bar{a}}{}^{\dot{2}} | \varPhi^{(a)} \rangle 
&=\Sumprime_{k, l \in \Bbb{N}_{0}, \;\! m, n \in  \Bbb{Z}^{-}} 
(-1)  \sqrt{-m-1} \;\! c_{k,l,m+1,n} \varPhi_{k,l,m,n} (\alpha) \, , 
\label{6.13c}
\\
\langle \bar{\alpha} | \hat{\bar{a}}{}^{\dot{3}} | \varPhi^{(a)} \rangle 
&=\Sumprime_{k, l \in \Bbb{N}_{0}, \;\! m, n \in  \Bbb{Z}^{-}} 
(-1) \sqrt{-n-1} \;\! c_{k,l,m,n+1} \varPhi_{k,l,m,n} (\alpha) \, . 
\label{6.13d}
\end{align}
\end{subequations}
Here, it should be noted that  
$\langle \bar{\alpha} |  \hat{\bar{a}}{}^{\dot{0}} | \varPhi_{0,l,m,n} \rangle
=\langle \bar{\alpha} |  \hat{\bar{a}}{}^{\dot{1}} | \varPhi_{k,0,m,n} \rangle=0$. 
As seen from Eqs. (\ref{6.12}) and (\ref{6.13}), each of the   
$\langle \bar{\alpha} | \hat{a}{}^{A} | \varPhi^{(b)} \rangle $ and 
$\langle \bar{\alpha} | \hat{\bar{a}}{}^{\dot{A}} | \varPhi^{(b)} \rangle$ 
is expressed as a finite linear combination of the basis functions 
$\{\varPhi_{k,l,m,n}\}_{k, l \in  \Bbb{N}_{0}, \;\! m, n \in \Bbb{Z}^{-}}$ of  
$\mathsf{H}^{\prime \:\! (b)}$. 
This implies that $\hat{a}{}^{A}$ and $\hat{\bar{a}}{}^{\dot{A}}$ are 
well-defined operators on $\mathsf{H}^{\prime \:\! (b)}$. 
Also, using Eqs. (\ref{6.12}) and (\ref{6.13}), it can be verified that 
\begin{align}
\overline{\langle  \varPsi^{(b)} | \hat{a}{}^{A} | \varPhi^{(b)} \rangle} 
=\langle  \varPhi^{(b)} | \hat{\bar{a}}{}^{\dot{A}} | \varPsi^{(b)} \rangle \,, 
\qquad \varPhi^{(b)}, \varPsi^{(b)} \in \mathsf{H}^{\prime \:\! (b)} . 
\label{6.14}
\end{align}
We thus see that 
$\hat{\bar{a}}{}^{\dot{A}}$ is represented on $\mathsf{H}^{\prime \:\! (b)}$ as 
the adjoint operator of $\hat{a}{}^{A}$. 
[As mentioned above, we can define the Hilbert space $\mathsf{H}^{(b)}$ 
from $\mathsf{H}^{\prime \:\! (b)}$ with the aid of the metric $\mathcal{J}^{(b)}$. 
Then, $\hat{a}{}^{A}$ and $\hat{\bar{a}}{}^{\dot{A}}$ can be treated as well-defined operators on 
their common domain $\mathsf{D}^{(b)}(\hat{a}{}^{A})=\mathsf{D}^{(b)}(\hat{\bar{a}}{}^{\dot{A}})
\subset \mathsf{H}^{(b)}$, not on the whole of $\mathsf{H}^{(b)}$. 
Here, for instance, $\mathsf{D}^{(b)}(\hat{a}{}^{0})$ is given by 
$\mathsf{D}^{(b)}(\hat{a}{}^{0}):=\big\{ \varPhi^{(b)}(\alpha) \in \mathsf{H}^{(b)} \big| 
\sum_{k, l \in \Bbb{N}_{0}, \;\! m, n \in \Bbb{Z}^{-}} k\:\!|c_{k,l,m,n}|^{2} < \infty \big\}$.]

\subsection{Case (c1)}

In the case (c1), we consider the linear combination 
\begin{align}
\varPhi^{(c1)}(\alpha):= 
\sum_{k, n \in \Bbb{Z}^{-}, \;\! l, m \in  \Bbb{N}_{0}} 
c_{k,l,m,n} \varPhi_{k,l,m,n} (\alpha) \,, 
\quad c_{k,l,m,n} \in \Bbb{C} \,, 
\label{6.15}
\end{align}
whose norm squared is defined from Eq. (\ref{5.29}) as 
\begin{align}
\langle \varPhi^{(c1)} | \varPhi^{(c1)} \rangle 
=\sum_{k, n \in \Bbb{Z}^{-}, \;\! l, m \in  \Bbb{N}_{0}}
(-1)^{l-n} |c_{k,l,m,n}|^{2} \,.
\label{6.16}
\end{align}
This is an indefinite norm squared. Therefore, as in the case (b),  
we now make do with the pre-Hilbert space
\begin{align}
\mathsf{H}^{\prime \:\! (c1)}:=\Bigg\{ \varPhi^{(c1)}(\alpha) \,\Bigg| \; 
\varPhi^{(c1)}(\alpha)= 
\Sumprime_{k, n \in \Bbb{Z}^{-}, \;\! l, m \in  \Bbb{N}_{0}} 
c_{k,l,m,n} \varPhi_{k,l,m,n} (\alpha) \Bigg\} 
\label{6.17}
\end{align}
equipped with the inner product 
$\langle \varPsi^{(c1)} | \varPhi^{(c1)} \rangle  
=\Sumprime_{k, n \in \Bbb{Z}^{-}, \;\! l, m \in  \Bbb{N}_{0}} 
(-1)^{l-n}\;\! \overline{b_{k,l,m,n}} 
\;\! c_{k,l,m,n}$. Here, the $b_{k,l,m,n}\in \Bbb{C}$ are coefficients of 
$\varPsi^{(c1)} \in \mathsf{H}^{\prime \:\! (c1)}$.  
[In common with the case (b), we can define a Hilbert space $\mathsf{H}^{(c1)}$ 
as the completion of $\mathsf{H}^{\prime \:\! (c1)}$ with respect to 
the norm $\sqrt{\langle \varPsi^{(c1)} |\mathcal{J}^{(c1)}| \varPhi^{(c1)} \rangle}$, 
provided that the metric $\mathcal{J}^{(c1)}$ satisfying 
$\langle \varPhi_{k,l,m,n} |\mathcal{J}^{(c1)}|\varPhi_{k^{\prime},l^{\prime},m^{\prime},n^{\prime}} \rangle 
=(-1)^{l-n} \langle \varPhi_{k,l,m,n} | \varPhi_{k^{\prime},l^{\prime},m^{\prime},n^{\prime}} \rangle$ 
is given to $\mathsf{H}^{\prime \:\! (c1)}$.]

Application of the operators $\hat{a}{}^{A}$ to $\varPhi^{(c1)}$ 
can be evaluated by using Eqs. (\ref{4.11}) and (\ref{5.28}) as
\begin{subequations}
\label{6.18}
\begin{align}
\langle \bar{\alpha} | \hat{a}{}^{0} | \varPhi^{(c1)} \rangle
&=\Sumprime_{k, n \in \Bbb{Z}^{-}, \;\! l, m \in  \Bbb{N}_{0}} 
\sqrt{-k} \;\! 
c_{k-1,l,m,n} \varPhi_{k,l,m,n} (\alpha) \, , 
\label{6.18a}
\\
\langle \bar{\alpha} | \hat{a}{}^{1} | \varPhi^{(c1)} \rangle
&=\Sumprime_{k, n \in \Bbb{Z}^{-}, \;\! l, m \in  \Bbb{N}_{0}} 
\sqrt{l} \;\! 
c_{k,l-1,m,n} \varPhi_{k,l,m,n} (\alpha) \, , 
\label{6.18b}
\\
\langle \bar{\alpha} | \hat{a}{}^{2} | \varPhi^{(c1)} \rangle
&=\Sumprime_{k, n \in \Bbb{Z}^{-}, \;\! l, m \in  \Bbb{N}_{0}} 
\sqrt{m} \;\!  c_{k,l,m-1,n} \varPhi_{k,l,m,n} (\alpha) \, , 
\label{6.18c}
\\
\langle \bar{\alpha} | \hat{a}{}^{3} | \varPhi^{(c1)} \rangle
&=\Sumprime_{k, n \in \Bbb{Z}^{-}, \;\! l, m \in  \Bbb{N}_{0}} 
\sqrt{-n} \;\!  c_{k,l,m,n-1} \varPhi_{k,l,m,n} (\alpha) \,. 
\label{6.18d}
\end{align}
\end{subequations}
Here, we should note that  
$\langle \bar{\alpha} | \hat{a}{}^{0} | \varPhi_{-1,l,m,n} \rangle
=\langle \bar{\alpha} | \hat{a}{}^{3} | \varPhi_{k,l,m,-1} \rangle=0$. 
Similarly, using Eqs. (\ref{4.12}) and (\ref{5.28}), we have  
\begin{subequations}
\label{6.19}
\begin{align}
\langle \bar{\alpha} | \hat{\bar{a}}{}^{\dot{0}} | \varPhi^{(c1)} \rangle 
&=\Sumprime_{k, n \in \Bbb{Z}^{-}, \;\! l, m \in  \Bbb{N}_{0}} 
\sqrt{-k-1} \;\!  c_{k+1,l,m,n} \varPhi_{k,l,m,n} (\alpha) \, , 
\label{6.19a}
\\
\langle \bar{\alpha} | \hat{\bar{a}}{}^{\dot{1}} | \varPhi^{(c1)} \rangle 
&=\Sumprime_{k, n \in \Bbb{Z}^{-}, \;\! l, m \in  \Bbb{N}_{0}} 
(-1) \sqrt{l+1} \;\!  c_{k,l+1,m,n} \varPhi_{k,l,m,n} (\alpha) \, , 
\label{6.19b}
\\
\langle \bar{\alpha} | \hat{\bar{a}}{}^{\dot{2}} | \varPhi^{(c1)} \rangle 
&=\Sumprime_{k, n \in \Bbb{Z}^{-}, \;\! l, m \in  \Bbb{N}_{0}} 
\sqrt{m+1} \;\! c_{k,l,m+1,n} \varPhi_{k,l,m,n} (\alpha) \, , 
\label{6.19c}
\\
\langle \bar{\alpha} | \hat{\bar{a}}{}^{\dot{3}} | \varPhi^{(c1)} \rangle 
&=\Sumprime_{k, n \in \Bbb{Z}^{-}, \;\! l, m \in  \Bbb{N}_{0}} 
(-1) \sqrt{-n-1} \;\! c_{k,l,m,n+1} \varPhi_{k,l,m,n} (\alpha) \, . 
\label{6.19d}
\end{align}
\end{subequations}
Here, it should be noted that  
$\langle \bar{\alpha} |  \hat{\bar{a}}{}^{\dot{1}} | \varPhi_{k,0,m,n} \rangle
=\langle \bar{\alpha} |  \hat{\bar{a}}{}^{\dot{2}} | \varPhi_{k,l,0,n} \rangle=0$. 
As seen from Eqs. (\ref{6.18}) and (\ref{6.19}),  
each of the $\langle \bar{\alpha} | \hat{a}{}^{A} | \varPhi^{(c1)} \rangle $ and 
$\langle \bar{\alpha} | \hat{\bar{a}}{}^{\dot{A}} | \varPhi^{(c1)} \rangle$ 
is expressed as a finite linear combination of the basis functions 
$\{\varPhi_{k,l,m,n}\}_{k, n \in \Bbb{Z}^{-}, \;\! l, m \in  \Bbb{N}_{0}}$  
of $\mathsf{H}^{\prime \:\! (c1)}$. 
In this way, $\hat{a}{}^{A}$ and $\hat{\bar{a}}{}^{\dot{A}}$ are verified to be 
well-defined operators on $\mathsf{H}^{\prime \:\! (c1)}$. 
By using Eqs. (\ref{6.18}) and (\ref{6.19}), it can be proved that 
\begin{align}
\overline{\langle  \varPsi^{(c1)} | \hat{a}{}^{A} | \varPhi^{(c1)} \rangle} 
=\langle  \varPhi^{(c1)} | \hat{\bar{a}}{}^{\dot{A}} | \varPsi^{(c1)} \rangle \,, 
\qquad \varPhi^{(c1)}, \varPsi^{(c1)} \in \mathsf{H}^{\prime \:\! (c1)} . 
\label{6.20}
\end{align}
We thus see that $\hat{\bar{a}}{}^{\dot{A}}$ is represented on $\mathsf{H}^{\prime \:\! (c1)}$ as 
the adjoint operator of $\hat{a}{}^{A}$.

The other possible three cases, namely  
(c2), (c3), and (c4), can be discussed within the framework of the case (c1) 
by the interchange of $k$ and $l$ and/or that of $m$ and $n$. 
The pre-Hilbert spaces for these three cases are found from Eq. (\ref{6.17}) to be  
\begin{align}
\mathsf{H}^{\prime \:\! (c2)}:
&=\Bigg\{ \varPhi^{(c2)}(\alpha) \,\Bigg| \; \varPhi^{(c2)}(\alpha)= 
\Sumprime_{l, m \in \Bbb{Z}^{-}, \;\! k, n \in  \Bbb{N}_{0}} 
c_{k,l,m,n} \varPhi_{k,l,m,n} (\alpha) \Bigg\} \,, 
\label{6.21a}
\\
\mathsf{H}^{\prime \:\! (c3)}:
&=\Bigg\{ \varPhi^{(c3)}(\alpha) \,\Bigg| \; \varPhi^{(c3)}(\alpha)= 
\Sumprime_{k, m \in \Bbb{Z}^{-}, \;\! l, n \in  \Bbb{N}_{0}} 
c_{k,l,m,n} \varPhi_{k,l,m,n} (\alpha) \Bigg\} \,, 
\label{6.21b}
\\
\mathsf{H}^{\prime \:\! (c4)}: 
&=\Bigg\{ \varPhi^{(c4)}(\alpha) \,\Bigg| \; \varPhi^{(c4)}(\alpha)= 
\Sumprime_{l, n \in \Bbb{Z}^{-}, \;\! k, m \in  \Bbb{N}_{0}} 
c_{k,l,m,n} \varPhi_{k,l,m,n} (\alpha) \Bigg\} \,, 
\label{6.21c}
\end{align}
where each set is equipped with the inner product that is defined from 
$\langle \varPsi^{(c1)} | \varPhi^{(c1)} \rangle$ by an appropriate permutation 
of $k$, $l$, $m$, and $n$ under the summation symbol $\Sumprime$. The relation 
$\overline{\langle  \varPsi^{(c\:\! i)} | \hat{a}{}^{A} | \varPhi^{(c\:\! i)} \rangle} 
=\langle  \varPhi^{(c\:\! i)} | \hat{\bar{a}}{}^{\dot{A}} | \varPsi^{(c\:\! i)} \rangle$  
holds for arbitrary elements 
$\varPhi^{(c\:\! i)}$ and $\varPsi^{(c\:\! i)}$ of $\mathsf{H}^{\prime \:\! (c\:\!i)}$ $(i=2, 3, 4)$.  
This shows that $\hat{\bar{a}}{}^{\dot{A}}$ is represented on $\mathsf{H}^{\prime \:\! (c\:\!i)}$ 
as the adjoint operator of $\hat{a}{}^{A}$.

We have demonstrated that in 
the spaces $\mathsf{D}^{(a)}$, $\mathsf{H}^{\prime \:\! (b)}$, and $\mathsf{H}^{\prime \:\! (c\:\!i)}$,  
the operator $\hat{\bar{a}}{}^{\dot{A}}$ is represented    
as the adjoint operator of $\hat{a}{}^{A}$. 
Its proof has been accomplished by basically using Eqs. (\ref{3.8a}) and (\ref{3.8b}), 
via Eqs. (\ref{4.11}) and (\ref{4.12}). 
This fact implies that the operators represented as 
\begin{align}
\hat{a}{}^{A} \doteq \alpha^{A}  \,,
\quad \; 
\hat{\bar{a}}{}^{\dot{A}}
\doteq - \dfrac{\partial}{\partial \alpha{}^{B}} I{}^{B \dot{A}}
+ \dfrac{1}{2} \bar{\alpha}{}^{\dot{A}} 
\label{6.22}
\end{align}
are realized in $\mathsf{D}^{(a)}$, $\mathsf{H}^{\prime \:\! (b)}$, and 
$\mathsf{H}^{\prime \:\! (c\:\!i)}$ as an adjoint pair of operators. 
Correspondingly, the twistor operators represented  as 
\begin{align}
\hat{Z}{}^{A} \doteq Z^{A} \,, 
\quad \; 
\hat{\bar{Z}}{}_{A} \doteq 
-\dfrac{\partial}{\partial Z{}^{A}} + \frac{1}{2} \bar{Z}_{A}   
\label{6.23}
\end{align}
are also realized in $\mathsf{D}^{(a)}$, $\mathsf{H}^{\prime \:\! (b)}$, and 
$\mathsf{H}^{\prime \:\! (c\:\!i)}$ as an adjoint pair of operators. 
If we adopt the twistor functions $\{f_{k,l,m,n} \}$ as basis functions instead of 
$\{\varPhi_{k,l,m,n} \}$, the differential operators  
\begin{align}
\hat{\bar{a}}{}^{\dot{A}}
\doteq - \dfrac{\partial}{\partial \alpha{}^{B}} I{}^{B \dot{A}} \,, 
\quad \; 
\hat{\bar{Z}}{}_{A} \doteq 
-\dfrac{\partial}{\partial Z{}^{A}}
\label{6.23.1}
\end{align}
are recognized in 
$\mathsf{D}^{(a)}$, $\mathsf{H}^{\prime \:\! (b)}$, and $\mathsf{H}^{\prime \:\! (c\:\!i)}$  
as the adjoint operators of $\hat{a}{}^{A} \doteq \alpha^{A}$ 
and $\hat{Z}{}^{A} \doteq Z^{A}$, respectively.  
Thus, it turns out that the representation (\ref{2.3}) is valid in
$\mathsf{D}^{(a)}$, $\mathsf{H}^{\prime \:\! (b)}$, and $\mathsf{H}^{\prime \:\! (c\:\!i)}$, 
provided $\{f_{k,l,m,n} \}$ are taken to be their basis functions.

Now, let \raisebox{-0.29ex}{\LARGE$\circ$} and \raisebox{-0.29ex}{\LARGE$\bullet$} 
be any of the symbols $a$, $b$ and $c$\:\!$i$ $(i=1, 2, 3, 4)$. 
Then, by virtue of the orthogonality condition for the eigenfunctions $\varPhi_{k,l,m,n}$, 
it follows that 
$\langle \varPhi^{(\mbox{\raisebox{-0.13ex}{\small$\circ$}})} | 
\varPhi^{(\mbox{\raisebox{-0.13ex}{\small$\bullet$}})} \rangle=0$  
for $\mbox{
\raisebox{-0.29ex}
{\LARGE$\circ$}}\neq \mbox{\raisebox{-0.29ex}
{\LARGE$\bullet$}}$. 
We can therefore consider the direct sum of the (pre-)Hilbert spaces:
\begin{align}
\mathsf{H}^{\prime}:=\mathsf{H}^{(a)} \oplus \mathsf{H}^{\prime \:\! (b)} \oplus 
\bigoplus_{i=1}^{4} \mathsf{H}^{\prime \:\! (c\:\!i)} .
\label{6.24}
\end{align}
This space contains all the eigenfunctions $\varPhi_{k,l,m,n}$ 
that have two subscript indices being negative integers  
and have two subscript indices being non-negative integers.  
The operators $\hat{a}{}^{A}$ and $\hat{\bar{a}}{}^{\dot{A}}$ are well-defined on 
\begin{align}
\mathsf{D}^{\prime}:=\mathsf{D}^{(a)} \oplus \mathsf{H}^{\prime \:\! (b)} \oplus 
\bigoplus_{i=1}^{4} \mathsf{H}^{\prime \:\! (c\:\!i)} 
\label{6.25}
\end{align}
and form an adjoint pair of operators there. 
Then it follows that the twistor operators represented as Eq. (\ref{6.22}), 
or equivalently as Eq. (\ref{6.23}), form, in $\mathsf{D}^{\prime}$,  
an adjoint pair of operators. 
By choosing $\{f_{k,l,m,n} \}$ as basis functions, 
the twistor operators represented as Eq. (\ref{2.3}) are recognized 
in $\mathsf{D}^{\prime}$ as an adjoint pair of operators. 
[If the positive-definite inner products are equipped with $\mathsf{H}^{\prime \:\! (b)}$ 
and $\mathsf{H}^{\prime \:\! (c\:\!i)}$ with the aid of the associated metrics $\mathcal{J}^{(b)}$ 
and $\mathcal{J}^{(c\:\!i)}$,  we can define the total Hilbert space 
$\mathsf{H}:=\mathsf{H}^{(a)} \oplus \mathsf{H}^{(b)} \oplus 
\bigoplus_{i=1}^{4} \mathsf{H}^{(c\:\!i)}$. This is precisely the completion of $\mathsf{H}^{\prime}$.]

\section{Penrose transforms of the simplest twistor functions} 

In Secs. V and VI, we have essentially treated the twistor functions $f_{k,l,m,n}$ 
on $\mathbf{T}^{+}$. 
The Penrose transforms of these functions yield positive-frequency massless fields 
in $\mathbb{C}\mathbf{M}$, or in other words, massless fields in the forward tube in 
$\mathbb{C}\mathbf{M}\:\!$:
\begin{align}
\mathbb{C}\mathbf{M}^{+} := \big\{\:\! (z^{\mu}) \in \mathbb{C}\mathbf{M} \, \big|\, 
z^{\mu}= x^{\mu} - i y^{\mu}, \, y_{\mu} y^{\mu} > 0\:\!, \, y^{0} > 0  \:\! \big\} \:\! , 
\label{7.1}
\end{align}
where $x^{\mu}$ and $y^{\mu}$ $(\mu=0, 1, 2, 3)$ are real numbers,  
and $y_{\mu} y^{\mu}:=(y^{0}){}^{2}-(y^{1}){}^{2}-(y^{2}){}^{2}-(y^{3}){}^{2}$.\cite{PM,PR,HT,Tak,Hug}  
[Strictly speaking, the forward and backward tubes are defined in the conformal compactification   
$\mathbb{C}\mathbf{M}^{\sharp}$ of $\mathbb{C}\mathbf{M}$. 
In this paper, however, we do not consider the $\alpha$-planes at infinity, and accordingly 
we use the (restricted) forward tube defined in Eq. (\ref{7.1}). 
This restriction is represented in $\mathbf{T}$ and $\mathbf{PT}$ as 
the condition $(\pi_{\dot{\alpha}})\ne 0$.]  
The bispinor notation $z^{\alpha \dot{\alpha}}$ and the 4-vector notation $z^{\mu}$  
are related by 
\begin{align}
\begin{pmatrix}
\: z^{0\dot{0}} &\, z^{0\dot{1}} \, \\
\: z^{1\dot{0}} &\, z^{1\dot{1}} \,
\end{pmatrix}
= \dfrac{1}{\sqrt{2}} \!
\begin{pmatrix}
\: z^{0} + z^{3}  &\: z^{1} + i z^{2} \, \\
\: z^{1} - i z^{2} &\: z^{0} - z^{3} \,
\end{pmatrix}.
\end{align}
Note that $z^{\alpha \dot{\alpha}}$ is Hermitian if and only if $z^{\mu}$ is real. 
As is known in twistor theory, a point $z=(z^{\alpha \dot{\alpha}})$ 
in $\mathbb{C}\mathbf{M}^{+}$ corresponds to the complex projective line 
\begin{align}
\mathbf{L}_{z} := \big\{ \:\! [(\alpha{}^{A})] \in \mathbf{PT} \, \big|\, 
\omega^{\alpha} = i z^{\alpha \dot{\alpha}} \pi_{\dot{\alpha}} \:\!,\, 
(\pi{}_{\dot{\alpha}})\ne 0 \:\! \big\}  
\label{7.2}
\end{align}
lying entirely in $\mathbf{PT}^{+}$. 
Here, recall that $\alpha{}^{A}$ and $(\omega^{\alpha}, \pi_{\dot{\alpha}})$ 
are related by Eq. (\ref{3.10}). 
The geometrical relation between $\mathbb{C}\mathbf{M}^{+}$ and $\mathbf{PT}^{+}$ 
can elegantly be formulated in terms of the Klein correspondence.\cite{WW}   
In the following, we actually demonstrate the Penrose transform of 
the simplest twistor function in each of the cases (a), (b) and (c\:\!$i$) $(i=1,2,3,4)$.

\subsection{Case (a)}
The simplest twistor function in the case (a) is found from  
Eqs. (\ref{4.5}) and (\ref{5.20}) to be 
\begin{align}
f_{-1, -1, 0, 0} (\alpha) 
= \frac{1}{\sqrt{2}\:\! (4\pi)^{2} \alpha^{0} \alpha^{1}} \,. 
\label{7.3}
\end{align}
The singularities of $f_{-1, -1, 0, 0}$ lie on the two hyperplanes in $\mathbf{T}$ 
that are specified by $\alpha^{0}=0$ and $\alpha^{1}=0$. 
These equations define the following two planes in $\mathbf{PT}\:\!$:  
\begin{align}
\mathbf{Q}^{0} := \big\{ \:\! [(\alpha^{A})] \in \mathbf{PT}\, \big|\,\alpha^{0}=0 \:\! \big\} \:\! , 
\quad
\mathbf{Q}^{1} := \big\{ \:\! [(\alpha^{A})] \in \mathbf{PT}\, \big|\,\alpha^{1}=0 \:\! \big\} \:\! .
\label{7.4}
\end{align} 
Obviously,   
they are not parallel.  Recalling Eq. (\ref{3.10}),  
we can write the simultaneous equations $\alpha^{0}=0$ and $\alpha^{1}=0$ 
in terms of the twistor variables $(\omega{}^{\alpha}, \pi{}_{\dot{\alpha}})$ as 
\begin{align}
\omega^{\alpha}=i u^{\alpha \dot{\alpha}} \pi_{\dot{\alpha}} \,, 
\quad \:
(u^{\alpha \dot{\alpha}}) := 
\begin{pmatrix}
\, i &\: 0 \, \\ \, 0 &\: i \,
\end{pmatrix}.
\label{7.5}
\end{align} 
With this expression, the intersection of $\mathbf{Q}^{0}$ and $\mathbf{Q}^{1}$ 
can be expressed as 
\begin{align}
\mathbf{L}_{u} = \big\{ \:\! [(\alpha{}^{A})] \in \mathbf{PT} \, \big|\, 
\omega^{\alpha} = i u^{\alpha \dot{\alpha}} \pi_{\dot{\alpha}} \:\!,\, 
(\pi{}_{\dot{\alpha}})\ne 0 \:\! \big\} \:\! . 
\label{7.6}
\end{align}
[Here, we mention necessity of the condition $(\pi{}_{\dot{\alpha}})\ne 0$ in Eq. (\ref{7.6}). 
Let $\boldsymbol{\mathcal{Q}}{}^{0}$ and $\boldsymbol{\mathcal{Q}}{}^{1}$ be the hyperplanes  
in $\mathbf{T}$ that are specified by $\alpha^{0}=0$ and $\alpha^{1}=0$, respectively. 
The intersection $\boldsymbol{\mathcal{Q}}{}^{0} \cap \boldsymbol{\mathcal{Q}}{}^{1}$ contains 
the origin $\boldsymbol{0}:=(0,0,0,0)$ in $\mathbf{T}$, and therefore $\boldsymbol{0}$ must be removed 
from $\boldsymbol{\mathcal{Q}}{}^{0} \cap \boldsymbol{\mathcal{Q}}{}^{1}$ to define 
the intersection of  $\mathbf{Q}^{0}$ and $\mathbf{Q}^{1}$. 
In fact, the intersection $\mathbf{Q}^{0} \cap \mathbf{Q}^{1}$ can be written as 
$\big( (\boldsymbol{\mathcal{Q}}{}^{0} \cap \boldsymbol{\mathcal{Q}}{}^{1}) 
\setminus \{\boldsymbol{0}\} \big)/\sim\,$,  
with the equivalence relation $(\alpha^{A}) \sim (\upsilon \alpha^{A})$ 
for all $\upsilon \in \Bbb{C}\setminus\{0\}$. The condition $(\pi_{\dot{\alpha}})\ne 0$ in 
$\mathbf{L}_{u}$ is thus necessary to state that $\boldsymbol{0}$ is removed in defining 
$\mathbf{Q}^{0} \cap \mathbf{Q}^{1}$.]   
The set $\mathbf{L}_{u} $ is precisely the complex projective line corresponding to the point 
$u=(u^{\alpha \dot{\alpha}}) \in \mathbb{C}\mathbf{M}$. 
Writing $(u^{\alpha \dot{\alpha}})$ as $(u^{\mu})=(\sqrt{2}$\:\!$ i, 0, 0, 0)$ in 4-vector notation,  
we immediately see that the point $u$ is in the backward tube in $\mathbb{C}\mathbf{M}\:\!$:
\begin{align}
\mathbb{C}\mathbf{M}^{-} := \big\{\:\! (z^{\mu}) \in \mathbb{C}\mathbf{M} \, \big|\, 
z^{\mu}= x^{\mu} - i y^{\mu}, \, y_{\mu} y^{\mu} > 0\:\!, \, y^{0} < 0  \:\! \big\} \:\! . 
\label{7.7}
\end{align}
Just as the forward tube $\mathbb{C}\mathbf{M}^{+}$ corresponds to $\mathbf{PT}^{+}$, 
the backward tube $\mathbb{C}\mathbf{M}^{-}$ corresponds to 
$\mathbf{PT}^{-}:=\big\{ \;\! [( \alpha^{A} )] \in \mathbf{PT} \, \big|\, \|\alpha\|^{2}<0  \big\}$.  
Then, since $u$ is in $\mathbb{C}\mathbf{M}^{-}$, 
it follows that $\mathbf{L}_{u}$ lies entirely in $\mathbf{PT}^{-}$. 
In fact, the condition $\|\alpha\|^{2}=-|\alpha^{2}|^{2}-|\alpha^{3}|^{2} <0$ is valid for 
an arbitrary element of $\mathbf{L}_{u}$.  
We thus see that $\mathbf{L}_{z}$ corresponding to an arbitrary point 
$z \in \mathbb{C}\mathbf{M}^{+}$ never meets $\mathbf{L}_{u}$.  
%
%
\begin{figure}
\begin{center}
\includegraphics[width=13cm,clip]{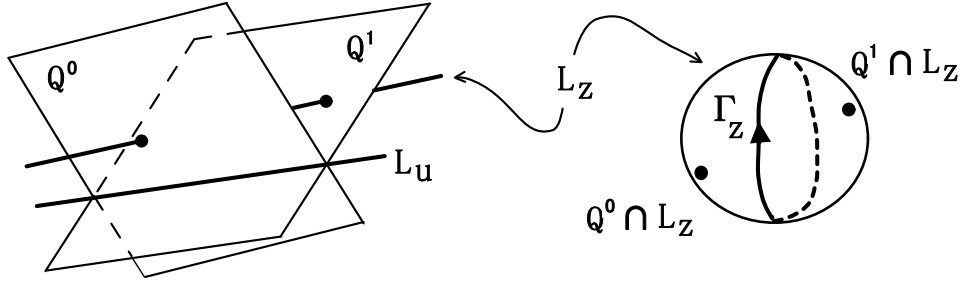} 
\end{center}
\vspace{-6mm}      
\caption{\small  
A layout drawing of geometrical objects in the case (a).  
The planes $\mathbf{Q}^{0}$ and $\mathbf{Q}^{1}$   
consist of the singularities of $f_{-1,-1,0,0}$ that are evaluated in $\mathbf{PT}$. 
The complex projective line $\mathbf{L}_{z}$ is homeomorphic to a sphere $S^{2}$,  
on which the contour $\varGamma_{z}$ is chosen in such a manner that one of 
the points $\mathbf{Q}^{0} \cap \mathbf{L}_{z}$ and $\mathbf{Q}^{1} \cap \mathbf{L}_{z}$ 
lies on either side of $\varGamma_{z}$.} 
\end{figure}

In terms of $z^{\alpha \dot{\alpha}}$, $\pi_{\dot{0}}$ and $\zeta := \pi_{\dot{1}}/\pi_{\dot{0}}\:\!$, 
the twistor function $f_{-1,-1,0,0}$ can be written as 
\begin{align}
& f_{-1,-1,0,0}(z, \pi_{\dot{0}}, \zeta) 
\notag 
\\
&=\dfrac{\sqrt{2}}{(4\pi)^{2} (\pi_{\dot{0}})^{2}\:\!  iz^{0 \dot{1}} (iz^{1 \dot{1}} + 1)
\! \left( \zeta + \dfrac{iz^{0 \dot{0}} + 1}{iz^{0 \dot{1}}} \right) \!
\left( \zeta + \dfrac{iz^{1 \dot{0}}}{iz^{1 \dot{1}} + 1} \right)} \,. 
\label{7.8}
\end{align}
This is a function on $\mathbf{T}^{+}$, as long as $z$ is a point in $\mathbb{C}\mathbf{M}^{+}$. 
The ratio $\zeta$ can be regarded as an inhomogeneous coordinate of a point 
on $\mathbf{L}_{z}\:\!(\;\!\cong \mathbb{C}\mathbf{P}^{1} \cong S^{2})$. 
Although $\mathbf{L}_{z}$ does not meet $\mathbf{L}_{u}$, it meets the planes 
$\mathbf{Q}^{0}$ and $\mathbf{Q}^{1}$ at distinct points in $\mathbf{PT}^{+}$; see Fig. 1.  
These intersection points, that is, 
$Q^{0}:=\mathbf{Q}^{0} \cap \mathbf{L}_{z}$ and 
$Q^{1}:=\mathbf{Q}^{1} \cap \mathbf{L}_{z}$ are poles of $f_{-1,-1,0,0}$. 
Noting this, now we consider the Penrose transform of $f_{-1,-1,0,0}$. 
Since $f_{-1,-1,0,0}$ is a twistor function in the case $s=0$, its Penrose transform is given by 
\begin{align}
\phi^{(a)}(z):= \frac{1}{2\pi i} \oint_{\varGamma_{z}}   f_{-1,-1,0,0}(z, \pi_{\dot{0}}, \zeta) 
\pi_{\dot{\alpha}} d\pi^{\dot{\alpha}} \,, 
\label{7.9}
\end{align}
where $\varGamma_{z}$ denotes a closed contour on $\mathbf{L}_{z}$. 
To carry out the contour integration so that it can yield a non-trivial result, 
we choose $\varGamma_{z}$ to be a topological circle 
such that only one of $Q^{0}$ and $Q^{1}$ lies on either side of $\varGamma_{z}$. 
Then, after using $\pi_{\dot{\alpha}} d\pi^{\dot{\alpha}}=(\pi_{\dot{0}})^{2} d\zeta$,  
Cauchy's theorem gives  
\begin{align}
\phi^{(a)}(z)= \frac{1}{4\sqrt{2}\;\! \pi^{2} (z_{\mu}-u_{\mu}) (z^{\mu}-u^{\mu})} \,.  
\label{7.10}
\end{align}
Here, $Q^{0}$ or $Q^{1}$ has been chosen as a simple pole 
surrounded by $\varGamma_{z}$, 
and accordingly an appropriate orientation of $\varGamma_{z}$ has been considered. 
Because $\mathbf{L}_{z}$ does not meet $\mathbf{L}_{u}$, 
the point $z$ is not null-separated from $u$; that is, 
$(z_{\mu}-u_{\mu})(z^{\mu}-u^{\mu}) \neq 0$ holds for 
$z \in \mathbb{C}\mathbf{M}^{+}$. 
[The null-separated condition $(z_{\mu}- z_{\mu}^{\prime})(z^{\mu}-z^{\prime \mu})=0$ 
($z, z^{\prime} \in \mathbb{C}\mathbf{M}^{\sharp}$) holds if and only if 
$\mathbf{L}_{z} \cap \mathbf{L}_{z^{\prime}} \neq \emptyset$ 
($\:\! \mathbf{L}_{z}, \mathbf{L}_{z^{\prime}} \subset \mathbf{PT} \:\!$). 
This implies that meeting lines in $\mathbf{PT}$ correspond to null-separated points in 
$\mathbb{C}\mathbf{M}^{\sharp}$, and vice versa.]  
In this way, $\phi^{(a)}$ is proven to be a regular function on $\mathbb{C}\mathbf{M}^{+}$. 
Noting the regularity of $\phi^{(a)}$, we can readily verify by a direct calculation  
that $\phi^{(a)}$ is a solution of the complexified Klein-Gordon equation 
$\partial_{\mu}\partial^{\mu} \phi (z)=0$ with $\partial_{\mu}:=\partial/\partial z^{\mu}$.

More generally, we can perform the Penrose transform of the twistor function 
\begin{align}
f_{k,l,m,n} (\alpha)
= C_{k,l,m,n} \frac{(\alpha^{2})^{m} (\alpha^{3})^{n}} {(\alpha^{0})^{-k} (\alpha^{1})^{-l}} \,, 
\quad   
k, l \in  \Bbb{Z}^{-}, \, m, n \in \Bbb{N}_{0} \,,
\label{7.11}
\end{align}
where $C_{k,l,m,n}$ is given in Eq. (\ref{5.20}). 
As is demonstrated in Appendix B, the massless field obtained by this transform 
takes the form of a sum of monomial functions each of which is proportional to a negative power of 
$(z_{\mu}-u_{\mu})(z^{\mu}-u^{\mu})$. 
Then the resulting massless field can be shown to be regular on 
$\mathbb{C}\mathbf{M}^{+}$.

\subsection{Case (b)}
Next we consider the simplest twistor function in the case (b): 
\begin{align}
f_{0, 0, -1, -1} (\alpha) 
= \frac{1}{\sqrt{2}\:\! (4\pi)^{2} \alpha^{2} \alpha^{3}} \,. 
\label{7.12}
\end{align}
The singularities of $f_{0, 0, -1, -1}$ lie on the two hyperplanes in $\mathbf{T}$ 
that are specified by $\alpha^{2}=0$ and $\alpha^{3}=0$. These equations define 
the following non-parallel planes in $\mathbf{PT}\:\!$:  
\begin{align}
\mathbf{Q}^{2} := \big\{ \:\! [(\alpha^{A})] \in \mathbf{PT}\, \big|\,\alpha^{2}=0 \:\! \big\} \:\! , 
\quad
\mathbf{Q}^{3} := \big\{ \:\! [(\alpha^{A})] \in \mathbf{PT}\, \big|\,\alpha^{3}=0 \:\! \big\} \:\! .  
\label{7.13}
\end{align} 
We can write the simultaneous equations $\alpha^{2}=0$ and $\alpha^{3}=0$ as 
\begin{align}
\omega^{\alpha}=i v^{\alpha \dot{\alpha}} \pi_{\dot{\alpha}} \,, 
\quad \:
(v^{\alpha \dot{\alpha}}) := 
\begin{pmatrix}
\, -i &\, 0 \, \\ \, 0 &\, -i \,
\end{pmatrix}.
\label{7.14}
\end{align} 
The intersection of $\mathbf{Q}^{2}$ and $\mathbf{Q}^{3}$ is the complex projective line 
corresponding to the point $v=(v^{\alpha \dot{\alpha}}) \in \mathbb{C}\mathbf{M}\:\!$: 
\begin{align}
\mathbf{L}_{v} = \big\{ \:\! [(\alpha{}^{A})] \in \mathbf{PT} \, \big|\, 
\omega^{\alpha} = i v^{\alpha \dot{\alpha}} \pi_{\dot{\alpha}} \:\!,\, 
(\pi{}_{\dot{\alpha}})\ne 0 \:\! \big\} \:\! . 
\label{7.15}
\end{align}
Here, the condition $(\pi_{\dot{\alpha}})\ne 0$ is necessary to state that 
the origin $\boldsymbol{0} \in \mathbf{T}$ is removed in defining 
the intersection $\mathbf{Q}^{2} \cap \mathbf{Q}^{3}$. 
Writing $(v^{\alpha \dot{\alpha}})$ as $(v^{\mu})=(-\sqrt{2}$\:\!$ i, 0, 0, 0)$ in 4-vector notation,  
we immediately see that the point $v$ is in the forward tube $\mathbb{C}\mathbf{M}^{+}$, 
and hence $\mathbf{L}_{v}$ lies entirely in $\mathbf{PT}^{+}$. 
In fact, the condition $\|\alpha\|^{2}=|\alpha^{0}|^{2}+|\alpha^{1}|^{2} >0$ is valid for 
an arbitrary element of $\mathbf{L}_{v}$.  
For this reason, $\mathbf{L}_{z}$ ($z \in \mathbb{C}\mathbf{M}^{+}$) may meet $\mathbf{L}_{v}$.

In terms of $z^{\alpha \dot{\alpha}}$, $\pi_{\dot{0}}$ and $\zeta$, 
the twistor function $f_{0,0,-1,-1}$ can be written as 
\begin{align}
& f_{0,0,-1,-1}(z, \pi_{\dot{0}}, \zeta) 
\notag 
\\
&=\dfrac{\sqrt{2}}{(4\pi)^{2} (\pi_{\dot{0}})^{2}\:\!  (-iz^{0 \dot{1}}) (-iz^{1 \dot{1}} + 1)
\! \left( \zeta + \dfrac{-iz^{0 \dot{0}} + 1}{-iz^{0 \dot{1}}} \right) \!
\left( \zeta + \dfrac{-iz^{1 \dot{0}}}{-iz^{1 \dot{1}} + 1} \right)} \,. 
\label{7.16}
\end{align}
Suppose now that $\mathbf{L}_{z}$ does not meet $\mathbf{L}_{v}$. 
Then the intersection points  
$Q^{2}:=\mathbf{Q}^{2} \cap \mathbf{L}_{z}$ and 
$Q^{3}:=\mathbf{Q}^{3} \cap \mathbf{L}_{z}$ are distinct, being two poles of $f_{0,0,-1,-1}$. 
To carry out the Penrose transform of $f_{0,0,-1,-1}$, 
we choose a closed contour $\varGamma_{z}$ on $\mathbf{L}_{z}$ in such a manner that 
only one of $Q^{2}$ and $Q^{3}$ lies on either side of $\varGamma_{z}$.   
Thereby, we have  
\begin{equation}
\begin{split}
\phi^{(b)}(z) &:= \frac{1}{2\pi i} \oint_{\varGamma_{z}}   f_{0,0,-1,-1}(z, \pi_{\dot{0}}, \zeta) 
\pi_{\dot{\alpha}} d\pi^{\dot{\alpha}} 
\\
& \; =\frac{1}{4\sqrt{2}\;\! \pi^{2} (z_{\mu}-v_{\mu}) (z^{\mu}-v^{\mu})} \,, 
\end{split}
\label{7.17}
\end{equation}
where an orientation of $\varGamma_{z}$ has been taken appropriately 
in accordance with the choice of a simple pole surrounded by $\varGamma_{z}$. 
If $\mathbf{L}_{z}$ meets $\mathbf{L}_{v}$, as seen in Fig. 2, then 
the two points $Q^{2}$ and $Q^{3}$ degenerate into the single point 
denoted by $\mathbf{L}_{z} \cap \mathbf{L}_{v}$.  
In this situation, 
the contour integral in Eq. (\ref{7.17}) is not well-defined, 
and correspondingly $\phi^{(b)}$ becomes infinite 
owing to the fact that $z$ is null-separated from $v$.  
Thus, $\phi^{(b)}$ turns out to have singularities in $\mathbb{C}\mathbf{M}^{+}$. 
%
%
\begin{figure}
\begin{center}
\includegraphics[width=13cm,clip]{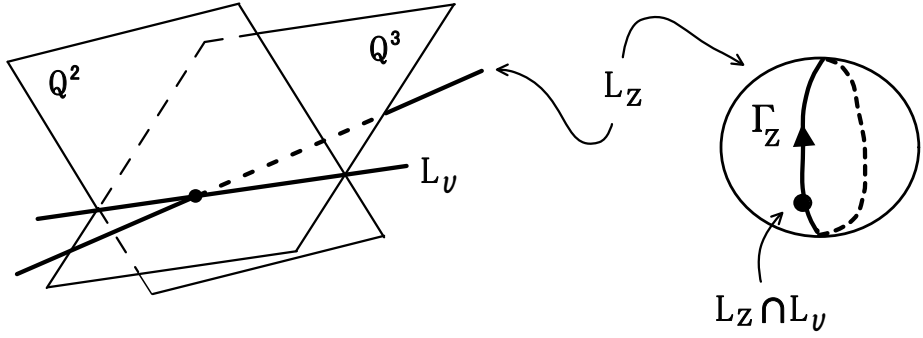} 
\end{center}
\vspace{-6mm}      
\caption{\small  
A layout drawing of geometrical objects in an exceptional situation in the case (b).  
The complex projective line $\mathbf{L}_{z}$ meets the intersection $\mathbf{L}_{v}$ of 
$\mathbf{Q}^{2}$ and $\mathbf{Q}^{3}$, and correspondingly 
the two points $\mathbf{Q}^{2} \cap \mathbf{L}_{z}$ and $\mathbf{Q}^{3} \cap \mathbf{L}_{z}$ 
on $S^{2}$ degenerate into the single point $\mathbf{L}_{z} \cap \mathbf{L}_{v}$ on $\varGamma_{z}$.  
As a result, the contour integral in Eq. (\ref{7.17}) becomes ambiguous.} 
\end{figure}

We can also perform the Penrose transform of the twistor function 
\begin{align}
f_{k,l,m,n} (\alpha)
= C_{k,l,m,n} \frac{(\alpha^{0})^{k} (\alpha^{1})^{l}}{(\alpha^{2})^{-m} (\alpha^{3})^{-n}}  \,, 
\quad   
k, l \in  \Bbb{N}_{0} \:\!, \: m, n \in  \Bbb{Z}^{-}, 
\label{7.18}
\end{align}
where $C_{k,l,m,n}$ is given in Eq. (\ref{5.24}).  
The massless field derived by this transform is a sum of monomial functions 
each of which is proportional to a negative power of 
$(z_{\mu}-v_{\mu})(z^{\mu}-v^{\mu})$. 
This implies that the resulting massless field has singularities in $\mathbb{C}\mathbf{M}^{+}$.  

\subsection{Case (c1)}

In the case (c1), the simplest twistor function is given by 
\begin{align}
f_{-1, 0, 0, -1} (\alpha) 
= \frac{1}{\sqrt{2}\:\! (4\pi)^{2} \alpha^{0} \alpha^{3}} \,.  
\label{7.19}
\end{align}
The singularities of $f_{-1, 0, 0, -1}$ constitute the planes 
$\mathbf{Q}^{0}$ and $\mathbf{Q}^{3}$ defined in Eqs. (\ref{7.4}) and (\ref{7.13}), 
respectively. These planes are not parallel, so that their intersection 
can be defined as 
\begin{align}
\mathbf{L}_{w} = \big\{ \:\! [(\alpha{}^{A})] \in \mathbf{PT} \, \big|\, 
\omega^{\alpha} = i w^{\alpha \dot{\alpha}} \pi_{\dot{\alpha}} \:\!,\, 
(\pi{}_{\dot{\alpha}})\ne 0 \:\! \big\} \:\! , 
\label{7.20}
\end{align}
with
\begin{align}
(w^{\alpha \dot{\alpha}}) := 
\begin{pmatrix}
\: i &\: 0 \, \\ \: 0 &\: -i \,
\end{pmatrix}.
\label{7.21}
\end{align}
Writing $(w^{\alpha \dot{\alpha}})$ as $(w^{\mu})=(0, 0, 0, \sqrt{2}$\:\!$ i )$ in 4-vector notation, 
we see that $\Im w_{\mu} \Im w^{\mu}=-2$ for the imaginary part of $(w^{\mu})$ 
specified by $w^{\mu}=\Re w^{\mu} -i\Im w^{\mu}$.   
The imaginary part $(\Im w^{\mu})$ is thus shown to be a spacelike vector. 
Then, it can be proven that $\mathbf{L}_{w}$ meets all three of $\mathbf{PT}^{+}$, $\mathbf{PT}^{-}$
and $\mathbf{PN}:=\big\{ \;\! [( \alpha^{A} )] \in \mathbf{PT} \, \big|\, \|\alpha\|^{2}=0  \big\}$.\cite{PR} 
This is also understood from the fact that  
$\|\alpha\|^{2}$ for an arbitrary element of $\mathbf{L}_{w}$ 
takes the indefinite form $|\alpha^{1}|^{2}-|\alpha^{2}|^{2}$. 
In this way, we see that 
$\mathbf{L}_{z}$ ($z \in \mathbb{C}\mathbf{M}^{+}$) may meet $\mathbf{L}_{w}$.

In terms of $z^{\alpha \dot{\alpha}}$, $\pi_{\dot{0}}$ and $\zeta$, 
the twistor function $f_{-1,0,0,-1}$ can be written as 
\begin{align}
& f_{-1,0,0,-1}(z, \pi_{\dot{0}}, \zeta) 
\notag 
\\
&=\dfrac{\sqrt{2}}{(4\pi)^{2} (\pi_{\dot{0}})^{2}\:\!  iz^{0 \dot{1}} (-iz^{1 \dot{1}} + 1)
\! \left( \zeta + \dfrac{iz^{0 \dot{0}} + 1}{iz^{0 \dot{1}}} \right) \!
\left( \zeta + \dfrac{-iz^{1 \dot{0}}}{-iz^{1 \dot{1}} + 1} \right)} \,. 
\label{7.22}
\end{align}
If $\mathbf{L}_{z}$ does not meet $\mathbf{L}_{w}$, 
then the intersection points $Q^{0}$ and $Q^{3}$ are distinct, being two poles of $f_{-1,0,0,-1}$. 
The Penrose transform of $f_{-1,0,0,-1}$ is carried out by choosing 
a closed contour $\varGamma_{z}$ such that only one of $Q^{0}$ and $Q^{3}$ lies on either side of 
$\varGamma_{z}$.   
Using Cauchy's theorem, we have  
\begin{equation}
\begin{split}
\phi^{(c1)}(z) &:= \frac{1}{2\pi i} \oint_{\varGamma_{z}}   f_{-1,0,0,-1}(z, \pi_{\dot{0}}, \zeta) 
\pi_{\dot{\alpha}} d\pi^{\dot{\alpha}} 
\\
& \; =\frac{1}{4\sqrt{2}\;\! \pi^{2} (z_{\mu}-w_{\mu}) (z^{\mu}-w^{\mu})} \,, 
\end{split}
\label{7.23}
\end{equation}
where an appropriate orientation of $\varGamma_{z}$ has been considered. 
If $\mathbf{L}_{z}$ meets $\mathbf{L}_{w}$, then the contour integral in Eq. (\ref{7.23}) 
is not well-defined, and correspondingly $\phi^{(c1)}$ becomes infinite.  
Hence, in common with $\phi^{(b)}$, 
the field $\phi^{(c1)}$ has singularities in $\mathbb{C}\mathbf{M}^{+}$.

We can perform the Penrose transform of the twistor function 
\begin{align}
f_{k,l,m,n} (\alpha)
= C_{k,l,m,n} \frac{(\alpha^{1})^{l} (\alpha^{2})^{m}}{(\alpha^{0})^{-k} (\alpha^{3})^{-n}}  \,, 
\quad   
k, n \in  \Bbb{Z}^{-}, \, l, m \in \Bbb{N}_{0} \,,
\label{7.24}
\end{align}
where $C_{k,l,m,n}$ is given in Eq. (\ref{5.28}).  
It turns out that the massless field derived by this transform has singularities in 
$\mathbb{C}\mathbf{M}^{+}$ that are specified by 
$(z_{\mu}-w_{\mu}) (z^{\mu}-w^{\mu})=0$.

\subsection{Case (c2)}

In the case (c2), the simplest twistor function is given by 
\begin{align}
f_{0, -1, -1, 0} (\alpha) 
= \frac{1}{\sqrt{2}\:\! (4\pi)^{2} \alpha^{1} \alpha^{2}} \,.  
\label{7.25}
\end{align}
The singularities of $f_{0, -1, -1, 0}$ constitute the non-parallel planes 
$\mathbf{Q}^{1}$ and $\mathbf{Q}^{2}$, whose intersection can be expressed as 
\begin{align}
\mathbf{L}_{-w} = \big\{ \:\! [(\alpha{}^{A})] \in \mathbf{PT} \, \big|\, 
\omega^{\alpha} = -i w^{\alpha \dot{\alpha}} \pi_{\dot{\alpha}} \:\!,\, 
(\pi{}_{\dot{\alpha}})\ne 0 \:\! \big\} \:\! , 
\label{7.26}
\end{align}
with $(w^{\alpha \dot{\alpha}})$ given in Eq. (\ref{7.21}). 
Now it is clear that just like $\mathbf{L}_{w}$ in the case (c1),  
the line $\mathbf{L}_{-w}$ meets all three of $\mathbf{PT}^{+}$, $\mathbf{PT}^{-}$ and $\mathbf{PN}$. 
In fact, $\|\alpha\|^{2}$ for an arbitrary element of $\mathbf{L}_{-w}$ 
takes the indefinite form $|\alpha^{0}|^{2}-|\alpha^{3}|^{2}$.  
For this reason, $\mathbf{L}_{z}$ ($z \in \mathbb{C}\mathbf{M}^{+}$) may meet $\mathbf{L}_{-w}$.

If $\mathbf{L}_{z}$ does not meet $\mathbf{L}_{-w}$, 
then the intersection points $Q^{1}$ and $Q^{2}$ are distinct, 
and the Penrose transform of $f_{0,-1,-1,0}$ can be carried out by choosing 
a closed contour $\varGamma_{z}$ and its orientation appropriately:            
\begin{equation}
\begin{split}
\phi^{(c2)}(z) &:= \frac{1}{2\pi i} \oint_{\varGamma_{z}}   f_{0,-1,-1,0}(z, \pi_{\dot{0}}, \zeta) 
\pi_{\dot{\alpha}} d\pi^{\dot{\alpha}} 
\\
& \; =\frac{1}{4\sqrt{2}\;\! \pi^{2} (z_{\mu}+w_{\mu}) (z^{\mu}+w^{\mu})} \,. 
\end{split}
\label{7.27}
\end{equation}
If $\mathbf{L}_{z}$ meets $\mathbf{L}_{-w}$, then the contour integral in Eq. (\ref{7.27}) 
is not well-defined, and correspondingly $\phi^{(c2)}$ becomes infinite.  
Hence, $\phi^{(c2)}$ also has singularities in $\mathbb{C}\mathbf{M}^{+}$.

We can also show that the Penrose transform of $f_{k,l,m,n}$ in the case (c2) 
yields a massless field possessing singularities in $\mathbb{C}\mathbf{M}^{+}$ that are  
specified by $(z_{\mu}+w_{\mu}) (z^{\mu}+w^{\mu})$=0.

\subsection{Case (c3)}

In the case (c3), the simplest twistor function is given by 
\begin{align}
f_{-1, 0, -1, 0} (\alpha) 
= \frac{1}{\sqrt{2}\:\! (4\pi)^{2} \alpha^{0} \alpha^{2}} \,.  
\label{7.28}
\end{align}
The singularities of $f_{-1, 0, -1, 0}$ constitute the non-parallel planes 
$\mathbf{Q}^{0}$ and $\mathbf{Q}^{2}$. Using Eq. (\ref{3.10}), we see that 
$\alpha^{0}=\alpha^{2}=0$ is equivalent to $\omega^{0}=\pi_{\dot{0}}=0$. 
Then the intersection of $\mathbf{Q}^{0}$ and $\mathbf{Q}^{2}$ is found to be 
\begin{align} 
\mathbf{L}^{02}: = \big\{ \:\! [\;\! 0, \omega^{1}, 0, \pi_{\dot{1}} \:\!] \in \mathbf{PT} \, \big|\, 
( \omega^{1}, \pi_{\dot{1}})\ne (0, 0) \big\} \:\! . 
\label{7.29}
\end{align}
This cannot be regarded as a complex projective line corresponding to a point in $\mathbb{C}\mathbf{M}$, 
because the relation $\omega^{\alpha} = i \tilde{w}^{\alpha \dot{\alpha}} \pi_{\dot{\alpha}}$ applied to 
$\mathbf{L}^{02}$ does not uniquely determine a point $\tilde{w}$ in $\mathbb{C}\mathbf{M}$. 
More precisely, this relation leaves $\tilde{w}^{0 \dot{0}}$ and $\tilde{w}^{1 \dot{0}}$ undetermined, 
whereas it determines $\tilde{w}^{0 \dot{1}}$ and $\tilde{w}^{1 \dot{1}}$ 
to be $0$ and $-i \omega^{1} / \pi_{\dot{1}}$, respectively. 
The norm squared $\|\alpha\|^{2}$ for an arbitrary element of $\mathbf{L}^{02}$ 
takes the indefinite form $|\alpha^{1}|^{2}-|\alpha^{3}|^{2}$, which fact implies that 
$\mathbf{L}^{02}$ meets all three of $\mathbf{PT}^{+}$, $\mathbf{PT}^{-}$, and $\mathbf{PN}$. 
For this reason, $\mathbf{L}_{z}$ ($z \in \mathbb{C}\mathbf{M}^{+}$) may meet $\mathbf{L}^{02}$.

If $\mathbf{L}_{z}$ does not meet $\mathbf{L}^{02}$, 
then the intersection points $Q^{0}$ and $Q^{2}$ are distinct, 
and the Penrose transform of $f_{-1,0,-1,0}$ can be carried out 
by choosing a closed contour $\varGamma_{z}$ and its orientation appropriately:            
\begin{equation}
\begin{split}
\phi^{(c3)}(z) &:= \frac{1}{2\pi i} \oint_{\varGamma_{z}}   f_{-1,0,-1,0} (z, \pi_{\dot{0}}, \zeta) 
\pi_{\dot{\alpha}} d\pi^{\dot{\alpha}} 
\\
& \; =\frac{1}{\sqrt{2} \:\! (4\pi)^{2} \:\! i z^{0\dot{1}}} \,. 
\end{split}
\label{7.30}
\end{equation}
Although $\phi^{(c3)}$ has an unusual form, it indeed satisfies 
the complexified Klein-Gordon equation provided that $z^{0\dot{1}} \neq 0$. 
If $\mathbf{L}_{z}$ meets $\mathbf{L}^{02}$, then the contour integral in Eq. (\ref{7.30}) 
is not well-defined, and correspondingly $\phi^{(c3)}$ becomes infinite. 
In fact, $z^{0 \dot{1}}=0$ holds at the points corresponding to 
$\mathbf{L}_{z} \cap \mathbf{L}^{02}$. 
Thus we see that $\phi^{(c3)}$ has singularities in $\mathbb{C}\mathbf{M}^{+}$.

We can also show that the Penrose transform of $f_{k,l,m,n}$ in the case (c3) 
yields a massless field possessing singularities in $\mathbb{C}\mathbf{M}^{+}$ that are  
specified by $z^{0 \dot{1}}=0$.

\subsection{Case (c4)}

In the case (c4), the simplest twistor function is given by 
\begin{align}
f_{0, -1, 0, -1} (\alpha) 
= \frac{1}{\sqrt{2}\:\! (4\pi)^{2} \alpha^{1} \alpha^{3}} \,.  
\label{7.31}
\end{align}
The singularities of $f_{0, -1, 0, -1}$ constitute the non-parallel planes 
$\mathbf{Q}^{1}$ and $\mathbf{Q}^{3}$, whose intersection is found to be 
\begin{align}
\mathbf{L}^{13}= \big\{ \:\! [\;\! \omega^{0}, 0, \pi_{\dot{0}}, 0 \:\!] \in \mathbf{PT} \, \big|\, 
( \omega^{0}, \pi_{\dot{0}})\ne (0, 0) \big\} \:\! . 
\label{7.32}
\end{align}
In common with $\mathbf{L}^{02}$, 
the intersection $\mathbf{L}^{13}$ cannot be regarded as a complex projective line 
corresponding to a point in $\mathbb{C}\mathbf{M}$. In the present case, 
the relation $\omega^{\alpha} = i \tilde{w}^{\alpha \dot{\alpha}} \pi_{\dot{\alpha}}$ 
leaves $\tilde{w}^{0 \dot{1}}$ and $\tilde{w}^{1 \dot{1}}$ undetermined, 
whereas it determines $\tilde{w}^{1 \dot{0}}$ and $\tilde{w}^{0 \dot{0}}$ to be 
$0$ and $-i \omega^{0} / \pi_{\dot{0}}$, respectively. 
The norm squared $\|\alpha\|^{2}$ for an arbitrary element of $\mathbf{L}^{13}$ 
takes the indefinite form $|\alpha^{0}|^{2}-|\alpha^{2}|^{2}$, which fact implies that 
$\mathbf{L}^{13}$ meets all three of $\mathbf{PT}^{+}$, $\mathbf{PT}^{-}$, and $\mathbf{PN}$.    
For this reason, $\mathbf{L}_{z}$ ($z \in \mathbb{C}\mathbf{M}^{+}$) may meet $\mathbf{L}^{13}$.

If $\mathbf{L}_{z}$ does not meet $\mathbf{L}^{13}$, 
then the intersection points $Q^{1}$ and $Q^{3}$ are distinct, 
and the Penrose transform of $f_{0,-1,0,-1}$ can be carried out 
by choosing a closed contour $\varGamma_{z}$ and its orientation appropriately:            
\begin{equation}
\begin{split}
\phi^{(c4)}(z) &:= \frac{1}{2\pi i} \oint_{\varGamma_{z}}   f_{0,-1,0,-1} (z, \pi_{\dot{0}}, \zeta) 
\pi_{\dot{\alpha}} d\pi^{\dot{\alpha}} 
\\
& \; =-\frac{1}{\sqrt{2} \:\! (4\pi)^{2} \:\! i z^{1\dot{0}}} \,. 
\end{split}
\label{7.33}
\end{equation}
It can readily be verified that $\phi^{(c4)}$ satisfies the complexified Klein-Gordon equation 
provided that $z^{1\dot{0}} \neq 0$. 
If $\mathbf{L}_{z}$ meets $\mathbf{L}^{13}$, then the contour integral in Eq. (\ref{7.33}) 
is not well-defined, and correspondingly $\phi^{(c4)}$ becomes infinite. 
In fact, $z^{1 \dot{0}}=0$ holds at the points corresponding to 
$\mathbf{L}_{z} \cap \mathbf{L}^{13}$. 
Thus we see that $\phi^{(c4)}$ has singularities in $\mathbb{C}\mathbf{M}^{+}$. 

We can also show that the Penrose transform of $f_{k,l,m,n}$ in the case (c4) 
yields a massless field possessing singularities in $\mathbb{C}\mathbf{M}^{+}$ that are 
specified by $z^{1 \dot{0}}=0$.

We conclude this section with the following remarks: 
Recalling the Penrose transforms carried out in this section, 
we observe that only the twistor functions in the case (a) lead to massless fields 
without singularities in $\mathbb{C}\mathbf{M}^{+}$, 
while the twistor functions in the other cases always lead to massless fields 
with singularities in $\mathbb{C}\mathbf{M}^{+}$. 
In this situation, we should consider only the massless fields obtained in the case (a) to be 
{\em genuine} positive-frequency massless fields in $\mathbb{C}\mathbf{M}$.\cite{PM, PR, HT, Tak}

\section{Summary and discussion} 
We have accomplished our central goal of finding (pre-)Hilbert spaces 
in twistor quantization, showing that the twistor operators 
$\hat{a}{}^{A}$ and $\hat{\bar{a}}{}^{\dot{A}}$ 
(or equivalently, $\hat{Z}{}^{A}$ and $\hat{\bar{Z}}{}^{\dot{A}}$) 
form an adjoint pair of operators in all these spaces. 
We first provided a coherent state defined as a simultaneous eigenstate of 
the operators $\hat{\bar{a}}{}^{\dot{A}}$  
and gave an explicit representation of the twistor operators by choosing 
the coherent-state vectors $\langle \bar{\alpha} |$, satisfying 
$\langle \bar{\alpha} | \hat{a}{}^{A} = \alpha^{A} \langle \bar{\alpha} |$,  
as basis vectors.  
Then, solving the helicity eigenvalue equation written in terms of $\alpha^{A}$,  
we obtained the eigenfunctions $\varPhi_{k,l,m,n} (\alpha)$ of the helicity operator. 
Also, it was shown that $\varPhi_{k,l,m,n}$ are simultaneous eigenfunctions of the 
Cartan generators of $\mathrm{SU}(2,2)$. This fact made it possible to denote 
$\varPhi_{k,l,m,n}$ as $\varPhi_{s,K,L,M}$ using the combination of eigenvalues 
$(s, K, L, M/\sqrt{2} \;\!)$.

An appropriate inner product for the helicity eigenfunctions $\varPhi_{s,K,L,M}$ was 
precisely defined as an integral over $S^{1} \times\mathbf{PT}{}^{+}$ 
in the case that the twistor functions 
$f_{s,K,L,M}=\varPhi_{s,K,L,M} \exp \!\left( -\frac{1}{2} \| \alpha \|^{2} \right)$  
are realized as functions on the twistor subspace $\mathbf{T}^{+}$. 
We performed the integration in the inner product by expressing it in  
terms of hyperbolic polar-coordinate variables   
and obtained an expression of the inner product that includes 
a multiplicative factor consisting of gamma functions (cf. Eq. (\ref{5.14})). 
By analytic continuation of the gamma functions, 
it became possible to use this expression when  
some or all of $k$, $l$, $m$ and $n$ take negative integer values. 
We also saw that the orthogonality condition 
for the eigenfunctions $\varPhi_{k,l,m,n}$ is guaranteed with this inner product. 
In particular, the orthogonality with respect to different helicity eigenvalues 
is valid for the helicity eigenfunctions with different degrees of homogeneity. 
We actually examined the inner product in the particular cases 
in which two of $k$, $l$, $m$ and $n$ are negative integers 
and the other two are non-negative integers. 
This was done by classifying the permissible combinations of $(k,l,m,n)$ into 
six cases, namely, (a), (b) and (c\:\!$i$) $(i=1,2,3,4)$. 
It was shown that the eigenfunctions $\varPhi_{k,l,m,n}$ in the case (a) are normalized to unity, 
while $\varPhi_{k,l,m,n}$ in the cases (b) and (c\:\!$i$) are normalized to either $1$ or $-1$. 
As seen from Eq. (\ref{4.6}), the helicity eigenvalue $s$ can take an arbitrary integer or 
half-integer value in all the six cases. 
This is due to the twistor quantization procedure, because the helicity at the classical level, 
given in Eq. (\ref{2.1.1}), can take only positive values when $(Z^{A}) \in \mathbf{T}^{+}$.

(Pre-)Hilbert spaces appropriate for twistor quantization were 
defined in each of the cases (a), (b), and (c\:\!$i$) as function spaces consisting of 
linear combinations of $\varPhi_{k,l,m,n}$.  
We found that the linear combinations in the case (a) have positive-definite norm squared,  
and hence the function space in this case is established as a Hilbert space. 
In contrast, the linear combinations in the cases (b) and (c\:\!$i$)  
have indefinite norm squared, and therefore we had to make do with 
indefinite-metric pre-Hilbert spaces to proceed with our investigation. 
(As was mentioned in Sec. VI,  it is possible to define Hilbert spaces   
in the cases (b) and (c\:\!$i$) with the aid of the additional metrics 
$\mathcal{J}^{(b)}$ and $\mathcal{J}^{(c\:\!i)}$). 
In the case (a), we proved that  
$\hat{\bar{a}}{}^{\dot{A}}$ is represented on a linear subspace of 
the Hilbert space as the adjoint operator of $\hat{a}{}^{A}$. 
In each of the cases (b) and (c\:\!$i$), similar proof was given for 
$\hat{a}{}^{A}$ and $\hat{\bar{a}}{}^{\dot{A}}$ defined on the corresponding pre-Hilbert space.  
These results imply that the twistor operators represented as 
Eq. (\ref{6.22}), or equivalently as Eq. (\ref{6.23}), form, on the relevant function spaces, 
an adjoint pair of operators. 
Correspondingly, the twistor operators represented as Eq. (\ref{2.3})  
are realized as an adjoint pair of operators by taking the twistor functions 
$\{ f_{k,l,m,n} \}$ as basis functions. 
Thus, we accomplished our purpose of finding appropriate (pre-)Hilbert spaces 
in which the representations (\ref{6.22}), (\ref{6.23}) and (\ref{2.3}) hold true.

We also argued the Penrose transforms of twistor functions 
on $\mathbf{T}^{+}$ in each of the cases (a), (b), and (c\:\!$i$) and derived  
the corresponding massless fields in $\mathbb{C}\mathbf{M}^{+}$. 
In particular, the Penrose transforms of the simplest twistor functions 
were demonstrated in detail by examining singularities of these functions closely. 
(In Appendix B, the Penrose transform of a general twistor function  
in the case (a) is demonstrated.) 
Then we observed that only the twistor functions in the case (a) lead to massless fields 
without singularities in $\mathbb{C}\mathbf{M}^{+}$, 
while the twistor functions in the other cases lead to massless fields 
with singularities in $\mathbb{C}\mathbf{M}^{+}$. 
We should therefore consider only the massless fields derived in the case (a) to be    
genuine positive-frequency massless fields. 
Even if we treat only the case (a), the helicity eigenvalue $s$ can take 
an arbitrary integer or half-integer value.

It should be emphasized that 
only in the case (a), we can define a (positive-definite) Hilbert space 
and also can obtain positive-frequency massless fields without singularities. 
Although the case (a) has these two remarkable properties, 
it is not clear at present whether these two are related by some profound reason. 
It is also not clear whether twistor quantization involves  
the probabilistic interpretation of twistor (wave) functions.  
If the probabilistic interpretation is required to twistor quantization, 
only the case (a) would be allowed. 
Otherwise, all the cases should be considered on an equal footing, and accordingly  
the total pre-Hilbert space defined by Eq. (\ref{6.24}), or its completion,  
may be adopted as a function space appropriate for twistor quantization.

Now, it is still left to investigate whether the inner product defined in this paper 
reproduces the scalar product on massless fields. 
This investigation will lead to finding relationship between Penrose's inner product \cite{Pen6} and ours, 
because Penrose's inner product can be obtained from 
the scalar product of massless fields in $\mathbf{M}$. 
Another possible method for this investigation is to compare 
our approach to the cohomological approach,\cite{BE}  
because the cohomological approach ensures consistency with the Penrose transform. 
Also, comparing the two approaches is necessary for formulating our approach 
in terms of cohomologies. 
In particular, it is interesting to verify density of the bases twistor functions 
$\{ f_{k,l,m,n} \}$ in each of the cases (a), (b), and (c\:\!$i$) by means of  
the cohomological method.\cite{EP} 
We hope to clarify these points, together with the above-mentioned unclear points,  
in the near future.

Finally, we note that the present paper has mainly treated twistor functions on $\mathbf{T}^{+}$   
so that positive-frequency massless fields in $\mathbb{C}\mathbf{M}$  
can be obtained by the Penrose transform. 
Of course, it is possible to treat twistor functions on the lower half of twistor space, namely, 
$\mathbf{T}^{-}:= \big\{ ( \alpha^{A} ) \in \mathbf{T} \, \big|\, \| \alpha \|^{2} <0 \big\}$.  
The Penrose transforms of such twistor functions yield negative-frequency massless fields 
in $\mathbb{C}\mathbf{M}$. 
We can immediately apply the arguments provided in this paper to 
constructing the (pre-)Hilbert spaces that consist of linear combinations of  
the helicity eigenfunctions on $\mathbf{T}^{-}$.



%
%

%

\begin{acknowledgments}
The authors would like to thank Kazuo Fujikawa, Shigefumi Naka, Takeshi Nihei, and Akitsugu Miwa   
for their encouragement and useful comments.  
The work of S. D. is supported in part by  
Grant-in-Aid for Fundamental Scientific Research from   
College of Science and Technology, Nihon University.  
\end{acknowledgments}

\newpage

\appendix
\section{The Schwinger representation of the $\mathbf{SU}\boldsymbol{(2,2)}$ Lie algebra}

In this appendix, we provide the Schwinger representation of the Lie algebra of ${\rm SU}(2,2)$, 
in which $\hat{a}{}^{A}$ and $\hat{\bar{a}}{}^{\dot{A}}$ are used as constituent operators.

An orthonormal basis of the ${\rm SU}(2,2)$ Lie algebra, or a set of generators of ${\rm SU}(2,2)$, 
is given by 
\begin{alignat}{2}
&
\varLambda_{1}  = \dfrac{1}{2}
\begin{pmatrix} 
\; 0 \:& 1 \:& 0 \:& 0 \; \\
\; 1 \:& 0 \:& 0 \:& 0 \; \\
\; 0 \:& 0 \:& 0 \:& 0 \; \\
\; 0 \:& 0 \:& 0 \:& 0 \;
\end{pmatrix}
,
&
\: 
&
\varLambda_{2}  = \dfrac{1}{2}
\begin{pmatrix} 
\; 0 \:& -i \:& 0 \:& 0 \; \\
\; i  \:& 0 \:& 0 \:& 0 \; \\
\; 0 \:& 0 \:& 0 \:& 0 \; \\
\; 0 \:& 0 \:& 0 \:& 0 \;
\end{pmatrix}
,
\notag
\\
&
\varLambda_{3} = \dfrac{1}{2}
\begin{pmatrix} 
\; 1 \:& 0 \:& 0 \:& 0 \; \\
\; 0 \:& -1 \:& 0 \:& 0 \; \\
\; 0 \:& 0 \:& 0 \:& 0 \; \\
\; 0 \:& 0 \:& 0 \:& 0 \;
\end{pmatrix}
,
&
\:
&
\varLambda_{4}  = \dfrac{1}{2}
\begin{pmatrix} 
\; 0 \:& 0 \:& 0 \:& 0 \; \\
\; 0 \:& 0 \:& 0 \:& 0 \; \\
\; 0 \:& 0 \:& 0 \:& 1 \; \\
\; 0 \:& 0 \:& 1 \:& 0 \;
\end{pmatrix}
,
\notag
\\
&
\varLambda_{5}  = \dfrac{1}{2}
\begin{pmatrix} 
\; 0 \:& 0 \:& 0 \:& 0 \; \\
\; 0 \:& 0 \:& 0 \:& 0 \; \\
\; 0 \:& 0 \:& 0 \:& -i \, \\
\; 0 \:& 0 \:& i \:& 0 \;
\end{pmatrix}
,
&
\:
&
\varLambda_{6}  = \dfrac{1}{2}
\begin{pmatrix} 
\; 0 \:& 0 \:& 0 \:& 0 \; \\
\; 0 \:& 0 \:& 0 \:& 0 \; \\
\; 0 \:& 0 \:& 1 \:& 0 \; \\
\; 0 \:& 0 \:& 0 \:& -1 \:
\end{pmatrix}
,
\notag
\\
&
\varLambda_{7}  = \dfrac{1}{2}
\begin{pmatrix} 
\; 0 \:& 0 \:&  i \:& 0 \; \\
\; 0 \:& 0 \:& 0 \:& 0 \; \\
\; i \:& 0 \:& 0 \:& 0 \; \\
\; 0 \:& 0 \:& 0 \:& 0 \;
\end{pmatrix}
,
&
\:
&
\varLambda_{8}  = \dfrac{1}{2}
\begin{pmatrix} 
\; 0 \:& 0 \:& 0 \:& i \; \\
\; 0 \:& 0 \:& 0 \:& 0 \; \\
\; 0 \:& 0 \:& 0 \:& 0 \; \\
\; i \:& 0 \:& 0 \:& 0 \;
\end{pmatrix}
,
\notag
\\
&
\varLambda_{9}  = \dfrac{1}{2}
\begin{pmatrix} 
\; 0 \:& 0 \:& 0 \:& 0 \; \\
\; 0 \:& 0 \:& i \:& 0 \; \\
\; 0 \:& i \:& 0 \:& 0 \; \\
\; 0 \:& 0 \:& 0 \:& 0 \; 
\end{pmatrix}
,
&
\:
&
\varLambda_{10}  = \dfrac{1}{2}
\begin{pmatrix} 
\; 0 \:& 0 \:& 0 \:& 0 \; \\
\; 0 \:& 0 \:& 0 \:& i \; \\
\; 0 \:& 0 \:& 0 \:& 0 \; \\
\; 0 \:& i \:& 0 \:& 0 \;
\end{pmatrix}
,
\notag
\\
&
\varLambda_{11}  = \dfrac{1}{2}
\begin{pmatrix} 
\; 0 \:& 0 \:& 1 \:& 0 \; \\
\; 0 \:& 0 \:& 0 \:& 0 \; \\
\, -1 \:& 0 \:& 0 \:& 0 \; \\
\; 0 \:& 0 \:& 0 \:& 0 \;
\end{pmatrix}
,
&
\:\qquad
&
\varLambda_{12}  = \dfrac{1}{2}
\begin{pmatrix} 
\; 0 \:& 0 \:& 0 \:& 1 \; \\
\; 0 \:& 0 \:& 0 \:& 0 \; \\
\; 0 \:& 0 \:& 0 \:& 0 \; \\
\; -1 \:& 0 \:& 0 \:& 0 \;
\end{pmatrix}
,
\notag
\end{alignat}
\begin{alignat}{2}
&
\varLambda_{13}  = \dfrac{1}{2}
\begin{pmatrix} 
\; 0 \:& 0 \:& 0 \:& 0 \; \\
\; 0 \:& 0 \:& 1 \:& 0 \; \\
\; 0 \:& -1 \:& 0 \:& 0 \; \\
\; 0 \:& 0 \:& 0 \:& 0 \; 
\end{pmatrix}
,
&
\:
&
\varLambda_{14}  = \dfrac{1}{2}
\begin{pmatrix} 
\; 0 \:& 0 \:& 0 \:& 0 \; \\
\; 0 \:& 0 \:& 0 \:& 1 \; \\
\; 0 \:& 0 \:& 0 \:& 0 \; \\
\; 0 \:& -1 \:& 0 \:& 0 \; 
\end{pmatrix}
,
\notag
\\
&
\varLambda_{15}  = \dfrac{1}{2\sqrt{2}}
\begin{pmatrix} 
\; 1 \;& 0 \:& 0 \:& 0 \; \\
\; 0 \;& 1 \:& 0 \:& 0 \; \\
\; 0 \;& 0 \:& -1 \:& 0 \; \\
\; 0 \;& 0 \:& 0 \:& -1 \: 
\end{pmatrix}
.
&
\:
&
\label{A.1}
\end{alignat}
These generators actually satisfy the following two conditions necessary for them 
to be generators of ${\rm SU}(2,2)$: 
The pseudo-Hermitian condition 
\begin{align}
\varLambda_{b}{}^{\dagger} =I \varLambda_{b} I \,, 
\quad b=1, 2, \ldots, 15 \, , 
\label{A.2}
\end{align}
with $I:= \text{diag}(1,1,-1,-1)$, and the traceless condition 
\begin{align}
\text{Tr} \varLambda{}_{b} = 0 \,.  
\label{A.3} 
\end{align}
Also, the generators $\varLambda_{b}$ fulfill the orthonormality condition  
\begin{align}
\text{Tr} (\varLambda{}_{b} \varLambda{}_{c} )
=\frac{1}{2} \eta_{bc} \,, 
\label{A.4}
\end{align}
where 
$(\eta_{bc}):
=\text{diag} (\;\! \overbrace{1,\ldots,1}^{6}, \overbrace{-1,\ldots,-1}^{8},1)$. 
Since ${\rm SU}(2,2)$ has rank 3, it possesses 3 diagonal generators, namely,  
$\varLambda_{3}$, $\varLambda_{6}$, and $\varLambda_{15}$. 
These are precisely the Cartan generators of ${\rm SU}(2,2)$ in the present matrix representation.

Using the generators $\varLambda_{b}$ and 
the twistor operators $\hat{a}{}^{A}$ and $\hat{\bar{a}}{}^{\dot{A}}$, now we define 
the operators 
\begin{align}
\hat{\varLambda}{}_{b} := \hat{\bar{a}}{}^{\dot{A}} 
(\varLambda_{b}){}_{\dot{A}}{}^{\dot{B}}
I{}_{\dot{B} C}\:\! \hat{a}{}^{C}, 
\label{A.5}
\end{align}
which can be written more precisely as  
\begin{alignat}{2}
&
\hat{\varLambda}{}_{1} 
= \dfrac{1}{2} 
\!\left( \hat{\bar{a}}{}^{\dot{0}} \hat{a}{}^{1} + \hat{\bar{a}}{}^{\dot{1}} \hat{a}{}^{0} 
\right),
&
\!\!\!\!
&
\hat{\varLambda}{}_{2} = -\dfrac{i}{2}
\!\left( \hat{\bar{a}}{}^{\dot{0}} \hat{a}{}^{1} - \hat{\bar{a}}{}^{\dot{1}} \hat{a}{}^{0} 
\right),
\notag
\\
&
\hat{\varLambda}{}_{3} 
= \dfrac{1}{2}
\!\left( \hat{\bar{a}}{}^{\dot{0}} \hat{a}{}^{0} - \hat{\bar{a}}{}^{\dot{1}} \hat{a}{}^{1}
\right),
&
\!\!\!\!
&
\hat{\varLambda}{}_{4} = -\dfrac{1}{2}
\!\left( \hat{\bar{a}}{}^{\dot{2}} \hat{a}{}^{3} + \hat{\bar{a}}{}^{\dot{3}} \hat{a}{}^{2}
\right),
\notag
\\
&
\hat{\varLambda}{}_{5} 
= \dfrac{i}{2}
\!\left( \hat{\bar{a}}{}^{\dot{2}} \hat{a}{}^{3} - \hat{\bar{a}}{}^{\dot{3}} \hat{a}{}^{2}
\right),
&
\!\!\!\!
&
\hat{\varLambda}{}_{6} = -\dfrac{1}{2}
\!\left( \hat{\bar{a}}{}^{\dot{2}} \hat{a}{}^{2} - \hat{\bar{a}}{}^{\dot{3}} \hat{a}{}^{3}
\right),
\notag
\\
&
\hat{\varLambda}{}_{7} 
= -\dfrac{i}{2}
\!\left( \hat{\bar{a}}{}^{\dot{0}} \hat{a}{}^{2} - \hat{\bar{a}}{}^{\dot{2}} \hat{a}{}^{0}
\right),
&
\!\!\!\!
&
\hat{\varLambda}{}_{8} = -\dfrac{i}{2}
\!\left( \hat{\bar{a}}{}^{\dot{0}} \hat{a}{}^{3} - \hat{\bar{a}}{}^{\dot{3}} \hat{a}{}^{0}
\right),
\notag
\\
&
\hat{\varLambda}{}_{9} 
= -\dfrac{i}{2}
\!\left( \hat{\bar{a}}{}^{\dot{1}} \hat{a}{}^{2} - \hat{\bar{a}}{}^{\dot{2}} \hat{a}{}^{1}
\right),
&
\!\!\!\!
&
\hat{\varLambda}{}_{10} = -\dfrac{i}{2}
\!\left( \hat{\bar{a}}{}^{\dot{1}} \hat{a}{}^{3} - \hat{\bar{a}}{}^{\dot{3}} \hat{a}{}^{1}
\right), 
\notag
\\
&
\hat{\varLambda}{}_{11} 
= -\dfrac{1}{2}
\!\left( \hat{\bar{a}}{}^{\dot{0}} \hat{a}{}^{2} + \hat{\bar{a}}{}^{\dot{2}} \hat{a}{}^{0}
\right), 
&
\!\!\!\!
&
\hat{\varLambda}{}_{12} = -\dfrac{1}{2}
\!\left( \hat{\bar{a}}{}^{\dot{0}} \hat{a}{}^{3} + \hat{\bar{a}}{}^{\dot{3}} \hat{a}{}^{0}
\right),
\notag
\\
&
\hat{\varLambda}{}_{13} 
= -\dfrac{1}{2}
\!\left( \hat{\bar{a}}{}^{\dot{1}} \hat{a}{}^{2} + \hat{\bar{a}}{}^{\dot{2}} \hat{a}{}^{1}
\right),
&
\!\!\!\!
&
\hat{\varLambda}{}_{14} = -\dfrac{1}{2}
\!\left( \hat{\bar{a}}{}^{\dot{1}} \hat{a}{}^{3} + \hat{\bar{a}}{}^{\dot{3}} \hat{a}{}^{1}
\right),
\notag
\\
&
\hat{\varLambda}{}_{15} 
= \dfrac{1}{2 \sqrt{2}} 
\!\left(
\hat{\bar{a}}{}^{\dot{0}} \hat{a}{}^{0} + \hat{\bar{a}}{}^{\dot{1}} \hat{a}^{1}
+ \hat{\bar{a}}{}^{\dot{2}} \hat{a}{}^{2} + \hat{\bar{a}}{}^{\dot{3}} \hat{a}^{3}
\right).
\label{A.6}
\end{alignat}
Using the commutation relations (\ref{3.3a}) and (\ref{3.3b}), together with 
the commutation relations at the matrix level  
\begin{align}
[\:\! \varLambda{}_{b}, \varLambda{}_{c}\:\!] =i f_{bc}{}^{d} \varLambda_{d} \,, 
\label{A.7}
\end{align}
we can prove that 
\begin{align}
[\:\! \hat{\varLambda}{}_{b}, \hat{\varLambda}{}_{c}\:\!]
=i f_{bc}{}^{d} \hat{\varLambda}_{d} \,. 
\label{A.8}
\end{align}
Here, the $f_{bc}{}^{d}$ denote the structure constants 
of the ${\rm SU}(2,2)$ Lie algebra. 
The commutation relations in Eq. (\ref{A.8}) show that the operators $\hat{\varLambda}{}_{b}$ 
constitute a basis of the Schwinger representation of the ${\rm SU}(2,2)$ Lie algebra.  
In this way, the Schwinger representation of the ${\rm SU}(2,2)$ Lie algebra is 
established by the use of the Weyl-Heisenberg algebra of indefinite-metric type 
that is defined by Eq. (\ref{3.3}).

In the Schwinger representation of the ${\rm SU}(2,2)$ Lie algebra,  
the quadratic Casimir operator is defined by 
\begin{align} 
\hat{C} := \hat{\Lambda}{}_{b} \eta^{bc} \hat{\Lambda}{}_{c} 
\label{A.9}
\end{align}
with $(\eta^{bc}):
=\text{diag} (\;\! \overbrace{1,\ldots,1}^{6}, \overbrace{-1,\ldots,-1}^{8},1)$.  
Substituting Eq. (\ref{A.5}) into Eq. (\ref{A.9}) and using Eq. (\ref{3.3}), we obtain 
\begin{align} 
\hat{C} = \dfrac{3}{2} \big( \hat{s}{}^{2} -1 \big) \,, 
\label{A.10}
\end{align}
where $\hat{s}$ is the helicity operator given in Eq. $(\ref{4.1})$, i.e.,  
\begin{align}
\hat{s} &= \frac{1}{2} \!\left(
\hat{\bar{a}}{}^{\dot{0}} \hat{a}{}^{0} + \hat{\bar{a}}{}^{\dot{1}} \hat{a}^{1}
- \hat{\bar{a}}{}^{\dot{2}} \hat{a}{}^{2} - \hat{\bar{a}}{}^{\dot{3}} \hat{a}^{3} \right) 
+1\,.
\label{A.11}
\end{align}
Because $\hat{s}$ commutes with all the operators $\hat{\varLambda}{}_{b}$, 
it follows that $\hat{C}$ commutes with all of the $\hat{\varLambda}{}_{b}$.  
It is obvious that $\hat{s}$, $\hat{\varLambda}_{3}$, $\hat{\varLambda}_{6}$, and 
$\hat{\varLambda}_{15}$ commute with each other. 
The operators $\hat{\varLambda}{}_{b}$ and $\hat{s}$ constitute  
a basis of the Schwinger representation of the ${\rm U}(2,2)$ Lie algebra.

\section{The Penrose transform of $\boldsymbol{f_{k,l,m,n}}$ in the case (a)} 

In this appendix, we demonstrate the Penrose transform \cite{PM, PR, HT, Tak} of 
the following twistor function in the case (a):
\begin{align}
f_{k,l,m,n} (\alpha)
= C_{k,l,m,n} \frac{(\alpha^{2})^{m} (\alpha^{3})^{n}} {(\alpha^{0})^{-k} (\alpha^{1})^{-l}} \,, 
\quad   
k, l \in  \Bbb{Z}^{-}, \, m, n \in \Bbb{N}_{0} \,, 
\label{B.1}
\end{align}
where $C_{k,l,m,n}$ is given in Eq. (\ref{5.20}). 
This function can be written, in terms of $\omega^{\alpha}$ and $\pi_{\dot{\alpha}}$, as 
\begin{align}
f_{k,l,m,n} (\omega^{\alpha}, \pi_{\dot{\alpha}})
= C_{k,l,m,n} \frac{2^{s+1} 
(-\omega^{0}+\pi_{\dot{0}})^{m} (-\omega^{1}+\pi_{\dot{1}})^{n}}
{(\omega^{0}+\pi_{\dot{0}})^{-k} (\omega^{1}+\pi_{\dot{1}})^{-l}} \,, 
\label{B.2}
\end{align}
or, in terms of $z^{\alpha \dot{\alpha}}$, $\pi_{\dot{0}}$ and $\zeta := \pi_{\dot{1}}/\pi_{\dot{0}}\:\!$, as 
\begin{align}
f_{k,l,m,n}(z, \pi_{\dot{0}}, \zeta) 
&= C_{k,l,m,n} 
\dfrac{2^{s+1}  (-iz^{0\dot{1}})^{m} (-iz^{1\dot{1}}+1)^{n}} 
{(\pi_{\dot{0}})^{2s+2}\:\!  (iz^{0 \dot{1}})^{-k} (iz^{1 \dot{1}} + 1)^{-l}}
\notag
\\
& \quad\; 
\times 
\frac{\left( \zeta + \dfrac{-iz^{0\dot{0}}+1}{-iz^{0\dot{1}}}\right)^{m} 
\left( \zeta + \dfrac{-iz^{1\dot{0}}}{-iz^{1\dot{1}}+1}\right)^{n} }
{\left( \zeta + \dfrac{iz^{0 \dot{0}} + 1}{iz^{0 \dot{1}}} \right)^{-k}
\left( \zeta + \dfrac{iz^{1 \dot{0}}}{iz^{1 \dot{1}} + 1} \right)^{-l} } \,. 
\label{B.3}
\end{align}
Here, Eq. (\ref{4.6}) has been used. 
In what follows, we individually perform the Penrose transform of $f_{k,l,m,n}$ 
in the cases of zero helicity ($s=0$), positive helicity ($s>0$), and negative helicity ($s<0$).

\subsection{Case $\boldsymbol{s=0}$}

In this case, the Penrose transform is readily carried out by using Cauchy's theorem:  
\begin{align}
\phi^{(a)}_{k,l,m,n} (z) &:= \frac{1}{2\pi i} \oint_{\varGamma_{z}}   f_{k,l,m,n}(z, \pi_{\dot{0}}, \zeta) 
\pi_{\dot{\beta}} d\pi^{\dot{\beta}} 
\notag
\\
&\; = C_{k,l,m,n} 
\dfrac{2(-iz^{0\dot{1}})^{m} (-iz^{1\dot{1}}+1)^{n}} 
{(iz^{0 \dot{1}})^{-k} (iz^{1 \dot{1}} + 1)^{-l}} \frac{1}{(-l-1)!}
\notag
\\
& \quad\; 
\times 
\left( \dfrac{d}{d\zeta} \right)^{-l-1} 
\frac{\left( \zeta + \dfrac{-iz^{0\dot{0}}+1}{-iz^{0\dot{1}}}\right)^{m} 
\left( \zeta + \dfrac{-iz^{1\dot{0}}}{-iz^{1\dot{1}}+1}\right)^{n} }
{\left( \zeta + \dfrac{iz^{0 \dot{0}} + 1}{iz^{0 \dot{1}}} \right)^{-k} } \Bigg|_{ \zeta=\zeta_1}  \,, 
\label{B.4}
\end{align}
where $\zeta_1:=-iz^{1 \dot{0}}/(iz^{1 \dot{1}} + 1)$. 
This expression has been found by choosing the intersection point $Q^{1}$ 
as the only pole surrounded by $\varGamma_{z}$.

\subsection{Case $\boldsymbol{s>0}$}

In this case, $s$ takes either positive integer or positive half-integer values, 
and accordingly the Penrose transform of $f_{k,l,m,n}$ is given by 
\begin{align}
\phi^{(a)}_{k,l,m,n; \;\! \dot{\alpha}_{1} \ldots \dot{\alpha}_{2s}} (z) 
&:= \frac{1}{2\pi i} \oint_{\varGamma_{z}} \pi_{\dot{\alpha}_{1}} \cdots \pi_{\dot{\alpha}_{2s}}
f_{k,l,m,n}(z, \pi_{\dot{0}}, \zeta) 
\pi_{\dot{\beta}} d\pi^{\dot{\beta}} \,.
\label{B.5}
\end{align}
Because $\phi^{(a)}_{k,l,m,n; \;\! \dot{\alpha}_{1} \ldots \dot{\alpha}_{2s}}$ is 
a totally symmetric spinor of rank $2s$, it is sufficient if we consider the components such that 
$\dot{\alpha}_{1}=\cdots=\dot{\alpha}_{2s-r}=\dot{0}$ 
and $\dot{\alpha}_{2s-r+1}=\cdots=\dot{\alpha}_{2s}=\dot{1}$ 
$(r=0, 1, \ldots, 2s)$. The integrand in Eq. (\ref{B.5}) can explicitly be written as
\begin{align}
& \pi_{\dot{\alpha}_{1}} \cdots \pi_{\dot{\alpha}_{2s}}
f_{k,l,m,n}(z, \pi_{\dot{0}}, \zeta) 
\notag
\\
&= C_{k,l,m,n} 
\dfrac{2^{s+1} \zeta^{r} (-iz^{0\dot{1}})^{m} (-iz^{1\dot{1}}+1)^{n}} 
{(\pi_{\dot{0}})^{2}\:\!  (iz^{0 \dot{1}})^{-k} (iz^{1 \dot{1}} + 1)^{-l}}
\notag
\\
& \quad\; 
\times 
\frac{\left( \zeta + \dfrac{-iz^{0\dot{0}}+1}{-iz^{0\dot{1}}}\right)^{m} 
\left( \zeta + \dfrac{-iz^{1\dot{0}}}{-iz^{1\dot{1}}+1}\right)^{n} }
{\left( \zeta + \dfrac{iz^{0 \dot{0}} + 1}{iz^{0 \dot{1}}} \right)^{-k}
\left( \zeta + \dfrac{iz^{1 \dot{0}}}{iz^{1 \dot{1}} + 1} \right)^{-l} } \,. 
\label{B.6}
\end{align}
Then, its contour integration around $Q^{1}$ leads to 
\begin{align}
&\phi^{(a)}_{k,l,m,n; \;\! \dot{\alpha}_{1} \ldots \dot{\alpha}_{2s}} (z) 
\notag
\\
&= C_{k,l,m,n} 
\dfrac{2^{s+1} (-iz^{0\dot{1}})^{m} (-iz^{1\dot{1}}+1)^{n}} 
{(iz^{0 \dot{1}})^{-k} (iz^{1 \dot{1}} + 1)^{-l}} \frac{1}{(-l-1)!}
\notag
\\
& \quad\; 
\times 
\left( \dfrac{d}{d\zeta} \right)^{-l-1} 
\frac{\zeta^{r} \left( \zeta + \dfrac{-iz^{0\dot{0}}+1}{-iz^{0\dot{1}}}\right)^{m} 
\left( \zeta + \dfrac{-iz^{1\dot{0}}}{-iz^{1\dot{1}}+1}\right)^{n} }
{\left( \zeta + \dfrac{iz^{0 \dot{0}} + 1}{iz^{0 \dot{1}}} \right)^{-k} } \Bigg|_{ \zeta=\zeta_1}  \,.  
\label{B.7}
\end{align}

\subsection{Case $\boldsymbol{s<0}$}

In this case, $s$ takes either negative integer or negative half-integer values, 
and accordingly the Penrose transform of $f_{k,l,m,n}$ is given by 
\begin{align}
\phi^{(a)}_{k,l,m,n; \;\! \alpha_{1} \ldots \alpha_{-2s}} (z) 
&:= \frac{1}{2\pi i} \oint_{\varGamma_{z}} 
\frac{\partial}{\partial \omega^{\alpha_1}} \cdots 
\frac{\partial}{\partial \omega^{\alpha_{-2s}}} 
f_{k,l,m,n}(z, \pi_{\dot{0}}, \zeta) 
\pi_{\dot{\beta}} d\pi^{\dot{\beta}} \,.
\label{B.8}
\end{align}
Because $\phi^{(a)}_{k,l,m,n; \;\! \alpha_{1} \ldots \alpha_{-2s}}$ is 
a totally symmetric spinor of rank $-2s$, it is sufficient if we consider the components such that 
$\alpha_{1}=\cdots=\alpha_{-2s-r}=0$ 
and ${\alpha}_{-2s-r+1}=\cdots={\alpha}_{-2s}=1$ 
$(r=0, 1, \ldots, -2s)$. 
For our calculation, it is convenient to exploit the formula 
\begin{align}
& \bigg( \frac{\partial}{\partial\omega} \bigg)^{\! h} \!\:
\frac{(-\omega+\pi)^{p}}{(\omega+\pi)^{q}} 
=(-1)^{h} h! \sum_{\hat{\imath}\:\!=0}^{h} \binom{p}{h-\hat{\imath}\:\!} \binom{q+\hat{\imath}-1}{\hat{\imath}} 
\frac{(-\omega+\pi)^{p-h+\hat{\imath}}}{(\omega+\pi)^{q+\hat{\imath}}} 
\label{B.9}
\end{align}
valid for $h, \;\! p\in \Bbb{N}_{0}$, and $q \in \Bbb{Z}^{+}$. 
Here 
\begin{align}
\binom{p}{h-\hat{\imath}\:\!}=0 \;\;\;\;  \mbox{for $\; p<h-\hat{\imath}$} 
\label{B.10}
\end{align}
is to be understood. Using Eq. (\ref{B.9}), we can obtain  
\begin{align}
& \frac{\partial}{\partial \omega^{\alpha_1}} \cdots 
\frac{\partial}{\partial \omega^{\alpha_{-2s}}} 
f_{k,l,m,n}(\omega^{\alpha}, \pi_{\dot{\alpha}}) 
\notag
\\
&= C_{k,l,m,n}\:\! 2^{s+1} 
\bigg( \frac{\partial}{\partial\omega^{0}} \bigg)^{\! -2s-r} \!\:
\frac{(-\omega^{0}+\pi_{\dot{0}})^{m}}{(\omega^{0}+\pi_{\dot{0}})^{-k}} 
\!\: \bigg( \frac{\partial}{\partial\omega^{1}} \bigg)^{\! r} \!\:
\frac{(-\omega^{1}+\pi_{\dot{1}})^{n}}{(\omega^{1}+\pi_{\dot{1}})^{-l}} 
\notag
\\
&=C_{k,l,m,n}\:\! 2^{s+1} (-1)^{-2s} (-2s-r)! \;\! r! 
\notag
\\
&\quad\;
\times 
\sum_{\hat{\imath}\:\!=0}^{-2s-r} \sum_{\hat{\jmath}\;\!=0}^{r}
\binom{m}{-2s-r-\hat{\imath}\:\!} \binom{-k+\hat{\imath}-1}{\hat{\imath}} 
\binom{n}{r-\hat{\jmath}\;\!} \binom{-l+\hat{\jmath}-1}{\hat{\jmath}} 
\notag
\\
&\quad\;  
\times
\frac{(-\omega^{0}+\pi_{\dot{0}})^{m+2s+r+\hat{\imath}} (-\omega^{1}+\pi_{\dot{1}})^{n-r+\hat{\jmath}}}
{(\omega^{0}+\pi_{\dot{0}})^{-k+\hat{\imath}} (\omega^{1}+\pi_{\dot{1}})^{-l+\hat{\jmath}}} \,. 
\label{B.11}
\end{align}
In terms of $z^{\alpha \dot{\alpha}}$, $\pi_{\dot{0}}$ and $\zeta$, 
Eq. (\ref{B.11}) can be written as  
\begin{align}
&\frac{\partial}{\partial \omega^{\alpha_1}} \cdots 
\frac{\partial}{\partial \omega^{\alpha_{-2s}}} 
f_{k,l,m,n}(z, \pi_{\dot{0}}, \zeta) 
\notag 
\\
&=C_{k,l,m,n}\:\! 2^{s+1} (-2s-r)! \;\! r! 
\notag
\\
&\quad\;
\times 
\sum_{\hat{\imath}\:\!=0}^{-2s-r} \sum_{\hat{\jmath}\;\!=0}^{r}
\binom{m}{-2s-r-\hat{\imath}\:\!} \binom{-k+\hat{\imath}-1}{\hat{\imath}} 
\binom{n}{r-\hat{\jmath}\;\!} \binom{-l+\hat{\jmath}-1}{\hat{\jmath}} 
\notag
\\
&\quad\; 
\times 
\dfrac{(-1)^{m+r+\hat{\imath}} (iz^{0\dot{1}})^{m+2s+r+k} (-iz^{1\dot{1}}+1)^{n-r+\hat{\jmath}}} 
{(\pi_{\dot{0}})^{2}\:\!  (iz^{1 \dot{1}} + 1)^{-l+\hat{\jmath}}}
\notag
\\
& \quad\; 
\times 
\frac{\left( \zeta + \dfrac{-iz^{0\dot{0}}+1}{-iz^{0\dot{1}}}\right)^{m+2s+r+\hat{\imath}} 
\left( \zeta + \dfrac{-iz^{1\dot{0}}}{-iz^{1\dot{1}}+1}\right)^{n-r+\hat{\jmath}} }
{\left( \zeta + \dfrac{iz^{0 \dot{0}} + 1}{iz^{0 \dot{1}}} \right)^{-k+\hat{\imath}}
\left( \zeta + \dfrac{iz^{1 \dot{0}}}{iz^{1 \dot{1}} + 1} \right)^{-l+\hat{\jmath}} } \,. 
\label{B.12}
\end{align}
Then, its contour integration around $Q^{1}$ leads to 
\begin{align}
&\phi^{(a)}_{k,l,m,n; \;\! \alpha_{1} \ldots \alpha_{-2s}} (z) 
\notag
\\
&=C_{k,l,m,n}\:\! 2^{s+1} (-2s-r)! \;\! r! 
\notag
\\
&\quad\;
\times 
\sum_{\hat{\imath}\:\!=0}^{-2s-r} \sum_{\hat{\jmath}\;\!=0}^{r}
\binom{m}{-2s-r-\hat{\imath}\:\!} \binom{-k+\hat{\imath}-1}{\hat{\imath}} 
\binom{n}{r-\hat{\jmath}\;\!} \binom{-l+\hat{\jmath}-1}{\hat{\jmath}} 
\notag
\\
&\quad\; 
\times 
\dfrac{(-1)^{m+r+\hat{\imath}} (iz^{0\dot{1}})^{m+2s+r+k} (-iz^{1\dot{1}}+1)^{n-r+\hat{\jmath}}} 
{(iz^{1 \dot{1}} + 1)^{-l+\hat{\jmath}}} 
\frac{1}{(-l+\hat{\jmath}-1)!}
\notag
\\
& \quad\; 
\times 
\left( \dfrac{d}{d\zeta} \right)^{-l+\hat{\jmath}-1} 
\frac{\left( \zeta + \dfrac{-iz^{0\dot{0}}+1}{-iz^{0\dot{1}}}\right)^{m+2s+r+\hat{\imath}} 
\left( \zeta + \dfrac{-iz^{1\dot{0}}}{-iz^{1\dot{1}}+1}\right)^{n-r+\hat{\jmath}} }
{\left( \zeta + \dfrac{iz^{0 \dot{0}} + 1}{iz^{0 \dot{1}}} \right)^{-k+\hat{\imath}} } \Bigg|_{ \zeta=\zeta_1}  \,.  
\label{B.13}
\end{align}

From Eqs. (\ref{B.4}), (\ref{B.7}), and (\ref{B.13}), we see that 
in the case (a), the massless field obtained by the Penrose transform of any arbitrary $f_{k,l,m,n}$ 
takes the form of a sum of monomial functions each of which is proportional to a negative power of 
$(z_{\mu}-u_{\mu})(z^{\mu}-u^{\mu})$. 
Then it can be shown that the resulting massless field possesses no singularities other than those specified by 
$(z_{\mu}-u_{\mu})(z^{\mu}-u^{\mu})=0$. 
Therefore, the massless fields derived here are recognized as a regular function on 
$\mathbb{C}\mathbf{M}^{+}$. 
An analysis similar to what has been done in this appendix 
can be performed in the cases (b) and (c\:\!$i$). 
In these cases, it can be shown that the Penrose transform of $f_{k,l,m,n}$ 
yields a massless field that possesses singularities in $\mathbb{C}\mathbf{M}^{+}$.


\newpage

\bibliography{your-bib-file}

\begin{thebibliography}{00}

\bibitem{Pen1}
R.~Penrose, ^^ ^^ Twistor algebra," J. Math. Phys. 8 (1967) 345.

\bibitem{PM}
R.~Penrose and M.~A.~H.~MacCallum, 
^^ ^^ Twistor theory: An approach to the quantisation of fields and space-time,"  
Phys. Rep. 6 (1973) 241. 

\bibitem{PR}
R.~Penrose and W.~Rindler, 
\textit{Spinors and Space-Time}, Vol.~2:  
Spinor and Twistor Methods in Space-Time Geometry, 
Cambridge Monographs on Mathematical Physics,  
(Cambridge University Press, Cambridge, 1986). 

\bibitem{HT}
S.~A.~Huggett and K.~P.~Tod, 
\textit{An Introduction to Twistor Theory}, Second Edition, 
London Mathematical Society, Student Texts 4 
(Cambridge University Press, Cambridge, 1994). 

\bibitem{Tak}
K.~Takasaki, 
\textit{The World of Twistors} 
(Kyoritsu Shuppan Co., Ltd., Tokyo, 2005, in Japanese). 

\bibitem{Hug}
L.~P.~Hughston, 
\textit{Twistors and Particles}, Lecture Notes in Physics 97 
(Springer-Verlag, Berlin, 1979). 

\bibitem{Pen2}
R.~Penrose, ^^ ^^ The twistor programme," Rep. Math. Phys. 12 (1977) 65.

\bibitem{Pen3}
R.~Penrose, ^^ ^^ The central programme of twistor theory,'' Chaos, Solitons and Fractals 10 (1999) 581. 

\bibitem{Wit}
E.~Witten, 
^^ ^^ Perturbative gauge theory as a string theory in twistor space,'' 
Commun. Math. Phys. 252 (2004) 189 [hep-th/0312171]. 

\bibitem{CSW1}
F.~Cachazo, P.~Svrcek and E.~Witten, 
^^ ^^ MHV vertices and tree amplitudes in gauge theory,''  
J. High Energy Phys. 09 (2004) 006 [hep-th/0403047]. 

\bibitem{CSW2}
F.~Cachazo, P.~Svrcek and E.~Witten, 
^^ ^^ Twistor space structure of one-loop amplitudes in
gauge theory,'' 
J. High Energy Phys. 10 (2004) 074 [hep-th/0406177]. 

\bibitem{Nai}
V.~P.~Nair, 
^^ ^^ A current algebra for some gauge theory amplitudes,"  
Phys. Lett. B 214 (1988) 215. 

\bibitem{ADHM}
M.~F.~Atiyah, V.~G.~Drinfeld, N.~J.~Hitchin and Yu.~I.~Manin, 
^^ ^^ Construction of instantons,"  
Phys. Lett. 65 A (1978) 185-187. 

\bibitem{AHS}
M.~F.~Atiyah, N.~J.~Hitchin and I.~M.~Singer, 
^^ ^^ Self-duality in four-dimensional Riemannian geometry,'' 
Proc. Roy. Soc. London A 362 (1978) 425.   

\bibitem{Pen4}
R.~Penrose, ^^ ^^ Nonlinear gravitons and curved twistor theory,''  
Gen. Rel. Grav. 7 (1976) 31-52. 

\bibitem{Hit}
N.~J.~Hitchin, ^^ ^^ Monopoles and geodesics,'' Commun. Math. Phys. 83 (1982) 579.

\bibitem{WW}
R.~S.~Ward and R.~O.~Wells,~Jr., 
\textit{Twistor Geometry and Field Theory}, 
Cambridge Monographs on Mathematical Physics   
(Cambridge University Press, Cambridge, 1990). 

\bibitem{BB}
T.~N.~Bailey and R.~J.~Baston (eds.), 
\textit{Twistors in Mathematics and Physics},  
London Mathematical Society, Lecture Note Series 156 
(Cambridge University Press, Cambridge, 1990). 

\bibitem{MW}
L.~J.~Mason and N.~M.~J.~Woodhouse, 
\textit{Integrability, Self-Duality and Twistor Theory}, 
London Mathematical Society Monographs, New Series 15 
(Oxford University Press, New York, 1996). 

\bibitem{Dun}
M.~Dunajski, 
\textit{Solitons, Instantons and Twistors},   
Oxford Graduate Texts in Mathematics 19 
(Oxford University Press, New York, 2010). 

\bibitem{Pen5}
R.~Penrose, ^^ ^^ Twistor quantisation and curved space-time,''  
Int. J. Theor. Phys. 1 (1968), 61. 

\bibitem{Pen6}
R.~Penrose, 
In \textit{Quantum Gravity}, an Oxford Symposium,
eds. C.~J.~Isham, R.~Penrose and D.~W.~Sciama,
(Oxford University Press, Oxford, 1975). 

\bibitem{EG}
M.~G.~Eastwood and M.~L.~Ginsberg, ^^ ^^ Duality in twistor theory,''
Duke Math. J. 48 (1981) 177. 

\bibitem{BE}
R.~J.~Baston and M.~G.~Eastwood, 
\textit{The Penrose Transform}: 
Its Interaction with Representation Theory, 
Oxford Mathematical Monographs 499 
(Clarendon Press, Oxford 1989); see also references therein. 

\bibitem{EP}
M.~G.~Eastwood and A. M. Pilato, 
^^ ^^ On the density of twistor elementary states," 
Pacific J. Math. 151 (1991) 201. 

\bibitem{Mul}
F.~M\"{u}ller, ``A minimum principle for the cohomological inner product on twistor space,'' 
Twistor Newsletter 39 (1995) 39. 

\bibitem{Kur}
H.~Kuratsuji, 
In \textit{Path Integrals and Coherent States of 
$\mathrm{SU}(2)$ and $\mathrm{SU}(1,1)$}, 
eds. A Inomata, H. Kuratsuji and C. C. Gerry, 
(World Scientific, Singapore, 1992). 

\bibitem{NO}
J.~W.~Negele and H.~Orland,
\textit{Quantum Many-Particle Systems},  
Frontiers in Physics 68 
(Addison-Wesley Publishing Company, Redwood City, California, 1987). 

\bibitem{HH}
A.~P.~Hodges and S.~A.~Huggett, ^^ ^^ Twistor diagrams,''
Surv. High Energy Phys. 1 (1980) 333. 

\bibitem{Hod1}
A.~P.~Hodges, ^^ ^^ Twistor diagrams," Physica 114A (1982) 157.

\bibitem{Hod2}
A.~P.~Hodges, ^^ ^^ Twistor diagrams and massless Moller scattering,'' 
Proc. Roy. Soc. London A 385 (1983) 207.

\bibitem{BP}
I.~Bars and M.~Pic\'{o}n, 
^^ ^^ Single twistor description of massless, massive, AdS, and other interacting particles,'' 
Phys. Rev. D {\bf 73} (2006) 064002 [hep-th/0512091]. 

\bibitem{Bar}
%
I.~Bars, 
^^ ^^ Lectures on twistors,"
 in Quantum Theory and Symmetries,
Proceedings of the 4th International Symposium, edited by V. K. Dobrev (Heron Press, Birmingham, 2006); e-print arXiv:hep-th/0601091.




\bibitem{DEN}
S.~Deguchi, T.~Egami and J.~Note,
^^ ^^ Spinor and twistor formulations of tensionless bosonic strings in four dimensions,'' 
Prog. Theor. Phys. 124 (2010) 969 [arXiv:1006.2438] 
(Appendix B). 

\bibitem{Yao}
T.~Yao, ^^ ^^ Unitary irreducible representations of $\mathrm{SU}(2,2)$. I,'' J. Math. Phys. 8 (1967) 1931. 

\bibitem{Pen7}
R.~Penrose, 
In \textit{Group Theory in Non-Linear Problems},
ed. A.~O.~Barut,
(D. Reidel Publishing Company, Dordrecht, 1974).

\bibitem{Wel} 
R.~O.~Wells,~Jr., 
^^ ^^ Complex manifolds and mathematical physics,"  
Bull. Amer. Math. Soc. (N.S.) 1 (1979) 296. 

\bibitem{Nak}
M.~Nakahara, 
\textit{Geometry, Topology and Physics}   
(IOP Publishing Ltd, Bristol, 1990).

\bibitem{IK}
I.~Bengtsson and K.~\.{Z}yczkowski, 
\textit{Geometry of Quantum States}: An Introduction to Quantum Entanglement 
(Cambridge University Press, New York, 1990).  

\bibitem{Ryd}
L.~H.~Ryder, 
\textit{Quantum Field Theory}, Second Edition 
(Cambridge University Press, Cambridge, 1996).

\bibitem{Kat}
T.~Kato, 
\textit{Perturbation Theory for Linear Operators}, 
(Springer-Verlag, Berlin, 1966). 

\bibitem{RS}
M.~Reed and B. Simon, 
\textit{Methods of Modern Mathematical Physics}, Vol.~1: 
Functional Analysis 
(Academic Press, New York, 1972). 

\bibitem{AE}
A.~Arai and H.~Ezawa, 
\textit{Mathematical Structure of Quantum Mechanics}, Vol.~1  
(Asakura Publishing Co., Ltd., Tokyo, 1999, in Japanese). 

\bibitem{Kre} 
M.~G.~Krein, ^^ ^^ Introduction to the geometry of indefinite J-spaces and to the theory of operators in those spaces,'' 
Amer. Math. Soc. Transl. 93 (1970) 103. 

\bibitem{AI}
T.~Ya.~Azizov and I.~S.~Iokhvidov, 
\textit{Linear operators in spaces with an indefinite metric}, 
(John Wiley \& Sons, Chichester, 1989). 

\bibitem{Rod}
L.~Rodman, ^^ ^^ Review: T. Ya. Azizov and I. S. Iokhvidov, Linear operators in spaces with an indefinite metric,'' 
Bull. Amer. Math. Soc. (N.S.) 25 (1991) 111. 

\end{thebibliography}

\end{document}